\documentclass[aps,prb,numerical,sd,amsmath,amssymb,reprint,graphicx, floatfix]{revtex4-2}

\usepackage{graphicx}
\usepackage{dcolumn}
\usepackage{bm}

\usepackage[utf8]{inputenc}
\usepackage[T1]{fontenc}
\usepackage{mathptmx}

\usepackage{xcolor}
\usepackage{soul}
\usepackage{amsmath}

\usepackage[unicode]{hyperref}

\hypersetup{colorlinks=true,
    linkcolor=purple,
    citecolor=purple, urlcolor=purple, breaklinks=true}

\usepackage{multirow}
\usepackage{colortbl}
\usepackage{booktabs}
\usepackage{longtable}
\usepackage{makecell,tabularx}
\usepackage{layout}

\usepackage[normalem]{ulem}

\usepackage{cancel}

\begin{document}


\title{Higher-order topological bound states in the continuum in a topoelectrical lattice with long-range coupling}



\author{Araceli \surname{Guti\'{e}rrez--Llorente}}
\email[]{araceli.gutierrez@urjc.es}
\affiliation{Universidad Rey Juan Carlos, Escuela Superior de Ciencias Experimentales y Tecnolog\'{i}a, Madrid 28933, Spain}



\begin{abstract}

Linear electric circuits composed of inductors and capacitors can serve as analogues of tight-binding models that describe the electronic band structure of materials. This mapping provides a versatile approach for exploring topological phenomena within engineered electrical lattices.  In this work, the two-dimensional Su-Schrieffer-Heeger model is examined through electric circuit analogues to study the interplay between higher-order topology, bound states in the continuum, and disorder.  Building upon this model, the effect of introducing next-nearest-neighbour interactions that preserve chiral and spatial symmetries of the system is analyzed.  The results reveal that even without Hamiltonian separability, corner-localized bound states in the continuum remain protected by symmetry in the long-range coupled lattice.  This robustness highlights the potential of circuit-based platforms for probing advanced topological phenomena in a highly controllable setting.

\end{abstract}


\maketitle 

\section{Introduction}

The behaviour of electrons in solids is determined by the topological properties of the energy bands formed by their wavefunctions under the periodic potential of the crystal lattice.\cite{berry:84, thouless:82} This topological perspective has profoundly transformed our understanding of electronic states and phases of matter,\cite{wen:17, tokura:19, wieder:22} driven by groundbreaking discoveries such as the quantum Hall effects and topological insulators.\cite{klitzing:80, haldane:88, kane:05, bernevig:06, konig:07, fu:07, moore:07, hsieh:08, roy:09, moore:10}

While topology first entered physics through the quantum mechanical description of electrons in solids, topological band structures are fundamentally a property of waves propagating through periodic media.\cite{haldane:08}  Notably, these structures do not depend on strong electron–electron interactions or low dimensionality, and they apply equally to both electrons and classical waves. This conceptual breadth has opened the door to exploring topological phenomena across a wide range of physical systems,\cite{ozawa:19, zhang:23, zhu:23, xue:22, ma:19, huber:16,shah:24, khanikaev:24} in which the primary goal is to harness the disorder-insensitive transport of classical waves, whether in the form of sound, light or mechanical vibrations.

Classical systems, such as photonic and phononic metamaterials, as well as electric circuits, designed as analogues of a variety of quantum phenomena encountered in condensed-matter physics, offer enhanced experimental control for emulating topological effects. These platforms make it possible to observe key features of topological excitations, including robustness against disorder or localization at system boundaries, more readily than in topological quantum materials, in which realizing the required symmetries can be very difficult and their experimental confirmation, technically challenging.

Among the classical platforms for emulating topological phenomena, electric inductor-capacitor (LC) circuits, referred to as topoelectrical circuits, have emerged as a particularly versatile medium.  These systems can reproduce tight-binding models with complex hopping configurations, enabling the realization of topological band structures within a purely classical framework. In topoelectrical circuits, each node is analogous to a lattice site, while capacitive and inductive connections between nodes of the circuit mimic hopping amplitudes between sites of a tight-binding Hamiltonian in condensed-matter physics, without direct physical equivalence to quantum tunneling.  This analogy captures the mathematical structure of the model without implying a direct physical equivalence to quantum tunneling, which does not occur in circuits.  For a comprehensive overview, readers may refer to recent review articles on this topic.\cite{zhao:18, dong:21, kotwal:21, yang:24, sahin:25, chen:25}

Topologically nontrivial band structures have been studied across a wide variety of electric circuits.  In 2015, works by Jia \textit{et al.} \cite{ningyuan:15} and Albert \textit{et al.} \cite{albert:15} implemented circuit-based analogues of quantum spin Hall states.  Since then, the field has expanded to include classical realizations of Chern insulators, in which active elements like operational amplifiers are employed to break time reversal symmetry,\cite{haenel:19, hofmann:19} and Weyl semimetal phases, featuring circuit-based Weyl band structures.\cite{lee:18, lu:19, Rafi-Ul-Islam:20}

Higher-order topological insulators (HOTIs), characterized by protected boundary states of reduced dimensionality, have significantly broadened the landscape of topological materials in condensed matter physics.\cite{benalcazar:17, schindler:18, xie:21} By extending the conventional bulk–boundary correspondence, HOTIs offer deeper insights into band topology. Remarkably, these phases have also been realized in electric circuits, such as the two-dimensional quadrupole\cite{imhof:18, serra-garcia:19, lv:21}, three-dimensional octupole\cite{bao:19, liu:20} and four-dimensional hexadecapole\cite{zhang:20} insulators exhibiting topologically protected corner states, non-Hermitian systems engineered through the inclusion of resistors,\cite{ezawa:19:a}, higher-order topological Anderson insulators induced by disorder,\cite{zhang:21}  and square-root topological insulators, superconductors and semimetals.\cite{ezawa:18,ezawa:19:b,song:20, rafi-ul-islam:24, luo:25, song:25}

Altogether, these developments underscore the versatility of topoelectrical circuits in capturing complex topological phenomena and in designing systems that support
unconventional edge and corner states. By tuning the inductive and capacitive parameters, these circuits offer an exceptional degree of flexibility and control.

Second-order topological states, localized at the corners in a quadrupole insulator emerge within the bulk energy gap.  These states have been experimentally realized across a variety of platforms, including lattices of coupled microwave resonators,\cite{peterson:18} topoelectrical circuits,\cite{imhof:18, lv:21} mechanical metamaterials,\cite{serra-garcia:18} nanophotonic silicon ring resonators,\cite{mittal:19} and acoustic metamaterials.\cite{xue:19}  In contrast, states fully localized at the corners can also appear embedded within the continuum of bulk modes, pinned to zero energy yet remaining uncoupled from the surrounding spectrum,\cite{cerjan:20} 
revealing a connection to bound states in the continuum (BICs), a phenomenon arising from distinct mechanisms and manifesting across both quantum and classical wave systems.\cite{zhen:14, hsu:16, kupriianov:19, bogdanov:19, benalcazar:20, azzam:21, koshelev:23}

On the other hand, the influence of next-nearest-neighbour coupling (NNN) on topological phases remains an active area of investigation. In particular, the introduction of NNN coupling in electrical circuits breaking chiral symmetry has been shown to give rise to an indirect gap phase.\cite{rafi-ul-islam:22}  Furthermore, NNN interactions have been reported to open a zero-energy band gap that hosts spectrally isolated corner states,\cite{olekhno:22} a behaviour closely resembling that of a quadrupole insulator, even when all couplings in the system share the same sign.

In this work, a previously unexplored configuration of NNN coupling in two-dimensional SSH models  is investigated within the framework of electrical circuit analogues, in which the NNN interactions are engineered to preserve both chiral and spatial symmetries. Within this framework, it is shown that the system hosts second-order localized corner modes that are embedded in the continuum yet remain decoupled from it.

The paper is organized as follows.  In section \ref{sec:2D-SSH} the effects of disorder on corner-localized BICs in a two-dimensional (2D) Su-Schrieffer-Heeger (SSH) model with nearest-neighbour (NN) interaction are studied.  Section \ref{sec:2D-SSH-NNN} is devoted to the analysis of symmetry-preserving NNN coupling, studying how these interactions shape the topological phases and affect the robustness of BIC corner modes against disorder.

\section{Results and discussion}

\subsection{2D SSH electric circuit}
\label{sec:2D-SSH}

We start by first studying the circuit implementation of the 2D analogue of the SSH model, which exhibits higher-order topology. This model has been studied in various contexts, such as topological insulators in condensed-matter physics,\cite{liu:17, benalcazar:20} thermal transport,\cite{liu:24:prl} photonic crystals,\cite{xie:18} and a non-Abelian lattice.\cite{qian:24}  Prior studies have also examined the emergence of edge states in such circuit configuration.\cite{liu:19} In contrast, the focus here is on zero-admittance modes localized at the corners of the system, which are the hallmark of higher-order topological phases in the 2D circuit.

\begin{figure}[htb]
 \includegraphics[keepaspectratio=true, width=1.0\linewidth]{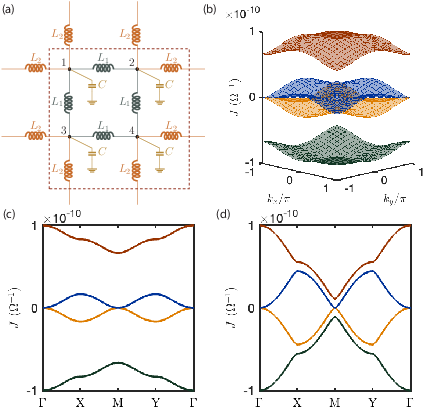}\caption{{\bf{Periodic 2D SSH circuit.}}{\bf{(a)}} Schematic of the unit cell of the 2D SSH circuit. Each unit cell hosts four sites, labeled 1-4. $L_1$ and $L_2$ represent the intracell and intercell couplings, respectively, corresponding to the NN hopping.  The ratio $\lambda = L_1/L_2$ defines the condition for a nontrivial topological phase, occurring when $\lambda <1$. Each node is grounded by an on-site capacitor $C$. {\bf{(b)}} Bulk band structure with open boundary conditions for $\lambda =0.1$ and $C=100\;\mbox{pF}$.  {\bf{(c, d)}} Bulk admittance bands along a high-symmetry path in the Brillouin zone for the circuit with periodic boundaries for $\lambda=0.1$ (c) and $\lambda=0.8$ (d).}\label{Fig_01}
\end{figure}

To implement the model a square lattice is used with real and positive nearest-neighbour couplings represented by lossless inductors $L_1$ and $L_2$ (Fig.~\ref{Fig_01}(a)).  Each node is connected to the ground via a lossless capacitor $C$.  These lumped elements allow for the definition of unique terminal voltages and currents, enabling analysis of the circuit’s topological properties.

Assuming the system is linear and time-invariant,  a sinusoidal drive that oscillates as $e^{-i\omega t}$ in time $t$ with angular frequency $\omega$, and steady-state analysis, the lumped LC circuit elements can be described by their frequency-dependent complex admittances, $Y_L=1/{i\omega L}$ for inductors, and $Y_C=i\omega C$ for capacitors. Sweeping across a range of frequency values is conceptually equivalent to evaluating the Fourier transform. Voltages and currents in the circuit can then be determined using standard linear circuit analysis techniques, such as node analysis based on Kirchhoff’s current law applied to the four nodes of the unit cell in Fig.~\ref{Fig_01}(a):

\begin{equation}\label{eq:001:sum}
I_{a} = \sum\limits_{b \in \mathcal{N}(a)} Y_{a,\:b} (V_a-V_b)+Y_{a}^{g}V_a
\end{equation}

where the sum extends over the neighboring nodes connected to $a$, denoted by $\mathcal{N}(a)$; $Y_{a,b}$ is the admittance between nodes $a$ and $b$; $Y_{a}^{g}$ is the admittance to ground at node $a$; $V_a$, $V_{b}$ are voltages at nodes $a$ and $b$, respectively. For the circuit shown in Fig.~\ref{Fig_01}, $Y_{a,b}=1/i\omega\;L_{k}$ with $L_{k} \in \{L_1, L_2\}$, and $Y_{a}^{g}=Y_{C}$ for all unit-cell nodes in the periodic boundary configuration.  Therefore, the response of the circuit at a given frequency $\omega$ is governed by the relation $\mathbf{I} = J(\omega) \mathbf{V}$, where $J(\omega)$, known as the circuit Laplacian or admittance matrix, links the external input current $I$ at each node to the vector voltage $V$ at each node with respect to the ground.  Under periodic boundary conditions and applying the Bloch's theorem, this matrix can be expressed as

\begin{equation}\label{eq_001}
J(\omega, {\bf{k}}) = i \omega \left[ \left(C - \frac{2}{\omega^2 L_1} - \frac{2}{\omega^2 L_2}\right) \mathbb{I} +  H(\omega, {\bf{k}}) \right]  
\end{equation}

where $\mathbb{I}$ is the $4\times 4$ identity matrix and $H(\omega, {\bf{k}})$ is a Hermitian matrix having zero diagonal (see Supplementary Material for a detailed derivation of the circuit Laplacian).

At the frequency 

\begin{equation}\label{eq_res_frq}
\omega_0=\sqrt{\frac{2(L_1 + L_2)}{C L_1 L_2}}
\end{equation}

the diagonal terms in Eq.~\ref{eq_001} vanish and the Laplacian becomes purely off-diagonal,

\begin{equation}\label{eq:laplacian_02}
J(\omega_0, {\bf{k}})=  i \: \sqrt{\frac{C}{2L_1(\lambda +1)}} \:
\begin{bmatrix}
0 & j_{12}   & j_{13} & 0\\[6pt]
 j_{12}^{*} & 0 & 0 & j_{13}\\[6pt]
j_{13}^{*} & 0 & 0 & j_{12}\\[6pt]
0 & j_{13}^{*} & j_{12}^{*} & 0\\
\end{bmatrix}
\end{equation}

where $j_{12} = 1 + \lambda e^{-ik_x}$ and $j_{13} = 1 + \lambda e^{-ik_y}$, with $\lambda=L_1/L_2$. This representation allows the electrical properties of the periodic circuit to be mapped onto the eigenstates of $J(\omega_0, {\bf{k}})$ in Eq.~\ref{eq:laplacian_02}, which define the circuit's admittance band structure and mimic the Hamiltonian of the SSH model. Furthermore, the matrix in Eq.~\ref{eq:laplacian_02} is separable, as it can be written as a sum of two components, each depending solely on one momentum component $(k_x, k_y)$.\cite{hsu:16, robnik:86}

This approach is validated through numerical calculations of the admittance bands for the periodic circuit (Fig.~\ref{Fig_01}(b,c,d)), which match the band structure obtained by diagonalizing the tight-binding Hamiltonian in Refs.~\cite{liu:17, benalcazar:20}.  The circuit exhibits gapless bulk bands at zero admittance. For $\lambda \neq \pm1$, two bulk gaps open (Fig.~\ref{Fig_01}(c,d)), which close at the high-symmetry points $\bf{X}$, $\bf{Y}$, and $\bf{M}$ when $\lambda = \pm1$.

\begin{figure}[htb]
\includegraphics[keepaspectratio=true, width=1.0\linewidth]{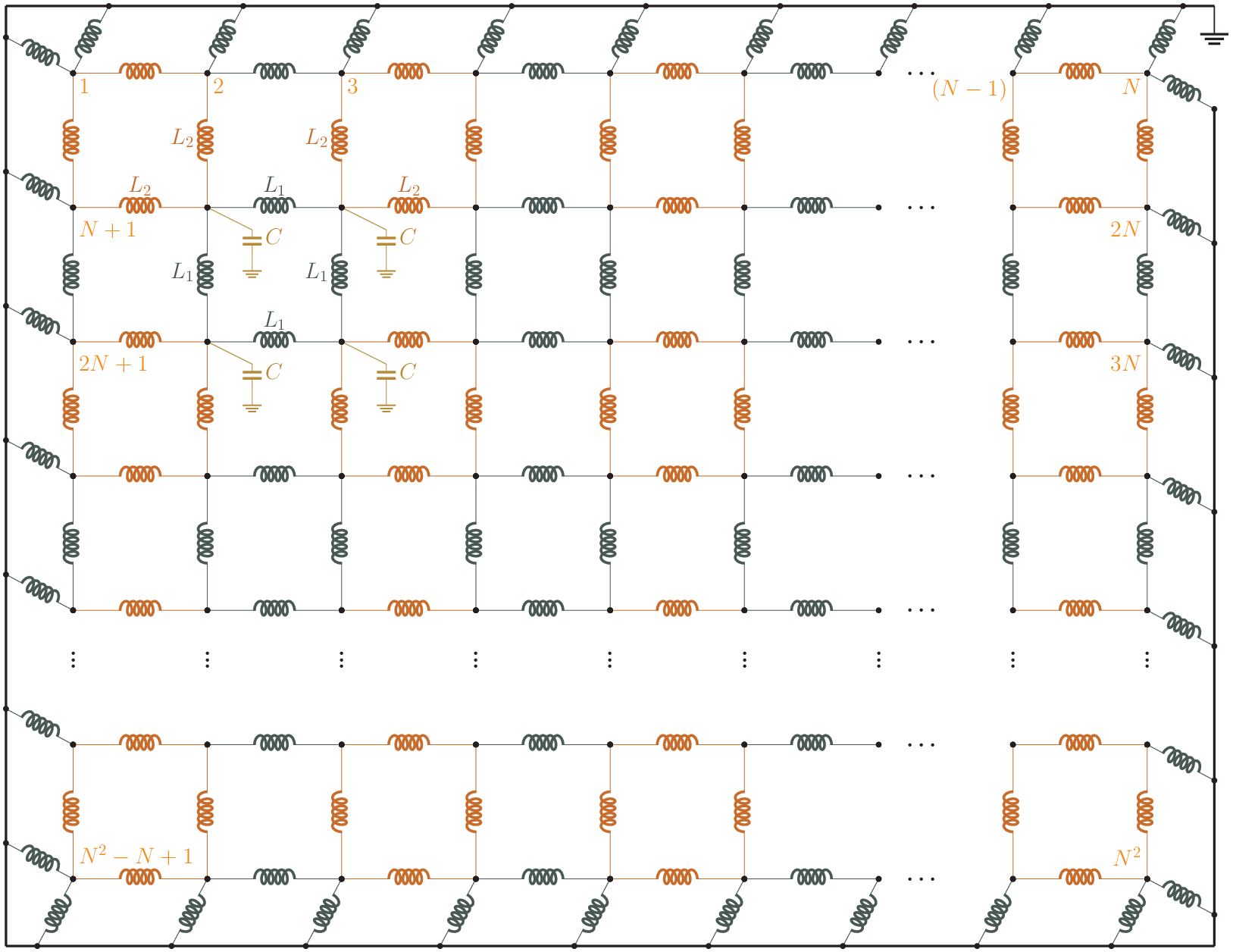}\caption{\textbf{Layout of the finite-sized 2D SSH circuit.} Inductors $L_1$ (green) and $L_2$ (brown) represent intracell and intercell couplings, respectively. All nodes (bulk, edge, and corner) are grounded via capacitors $C$ (yellow), which are omitted from the schematic for clarity except within the highlighted unit cell. The circuit contains $N^2$ nodes, with labels shown in orange.}\label{Fig_circuit_NN_finite}
\end{figure}

The study of the emergence of boundary phenomena, such as corner modes, requires a finite-sized configuration, which results in a discrete spectrum of eigenvalues. Fig.~\ref{Fig_circuit_NN_finite} displays a finite circuit with $N\times N$ nodes, where each node is labeled by an index $n$, where $n\in \{1,2,\dots, N^2\}$.  The open ends of the circuit are terminated with inductors $L_1$ in addition to capacitors $C$.  This arrangement ensures that the diagonal elements of the admittance matrix $J(\omega)$ vanish at the characteristic frequency $\omega = \omega_0$ (Eq.~\ref{eq_res_frq}).

In this configuration, the circuit Laplacian, $J(\omega)$, built using nodal analysis, is an $N^2\times N^2$ matrix. It can be expressed as $J(\omega)=D-A$, where $D$ is a diagonal matrix whose entries represent the total admittance from each node to ground and to the rest of the circuit,\cite{lee:18} analogous to on-site potentials in a tight-binding framework; and, the matrix $A$ is the admittance adjacency matrix, which is symmetric and has zeros along its diagonal. A detailed derivation of the admittance matrix for this configuration is provided in the Supplementary Material.

\begin{figure}[htpb!]
\includegraphics[keepaspectratio=true, width=1.0\linewidth]{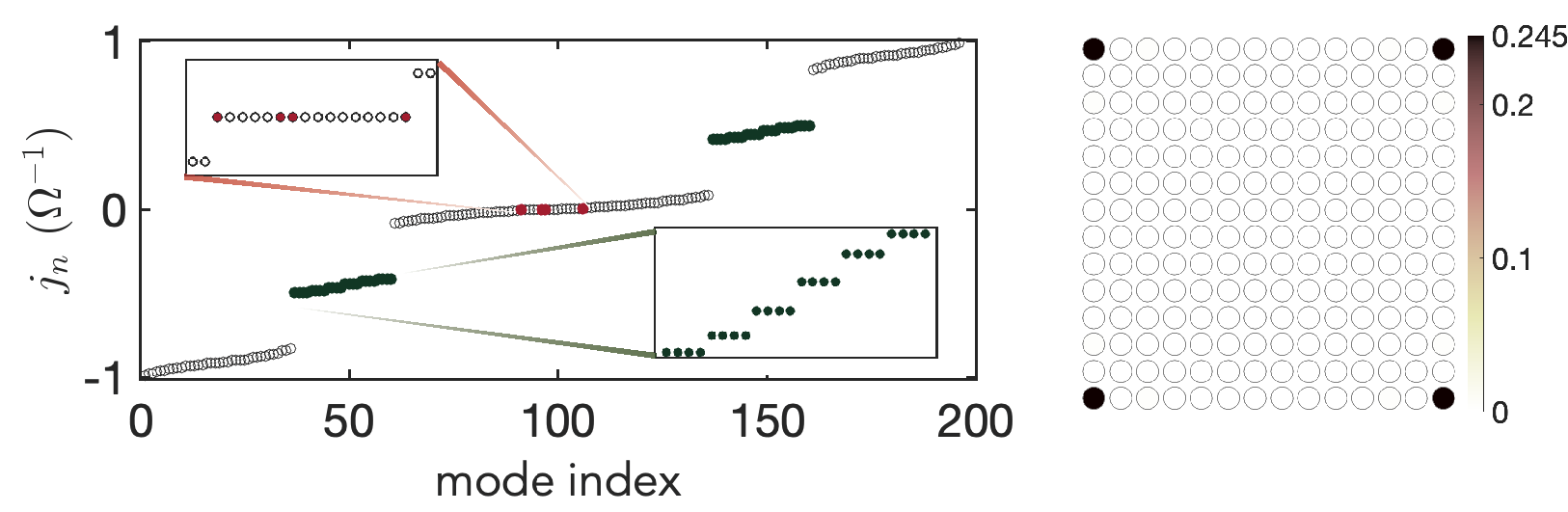}\caption{{\bf{Finite-sized 2D SSH circuit.}} Left panel: Numerically computed resonant admittance eigenvalues for the 2D SSH circuit comprising $196$ nodes ($14 \times 14$ node configuration), in the topological phase with $\lambda = 0.1$. In-gap edge-localized states are marked in green, while the four corner-localized modes in the continuum are highlighted in red (see also Fig.~S1-S2, and Table~S1, Supplementary Material).  Insets show enlarged views of these bound states.   Right panel: Normalized spatial distribution of the amplitude of a corner-localized eigenmode, $|\psi(\omega)|^2$. The colourmap quantifies the degree of localization across the circuit layout. The ground capacitance is set to $C=1\;\mbox{F}$ for numerical convenience, scaling the admittance spectrum to the range $[-1,\:1]$ without affecting the physical interpretation.}
\label{Fig_02}
\end{figure}

Let us consider a circuit with fully open boundaries comprising $196$ nodes, arranged in a $14 \times 14$ node configuration in its nontrivial topological phase ($\lambda = 0.1$).  Figure~\ref{Fig_02} displays the numerically computed admittance spectrum of eigenvalues.  By limiting interactions to nearest neighbours, the circuit preserves chiral symmetry, resulting in a spectrum symmetric around zero energy, with eigenvalues appearing in symmetric $\pm$ pairs. Furthermore, due to the presence of $C_{4v}$ symmetry, the spectrum in Fig.~\ref{Fig_02} has no bandgap at zero admittance.

This contrasts with the quadrupole insulator model, which features a bandgap at zero energy that arises from sign-alternating hopping terms in the Hamiltonian, enabling corner-localized modes within that gap.\cite{peterson:18, serra-garcia:18, serra-garcia:19, mittal:19, lv:21} In the present circuit, which does not include negative couplings, no bandgap exists at zero admittance; nevertheless, zero-admittance corner modes are observed.

In addition to the one-dimensional edge states within the bulk gap (marked as solid green circles in Fig.~\ref{Fig_02}), this system hosts corner-localized zero-admittance modes,highlighted in red. Figure~S1 (Supplementary Material) provides an extended discussion of these corner modes.  

Although these zero-dimensional corner modes are embedded within the bulk spectrum, they remain spatially confined and do not hybridize with bulk states at the same energy. As such, they are corner-localized BICs.\cite{benalcazar:20} To provide a visual comparison, Fig.~S2 shows the spatial distribution of the amplitude for selected bulk modes with zero admittance.

On the other hand, the observed localization length of the corner states is consistent with the expected exponential behaviour as a function of the coupling ratio $\lambda$. Fig.S3 (Supplementary Material) illustrates the decay of the wavefunction amplitude $\psi$ for a state localized at one of the corners, confined to a single sublattice within the bipartite lattice.  For comparison, Fig.~S4 (Supplementary Material) also displays the normalized spatial distribution of the amplitude for corner modes corresponding to different values of $\lambda$.

\begin{figure}[htpb!]
\includegraphics[keepaspectratio=true, width=0.91\linewidth]{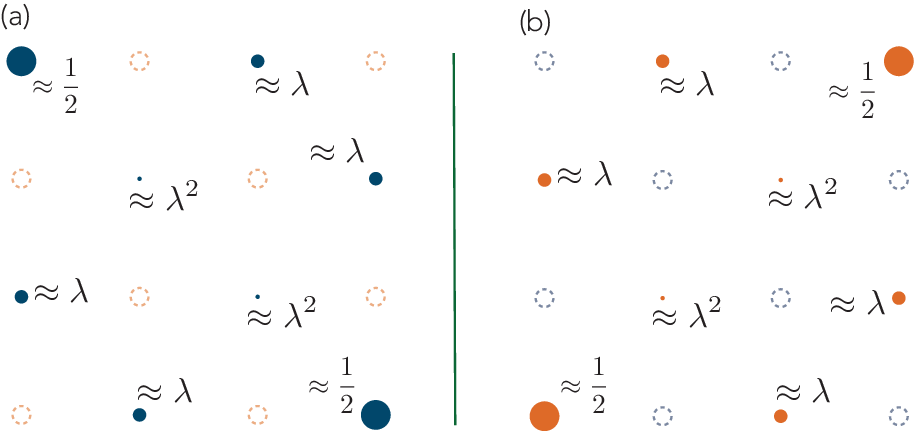}\caption{{\bf{Amplitude of the corner modes as a function of the coupling ratio $\lambda$ for $\lambda\ll1$.}} Panel (a) shows modes supported on sublattice $A$, and panel (b) shows modes on sublattice $B$ for a $4\times4$ lattice.  The amplitudes scale as $\lambda, \: \lambda^2$, consistent with exponential localization. The area of each circle representing a node is proportional to its amplitude for $\lambda=0.1$ , illustrating the strong localization at the corners. (Supplementary Material, section `Analytical derivation of corner modes')}\label{Fig_amplitudes_corner_modes_4x4}
\end{figure}

To complement the numerical results, an analytical derivation of the zero-energy corner modes for a finite lattice based on the 2D SSH model is provided in Supplementary Material.

The Laplacian can be brought into a block-off-diagonal form by reordering the nodes according to sublattice,

\begin{equation}\label{eq:J_4x4_chiral}
J=
\begin{pmatrix}
0 & Q\\
Q^{T} & 0
\end{pmatrix}
\end{equation}

The kernel subspaces of $Q$, $Q^{T}$ are then calculated, yielding basis vectors that span the set of zero-admittance states.  Corner-state wavefunctions are built as normalized linear combinations of these basis vectors.  Figure~\ref{Fig_amplitudes_corner_modes_4x4} shows the corner-mode amplitudes at each node for a $4\times4$ layout, approximated for a coupling ratio $\lambda \ll1$. These results exhibit the expected exponential decay away from the corner site, consistent with the strong localization characteristic of the nontrivial topological phase.  This analytical approach is extended to a $6\times 6$ layout in the Supplementary Material.

\begin{figure}[htpb!]
\includegraphics[keepaspectratio=true, width=1.0\linewidth]{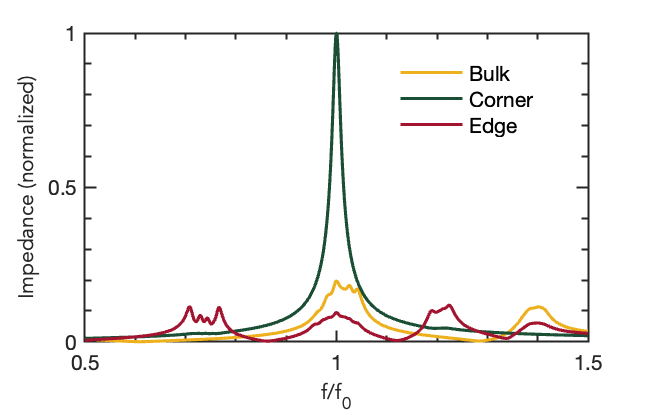}\caption{
{\bf{Normalized two-point impedance spectra.}} Calculated impedances for three representative configurations: \textit{Bulk}, between nodes $(N/2, N/2)$ and $(N/2, N/2+1)$; \textit{Corner}, between nodes $(1,1)$ and $(N,N)$; and, \textit{Edge}, between nodes $(1,N/2)$ and $(N/2, N/2)$. Computations used $L_1=10\: \mu \mbox{H}$, $L_2=100\: \mu \mbox{H}$, $C=10\:\mbox{nF}$, and $N=14$.}\label{Fig_impedance}
\end{figure}

To investigate circuit-specific signatures of corner states, two-point impedances $Z_{ab}(\omega)$ between nodes $a$ and $b$  are calculated over a frequency sweep from the frequency-dependent circuit Laplacian $J(\omega)$. Under current injection $I$, this impedance is expressed as $Z_{ab}(\omega)= (V_{a}(\omega)-V_{b}(\omega))/I$, where $V_a$, $V_b$ denote the node voltages at $a$ and $b$, respectively.  Alternatively, $Z_{ab}(\omega)$ can be obtained from the admittance eigenvalues $j_{k}$ with $k \in \{1, \: 2, \: \dots \: N^2\}$ in a $N\times N$ node configuration and eigenvectors $\psi_{k}(\omega)$ of the circuit Laplacian $J(\omega)$ (see Supplementary Information for details on its derivation):

\begin{equation}
Z_{a,b}(\omega) = \sum_k \frac{|\psi_k^{a}-\psi_k^{b}|^2}{j_k(\omega)}
\end{equation}

where $|\psi_{k}^{a}-\psi_{k}^{b}|$ represents the amplitude difference between nodes $a$ and $b$ in the admittance mode $\psi_{k}$.\cite{lee:18}  The two-point impedance diverges when an admittance eigenvalue $j_{k}$ approaches zero and the corresponding eigenmode $\psi_{k}$ has strong support at one probe but not the other, provided there is no dissipation.  To obtain finite impedance resonances and better approximate experimental peak widths in topoelectrical spectroscopy,\cite{franca:24} series resistances are introduced in inductors and capacitors, with admittances given by $Y_{L}(\omega) = 1/(R_{L}+i\omega L)$ and $Y_{C}(\omega)=i\omega C/(1+i\omega C R_C)$.  Figure~\ref{Fig_impedance} shows calculated impedances for three configurations: bulk, corner and edge nodes. The pronounced resonance observed for the corner-corner configuration (diagonal corners) reflects the presence of localized topological corner states in the circuit. The large impedance in this case originates from the eigenmode structure: both corners exhibit strong amplitude with opposite signs, as they belong to the same sublattice, which maximizes the voltage difference between the probes under current injection. This resonance provides a clear circuit-specific fingerprint of higher order topology, as the impedance peak reveals the spectral position and localization strength of the corner state.

\begin{figure}[htpb!]
 \includegraphics[keepaspectratio=true, width=1.0\linewidth]{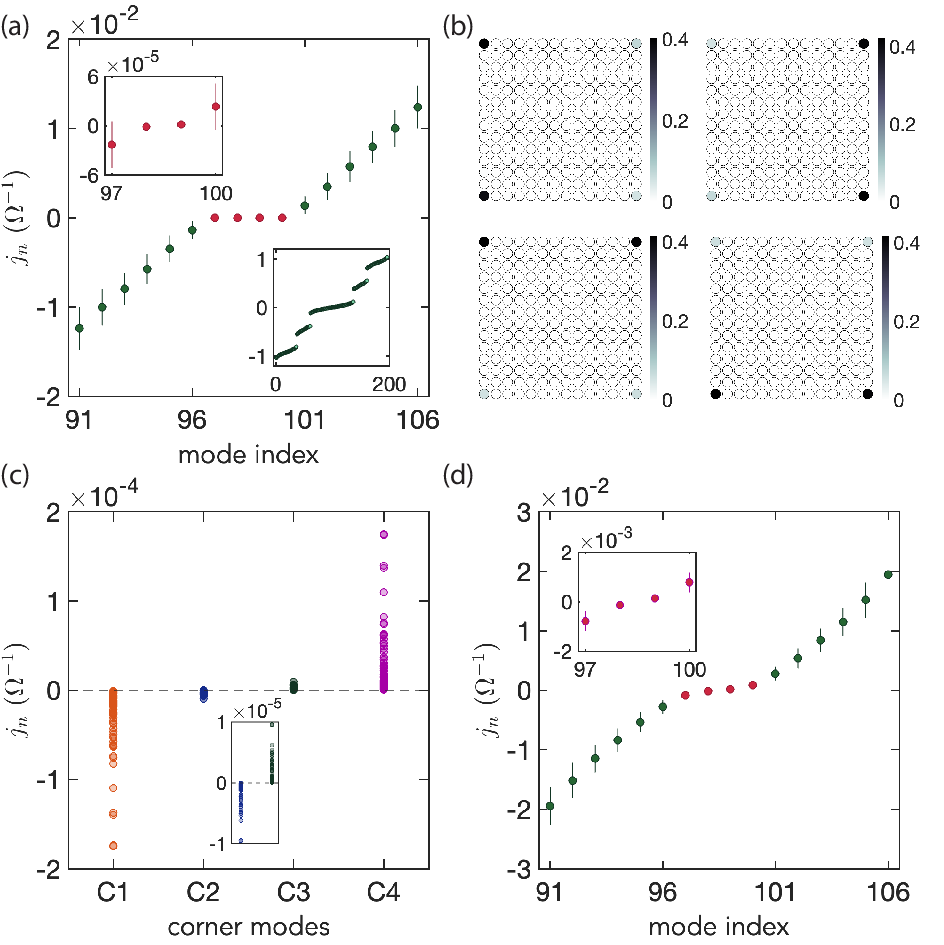}\caption{
 {\bf{Corner modes of the 2D SSH circuit under 15\% off-diagonal disorder.}} {\bf{(a)}} Numerically computed eigenvalues for $\lambda = 0.1$ (196 nodes). Each eigenvalue is averaged over 100 disorder simulations, with error bars indicating the standard deviation. Zero-admittance corner modes are highlighted in dark red. Insets: enlarged view of corner-localized states at zero-admittance (top left) and the full admittance spectrum (bottom right).  {\bf{(b)}} Normalized spatial distribution of the wave function amplitude, $|\psi(\omega)|^2$, for the four corner-localized modes, averaged over simulations exhibiting each localization pattern. The colourmap quantifies the degree of localization across the circuit layout. {\bf{(c)}}Eigenvalues corresponding to the four corner modes across 100 simulations; the inset shows an enlarged view of the two inner eigenvalues. {\bf{(d)}} Numerically computed eigenvalues for $\lambda = 0.35$; inset: enlarged view of the corner-localized states at zero-admittance.}\label{Fig_off_diagonal_disorder}
\end{figure}

To assess the robustness of corner-localized modes embedded within the continuum of bulk states, disorder is introduced into the system by randomly varying the inductance values of the links between nodes across the entire grid. This procedure modifies only the off-diagonal elements of the admittance matrix, as these inductances represent the couplings between nodes. A detailed description of the disorder implementation for this case is provided in the Supplementary Material.

Upon introducing off-diagonal disorder into the circuit, the spatial localization of these modes at the corners remains remarkably robust.  Figure~\ref{Fig_off_diagonal_disorder}(a) displays computed eigenvalues for a coupling ratio $\lambda=0.1$ for a disorder strength of 15\%.  Since this disorder affects only the off-diagonal elements of the admittance matrix, chiral symmetry is preserved, and the corner modes stay at zero admittance.  These modes remain embedded within the continuum, yet do not hybridize with bulk states.

Figure~\ref{Fig_off_diagonal_disorder}(b) shows the normalized spatial profiles of the corner mode amplitudes, $|\psi(\omega)|^2$, exhibiting strong confinement for $\lambda=0.1$ even under 15\% disorder. Figure~\ref{Fig_off_diagonal_disorder}(c) presents the detailed distribution of eigenvalues corresponding to the four corner modes observed for a 15\% off-diagonal disorder and $\lambda=0.1$ across 100 simulations. These eigenvalues exhibit a symmetric distribution around zero, forming two pairs of equal magnitude but opposite sign. Additional simulations for $\lambda=0.35$ confirm that the corner eigenvalues continue to lie at zero admittance under off-diagonal disorder (Fig.~\ref{Fig_off_diagonal_disorder}(d).

Furthermore, results for 40\% off-diagonal disorder are presented in Fig.~S9 (Supplementary Material), for which the bulk gap protecting edge states collapses, leading to the disappearance of edge modes. Despite this, the corner modes remain pinned at zero admittance, preserving their topological character.

The robustness of corner states in the 2D SSH topoelectrical circuit is consistent with the findings in Ref. \cite{cerjan:20}, which investigates the protection mechanisms of bound states in the continuum within the framework of crystalline topological phases. That study demonstrates that boundary-localized states can remain confined, even when their energies lie within the bulk continuum, provided certain symmetries are preserved. Specifically, both $C_{4v}$ symmetry (which encompasses fourfold rotational and mirror symmetries) and chiral symmetry are required to protect bound states in the continuum. Notably, this symmetry-based scheme does not rely on the separability of the Hamiltonian. In combination, chiral and $C_{4v}$ symmetries not only lead to gapless bulk bands at zero energy, but also pin the corner-localized states to zero energy, ensuring their degeneracy with bulk bands while simultaneously protecting them from hybridization.\cite{benalcazar:20}

\begin{figure}[htpb!]
\includegraphics[keepaspectratio=true, width=1.0\linewidth]{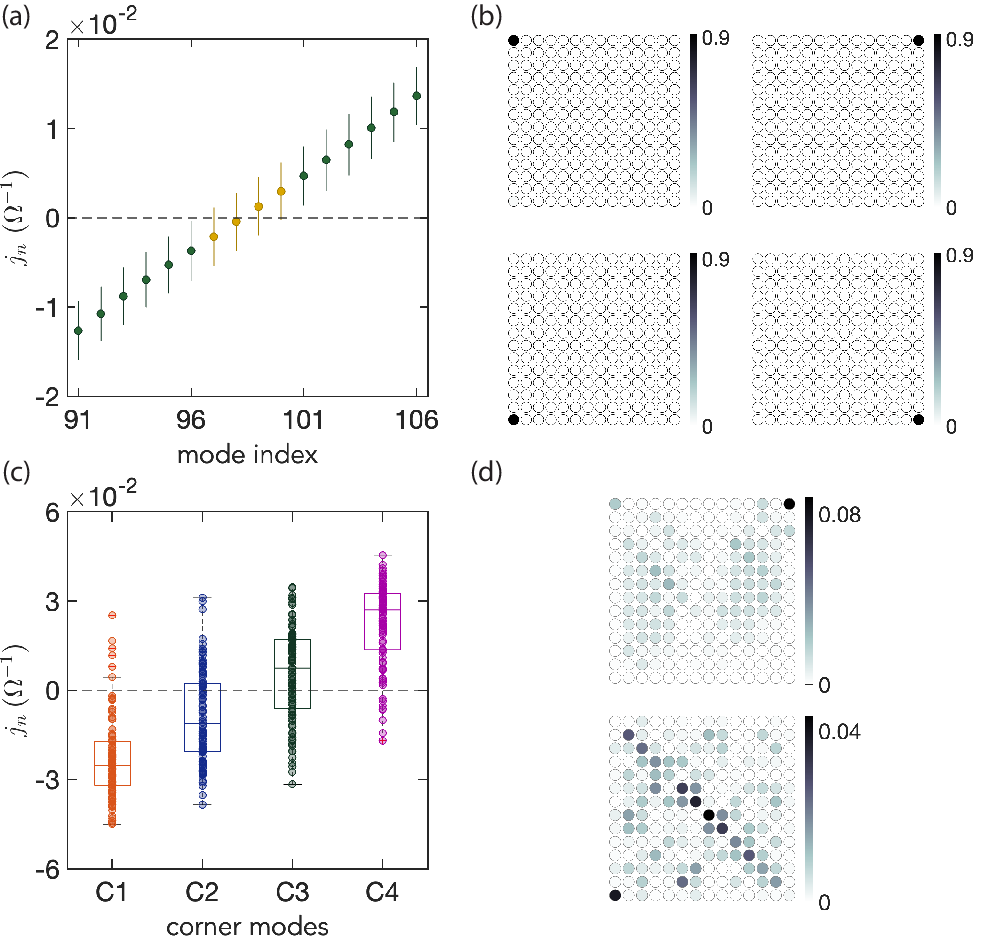}\caption{ {\bf{Corner modes of the 2D SSH circuit under 5\% off-diagonal and diagonal disorder.}} {\bf{(a)}} Numerically computed eigenvalues for $\lambda = 0.1$ with 196 nodes. Each eigenvalue is averaged over 100 disorder simulations; error bars indicate the standard deviation. Zero-admittance corner modes are highlighted in gold.  {\bf{(b)}} Normalized spatial distribution of the wavefunction amplitude, $|\psi(\omega)|^2$, for the four corner-localized modes at $\lambda = 0.1$, averaged over simulations exhibiting each localization pattern. The colourmap quantifies the degree of localization across the circuit layout. {\bf{(c)}} Eigenvalues corresponding to the four corner modes (C1-C4) for $\lambda = 0.35$. Each box represents the interquartile range (Q1-Q3) of the distribution of eigenvalues, with the center line marking the median across 100 simulations. {\bf{(d)}} Normalized spatial distribution of selected modes for $\lambda = 0.35$ illustrating hybridization between corner and bulk states.}\label{Fig_off_diagonal_and_diagonal_disorder}
\end{figure}

Diagonal disorder is implemented by randomizing the capacitance values associated with onsite terms (i.e., connections to ground) in the topoelectrical circuit. This procedure emulates imperfections inherent to physical components, thereby enabling a realistic assessment of the topological robustness of the system.  Figure~\ref{Fig_off_diagonal_and_diagonal_disorder} summarizes the effect of 5\% combined off-diagonal and diagonal disorder on corner modes.  Since diagonal disorder breaks chiral symmetry, the corner eigenvalues are no longer pinned to zero admittance (Fig.~\ref{Fig_off_diagonal_and_diagonal_disorder}(a)). Nevertheless, the four corner modes remain strongly confined  for a coupling ratio of $\lambda=0.1$ (Fig.~\ref{Fig_off_diagonal_and_diagonal_disorder}(b)).  Statistical distribution of the eigenvalues corresponding to the four corner modes across 100 simulations at $\lambda=0.35$ is shown in Fig.~\ref{Fig_off_diagonal_and_diagonal_disorder}(d), further illustrating the loss of zero-admittance pinning caused by chiral symmetry breaking when onsite disorder is introduced.  When the coupling ratio increases to $\lambda=0.35$, hybridization between corner and bulk states becomes evident.   While distinct corner-localized modes persist, some states exhibit mixed character, with their spatial profiles revealing non-negligible amplitude spread between corners and bulk nodes (Fig.~\ref{Fig_off_diagonal_and_diagonal_disorder}).

To date, experimental investigations of corner modes in systems governed by the Hamiltonian given in Eq.~\ref{eq:laplacian_02} have been limited to a 2D lattice of nanophotonic resonators featuring all positive couplings (analogous to inductors in the topoelectrical circuit) and a gapless spectrum.\cite{mittal:19} In that configuration, corner states were found to be sensitive to disorder in both coupling strengths and on-site potentials, exhibiting significant hybridization with bulk modes. This outcome contrasts with the numerical results presented here and those reported in Ref. \cite{benalcazar:20}, in which corner states remain robust despite spectral degeneracy with bulk bands under diagonal disorder. It is important to note, however, that in electromagnetic systems, long-range interactions are unavoidable and can break chiral symmetry, potentially compromising the experimental realization of topologically protected states. Nevertheless, chiral symmetry is preserved provided that interactions within the same sublattice remain negligible.

In the following section, we examine an extended 2D SSH model that incorporates NNN coupling while preserving chiral and spatial symmetries, yet breaks separability due to the long-range interactions.

\subsection{2D SSH electric circuit with NNN interaction}
\label{sec:2D-SSH-NNN}

\begin{figure}[htpb!]
\includegraphics[keepaspectratio=true, width=1.0\linewidth]{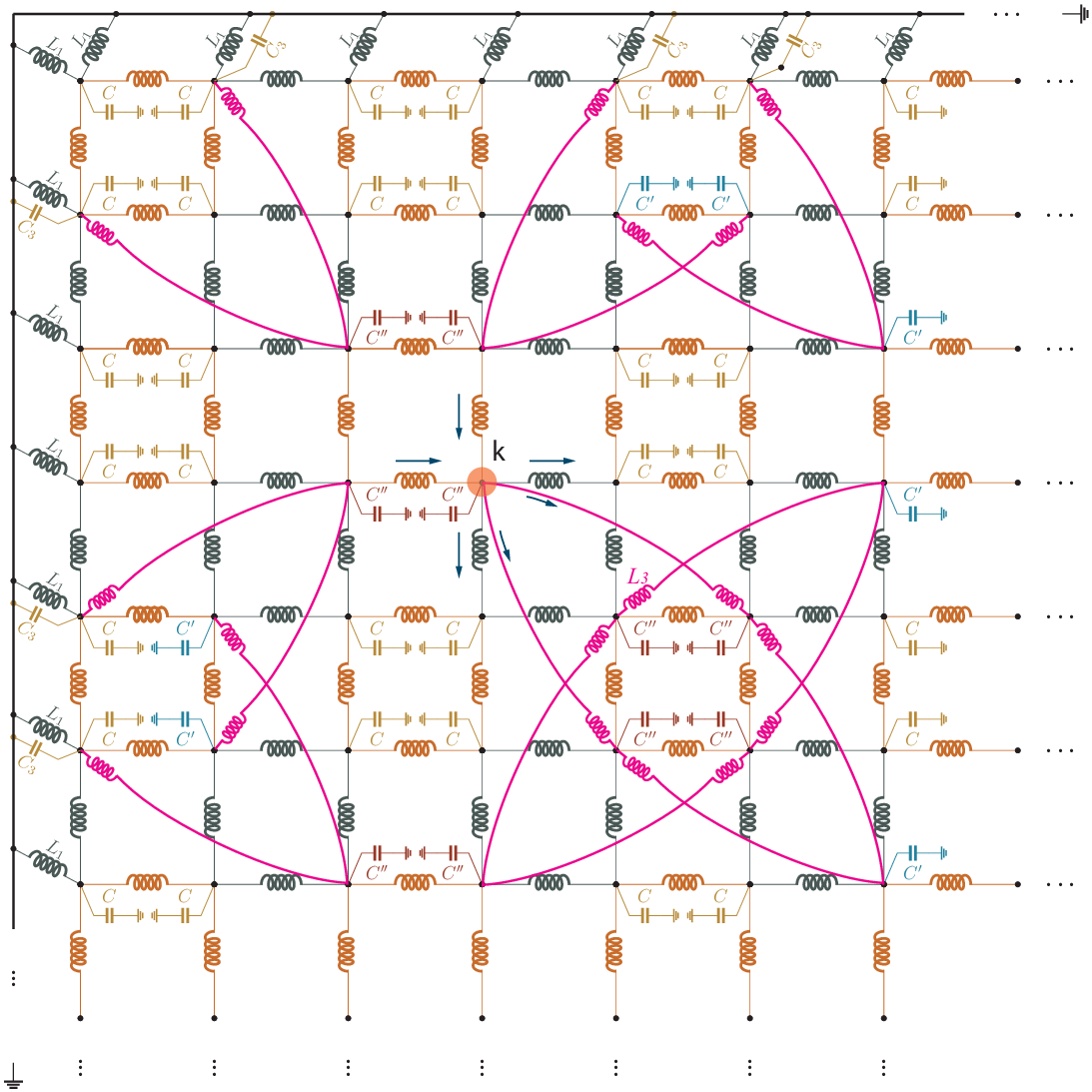}\caption{{\bf{Schematic of the 2D SSH electrical circuit with NNN coupling.}}  Intracell and intercell connections are implemented using inductors $L_1$ (green) and $L_2$ (brown),respectively, representing nearest-neighbour (NN) couplings. Long-range interactions are introduced via NNN inductors $L_3$ (purple).  In a circuit with $N\times N$ nodes, each node is labelled by an index $j\: \in{1,\dots, N^2}$. The $k\mbox{-th}$ node is highlighted. Blue arrows indicate current directions used in nodal analysis (currents leaving the node taken as positive).}\label{Fig_04}
\end{figure}

To investigate how the localization of corner modes is influenced by long-range interactions, the 2D SSH circuit is extended to include long-range coupling by connecting NNN nodes with lossless inductors $L_3$, as shown in Fig.~\ref{Fig_04}.  This circuit is parametrized by the dimensionless ratios $\alpha=L_1/L_2$ and $\beta=L_3/L_1$. The parameter $\alpha$ retains its definition as the coupling ratio ($\lambda$) in the 2D SSH without NNN coupling, so the topological phase transition is governed by $\alpha$, unaffected by the introduction of $L_3$. In contrast, $\beta$ controls the strength of NNN coupling introduced by $L_3$. Since inductance plays a role analogous to the inverse of an effective hopping amplitude in a tight-binding model, lower values of $\beta$ correspond to stronger long-range coupling, while higher values of $\beta$ indicate weaker long-range coupling.

Following the method previously described (Eq.~\ref{eq:001:sum}), by node analysis, the algebraic sum of currents at the $k\mbox{-th}$ node, which is highlighted in Fig.~\ref{Fig_04}, is given by

\begin{eqnarray}
I_{k} =&& -\frac{V_{k-1}-V_k}{i\omega L_2} - \frac{V_{k-N}-V_k}{i\omega L_2} + \frac{V_{k}-V_{k+1}}{i\omega L_1} + \frac{V_{k}-V_{k+N}}{i\omega L_1} \nonumber\\ &&+ \frac{V_{k}-V_{k+N+2}}{i\omega L_3} + \frac{V_{k}-V_{k+2N+1}}{i\omega L_3} + i \omega C^{\: \prime\prime}\: V_{k}
\end{eqnarray}

where $V_i$ is the voltage at the $i\mbox{-th}$ node with respect to the ground. This yields

\begin{eqnarray}\label{eq:JNN_02}
I_{k}=&& i\omega \left[ \left( C^{\: \prime\prime} - \frac{2}{\omega^2 L_1} - \frac{2}{\omega^2 L_2} - \frac{2}{\omega^2 L_3}\right)V_{k} \right. \nonumber\\ && \left.+ \frac{V_{k+1}+V_{k+N}}{\omega^2 L_1} + \frac{V_{k-1}+V_{k-N}}{\omega^2 L_2} \right. \nonumber\\ && \left.+ \frac{V_{k+N+2}+V_{k+2N+1}}{\omega^2 L_3}\right]
\end{eqnarray}

Each node is connected to the ground by a capacitor $C$, $C^{\:\prime}$ or $C^{\: \prime\prime}$ and the grounding at edge nodes consists of a resonator $L_1C_3$ (Fig. \ref{Fig_04}), with grounding capacitors satisfying  

\begin{align}
C&=(2/\omega_0^2)(1/L_1 + 1/L_2)\\
C_3&=1/{\omega_0}^2L_3\\
C^{\:\prime}&=C+C_3\\
C^{\: \prime\prime}&=C+2C_3
\end{align} 

This ensures that the diagonal term (coefficient of node voltage $V_{k}$) in Eq.~\ref{eq:JNN_02} cancels out at the frequency $\omega_0$ (Eq.\ref{eq_res_frq}).

\begin{figure*}[htb]
\includegraphics[keepaspectratio=true, width=1.0\linewidth]{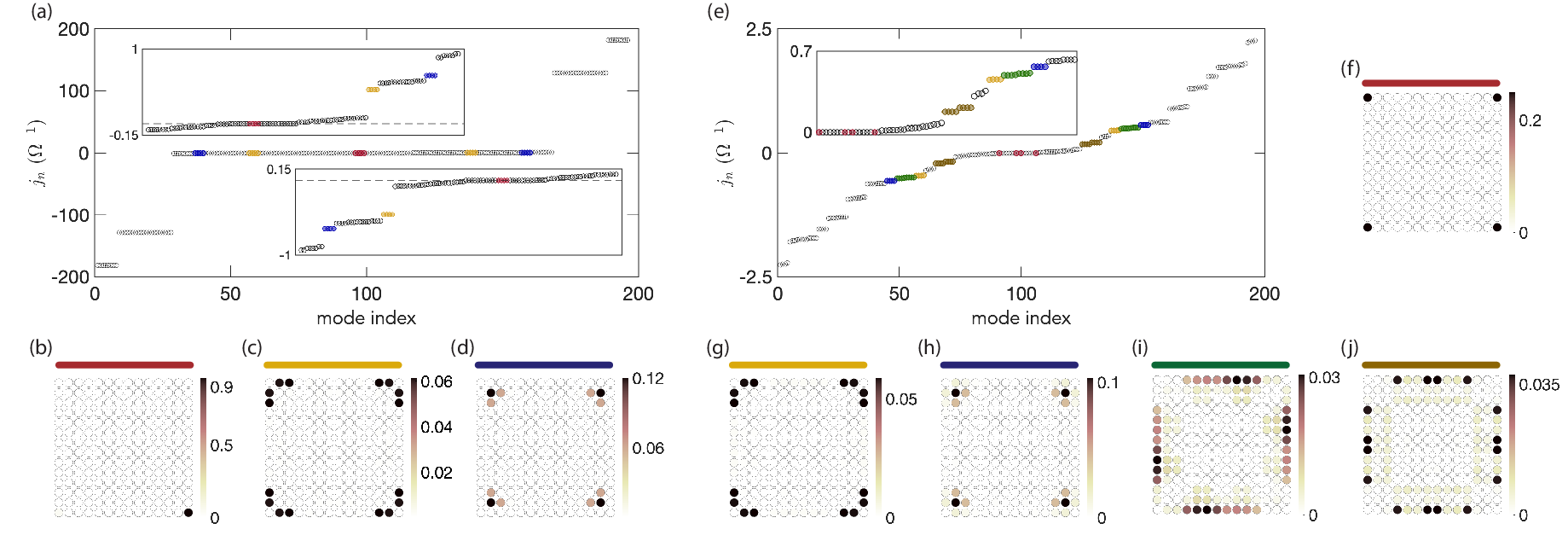}\caption{{\bf{Finite-sized extended 2D SSH circuit with NNN coupling.}}  {\bf{(a, b)}} Calculated eigenvalue spectra for a circuit with $14 \times 14$ sites including NNN coupling (schematic in Fig.~\ref{Fig_04}). Parameters are set to $\alpha=0.1$, $\beta=5\cdot10^{-3}$ in (a), and $\alpha=0.1$, $\beta=5\cdot10^{-1}$ in (b).  Insets in (a) show symmetric enlarged views of localized states near zero admittance, highlighting corner states (red), and off-site corner states, referred to as type II (ochre) and type III (blue).  The inset in (b) additionally highlights new  emerging localized states, precursors to edge states, marked in green and brown. {\bf{(c-e)}}  Normalized spatial distributions of the wavefunction amplitude $|\psi(\omega)|^2$ for the localized modes shown in (a), with corresponding color coding (red, ochre, blue) indicating the type of corner state. {\bf{(f-j)}} Spatial distributions of the localized modes shown in (b), including both previously identified corner states (red, ochre, blue) and the new precursor edge states (green, brown). The colormap in panels (c-e) and (f-j) quantifies the degree of localization across the circuit layout for different modes.  See also Fig.~S10 (Supplementary Material) for additional analysis at lower NNN coupling (increased $\beta$ parameter, $\alpha=0.1$, $\beta=1$).}\label{Fig_05}
\end{figure*}

Thus, nodal equations such as Eq.~\ref{eq:JNN_02} lead to the Laplacian representation of the circuit,  $\mathbf{I}=J_{NNN}(\omega)\mathbf{V}$, as discussed in the previous section, where $J_{NNN}$ becomes purely off-diagonal at the frequency $\omega_{0}$, analogous to a Hamiltonian in a tight-binding model.

The long-range couplings introduced via the $L_3$ inductors (purple, Fig. \ref{Fig_04}) preserve chiral symmetry, as they connect diagonally between nodes belonging to opposite sublattices.  Consequently, the Laplacian $J_{NNN}$ can be expressed in a block off-diagonal form (Eq.~\ref{eq:J_4x4_chiral}). Moreover, the circuit layout maintains the $C_{4v}$ point group symmetry, since it remains invariant under $90^{\circ}$ rotations and mirror reflections.  However, the introduction of diagonal NNN couplings breaks the separability of the Hamiltonian into independent components along the $x-$ and $y-$ directions, due to the coupling between momentum components in both directions.

The response of the circuit is governed by the eigenstates $j_n$ ($n\in\{1,\dots,N^2\}$) of the Laplacian $J_{NNN}$.  Figure~\ref{Fig_05} displays the calculated eigenvalue spectra for a finite-sized circuit ($14 \times 14$ sites) with NNN coupling, along with spatial distributions of several localized mode patterns for two values of the parameter $\beta$: $\beta=0.005$ (corresponding to strong NNN coupling) in Fig.~\ref{Fig_05}(a)-(d), and $\beta=0.5$ (weaker NNN coupling) in Fig.~\ref{Fig_05}(e)-(j)), while keeping constant $\alpha=0.1$.  This parameter choice reveals a rich variety of localized modes.

Regardless the value of $\beta$, conventional corner states embedded in the continuum and pinned to zero admittance are preserved upon introducing NNN coupling.  These are highlighted in red in Fig.~\ref{Fig_05}(a) for $\beta=0.005$, with the spatial distribution of one example shown in  Fig.~\ref{Fig_05}(b)). Similarly, for $\beta=0.5$, the corresponding corner states are marked in red in Fig.~\ref{Fig_05}(e), with one representative spatial distribution of the amplitude displayed in Fig.~\ref{Fig_05}(f)).  This persistence reflects that topological corner modes remain unaffected by the inclusion of long-range interactions as implemented in this system.  To the best of my knowledge, corner-localized modes embedded in the continuum have not previously been reported in systems featuring NNN coupling.

Beyond the conventional corner states, the inclusion of NNN coupling gives rise to additional localized modes. Although concentrated near the corners, these modes do not peak at the corner site itself.

For $\beta=0.005$, some modes exhibit a characteristic localization pattern in which the maximum amplitude occurs at the two sites adjacent to the corner along the edges, forming a distinctive V-shaped configuration that straddles the corner.  Their corresponding eigenvalues are marked in ochre in Fig.~\ref{Fig_05}(a), and the spatial distribution of one representative mode across the circuit layout is shown in Fig.~\ref{Fig_05}(c). These are referred to as type-II corner states.  

Additionally, certain modes display their highest amplitude at the corner sites of the innermost square nested within the entire lattice. These are referred to as type-III corner states, with eigenvalues marked in blue in Fig.~\ref{Fig_05}(a) and their localization pattern illustrated in Fig.~\ref{Fig_05}(d).  Notably, no edge states are observed for $\beta=0.005$ (strong NNN coupling).

Upon decreasing the strength of NNN coupling, for $\beta=0.5$, not only conventional corner states (Fig.~\ref{Fig_05}(f)), but also type-II (Fig.~\ref{Fig_05}(g)) and type-III (Fig.~\ref{Fig_05}(h)) off-site corner states are observed. In addition, new localization patterns emerge, characterized by wavefunction distributions with maximum amplitude along sites located at the edges of the lattice.  These modes exhibit elongated spatial profiles that stretch along the edges, suggesting a transition toward edge-like behaviour. Their spectral signatures are highlighted in green and brown in Fig.~\ref{Fig_05}(e), and the corresponding spatial distributions are shown in Fig.~\ref{Fig_05}(i) and Fig.~\ref{Fig_05}(j).

Type-II corner states exhibit a distinctive localization pattern that fundamentally differs from conventional corner modes. Instead of concentrating their amplitude on a single corner site belonging to one sublattice (as occurs for conventional corner states), these modes display their maximum amplitude on two adjacent sites along the edges. Importantly, these two sites belong to different sublattices, indicating that the localization mechanism is not purely corner-driven.  This spatial distribution suggests that type-II states originate from the splitting of edge states rather than from the conventional corner-state topology. As NNN coupling increases, edge-localized modes are progressively pulled toward the corners, creating hybrid states that retain edge-like character while exhibiting enhanced confinement near the corner region. The presence of amplitude on both sublattices reflects this mixed origin: these states are not strictly higher-order corner modes but rather edge-derived states reshaped by long-range interactions under preserved chiral and spatial symmetries.

This reasoning is consistent with the analysis for a further decrease in NNN coupling. Figure S10 (Supplementary Material) shows that type II and type III modes acquire an increasingly pronounced edge-like character when the NNN coupling is reduced to $\beta=1$.  At this value edge modes clearly emerge in the eigenvalue spectrum, while conventional higher-order corner states in the continuum (also observed under strong NNN coupling, Fig.~\ref{Fig_05}(b) and Fig.~\ref{Fig_05}(f)) continue to persist, owing to their protection by the $C_{4v}$ and chiral symmetries.

Type II corner states were initially reported in photonic Kagome crystals, in which conventional zero-energy corner states are spectrally isolated within the bulk gap.\cite{li:2020}  Tight-binding model simulations revealed that type II corner modes split off from the edge states and arise due to long-range interactions.  Off-site corner states have also been reported in sonic crystals, in which NNN interactions were implemented within a Kagome lattice.\cite{xiong:24}  Our results are consistent with these findings and show that NNN coupling modifies edge states, causing them to redistribute toward—but not directly at—the corners of the lattice.

\begin{figure}[htb]
\includegraphics[keepaspectratio=true, width=1.0\linewidth]{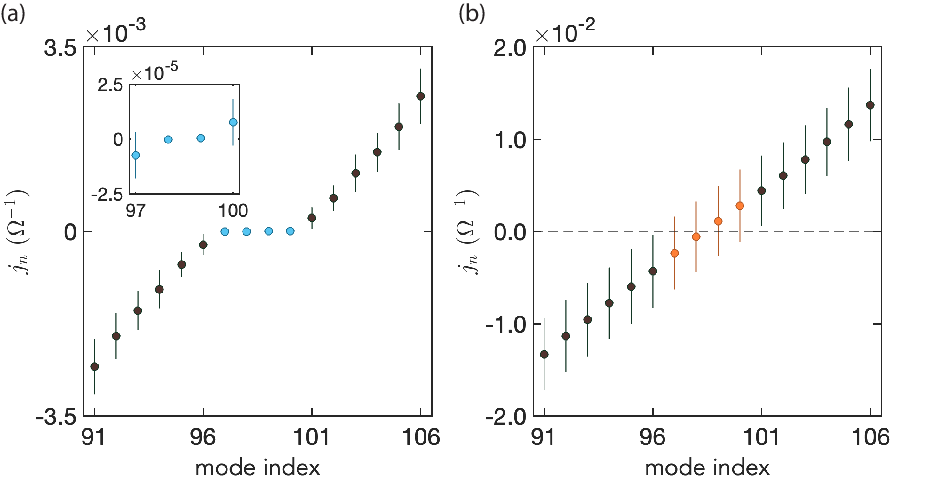}\caption{{\bf{Corner eigenvalues in the long-range coupled circuit under disorder.}} {\bf{(a)}} Numerically computed eigenvalues near zero for the 2D SHH circuit with long-range interactions (196 nodes), subject to 15\% off-diagonal disorder introduced via randomly variations in the coupling elements $L_1$, $L_2$, and $L_3$. Each eigenvalue is averaged over 100 disorder realizations, with error bars indicating the standard deviation. Zero-admittance corner eigenvalues are highlighted in blue; the inset shows an enlarged view of these corner-localized states.  {\bf{(b)}} Eigenvalues computed under 5\% diagonal and off-diagonal disorder.  Parameters for both pannels: $\alpha=0.1$, $\beta=0.005$.}\label{Fig_disorder_NNN}
\end{figure}

We now turn our attention to the effect of disorder in the presence of NNN interactions. This reproduces the qualitative behavior previously observed in the 2D SSH circuit with only NN interactions.  As illustrated in Fig.~\ref{Fig_disorder_NNN}(a), the corner eigenvalues remain pinned at zero-admittance under significant off-diagonal disorder (up to 15\%), introduced by random variations in the hopping amplitudes $L_1$, $L_2$, and $L_3$ of the topoelectrical lattice for $\alpha=0.1$, $\beta=0.005$. This type of disorder preserves chiral symmetry by avoiding any on-site terms.  As a result, the corner states persist as two pairs of chiral partners. Conversely, when diagonal disorder is added together with off-diagonal disorder (Fig.~\ref{Fig_disorder_NNN}(b)), chiral symmetry is broken and the corner eigenvalues shift away from zero admittance.

In contrast to the initial system discussed in this paper (section~\ref{sec:2D-SSH}), which features a separable Hamiltonian (Eq. \ref{eq:laplacian_02}), the long-range coupled system shown in Fig.\ref{Fig_04} breaks separability. Nevertheless, despite the lack of separability, corner bound states in the continuum remain protected by symmetry in the long-range coupled lattice, owing to the combined presence of $C_{4v}$ and chiral symmetries. This finding aligns with previous studies demonstrating that symmetry-protected boundary-localized states can exist within topological bands, remaining degenerate with bulk states of the lattice while avoiding hybridization.\cite{cerjan:20}

\section{Conclusion}

In summary, this work has explored the topological behaviour of 2D SSH models under varying symmetry constraints and coupling schemes. Numerical analysis reveals that, even in lattices with long-range interactions, in which Hamiltonian separability is absent, corner-localized bound states in the continuum remain robustly protected.  This robustness is preserved under off-diagonal disorder due to the underlying chiral and $C_{4v}$ symmetries.

The circuit architecture presented here, enriched by NNN interactions, offers a compelling and experimentally promising framework for investigating the interplay between higher-order topology and bound states in the continuum.  These findings pave the way for future experimental implementations, in which impedance spectroscopy\cite{franca:24} could serve as a powerful tool to validate theoretical predictions and further elucidate the role of symmetry in protecting topological states. 

Nevertheless, implementing robust corner states in real circuit platforms presents significant challenges, particularly when introducing NNN couplings. The robustness of these states relies on chiral and spatial symmetries, yet real components exhibit manufacturing tolerances that can introduce asymmetries and cause deviations from theory. Moreover, wiring NNN links requires connections beyond adjacent nodes across the lattice, which introduces long interconnects prone to parasitic inductance and capacitance, creating unintended couplings that distort the effective Hamiltonian. Careful printed circuit board (PCB) design is therefore essential to mitigate these effects. In addition, since circuit analogues rely on resonance frequencies to emulate lattice eigenstates, precise impedance matching becomes critical to preserve sharp spectral features. 

To minimize parasitic modes in experimental realizations, prior studies recommend multilayer PCB designs with ground layers placed between signal layers, wide traces to minimize inductance, and sufficient spacing between components to avoid spurious coupling.\cite{zhang:20, zhang:21, imhof:18, helbig:20, olekhno:22} Furthermore, using components with low tolerance helps prevent detuning of corner resonances.  Overcoming these challenges experimentally would mark a significant step toward leveraging topoelectrical circuits for the design of robust, tuneable topological devices in classical systems.\\

\section{Supplementary Material}

The supplementary material provides a detailed description of corner-localized bound states in the continuum,  covering their spatial localization, amplitude decay, and analytical derivation under chiral symmetry, as well as derivations of the circuit Laplacian for both periodic and open boundary conditions. It explains how disorder is implemented in both models, and offers further results for the extended 2D SSH model with NNN interactions under reduced long-range coupling.\\

\section*{Conflicts of interest}

There are no conflicts to declare.




\begin{thebibliography}{78}%
\makeatletter
\providecommand \@ifxundefined [1]{%
 \@ifx{#1\undefined}
}%
\providecommand \@ifnum [1]{%
 \ifnum #1\expandafter \@firstoftwo
 \else \expandafter \@secondoftwo
 \fi
}%
\providecommand \@ifx [1]{%
 \ifx #1\expandafter \@firstoftwo
 \else \expandafter \@secondoftwo
 \fi
}%
\providecommand \natexlab [1]{#1}%
\providecommand \enquote  [1]{``#1''}%
\providecommand \bibnamefont  [1]{#1}%
\providecommand \bibfnamefont [1]{#1}%
\providecommand \citenamefont [1]{#1}%
\providecommand \href@noop [0]{\@secondoftwo}%
\providecommand \href [0]{\begingroup \@sanitize@url \@href}%
\providecommand \@href[1]{\@@startlink{#1}\@@href}%
\providecommand \@@href[1]{\endgroup#1\@@endlink}%
\providecommand \@sanitize@url [0]{\catcode `\\12\catcode `\$12\catcode
  `\&12\catcode `\#12\catcode `\^12\catcode `\_12\catcode `\%12\relax}%
\providecommand \@@startlink[1]{}%
\providecommand \@@endlink[0]{}%
\providecommand \url  [0]{\begingroup\@sanitize@url \@url }%
\providecommand \@url [1]{\endgroup\@href {#1}{\urlprefix }}%
\providecommand \urlprefix  [0]{URL }%
\providecommand \Eprint [0]{\href }%
\providecommand \doibase [0]{https://doi.org/}%
\providecommand \selectlanguage [0]{\@gobble}%
\providecommand \bibinfo  [0]{\@secondoftwo}%
\providecommand \bibfield  [0]{\@secondoftwo}%
\providecommand \translation [1]{[#1]}%
\providecommand \BibitemOpen [0]{}%
\providecommand \bibitemStop [0]{}%
\providecommand \bibitemNoStop [0]{.\EOS\space}%
\providecommand \EOS [0]{\spacefactor3000\relax}%
\providecommand \BibitemShut  [1]{\csname bibitem#1\endcsname}%
\let\auto@bib@innerbib\@empty
\bibitem [{\citenamefont {Berry}(1984)}]{berry:84}%
  \BibitemOpen
  \bibfield  {author} {\bibinfo {author} {\bibfnamefont {M.~V.}\ \bibnamefont
  {Berry}},\ }\bibfield  {title} {\bibinfo {title} {Quantal phase factors
  accompanying adiabatic changes},\ }\href
  {https://doi.org/10.1098/rspa.1984.0023} {\bibfield  {journal} {\bibinfo
  {journal} {Proceedings of the Royal Society of London A - Mathematical and
  Physical Sciences}\ }\textbf {\bibinfo {volume} {392}},\ \bibinfo {pages}
  {45} (\bibinfo {year} {1984})}\BibitemShut {NoStop}%
\bibitem [{\citenamefont {Thouless}\ \emph {et~al.}(1982)\citenamefont
  {Thouless}, \citenamefont {Kohmoto}, \citenamefont {Nightingale},\ and\
  \citenamefont {den Nijs}}]{thouless:82}%
  \BibitemOpen
  \bibfield  {author} {\bibinfo {author} {\bibfnamefont {D.~J.}\ \bibnamefont
  {Thouless}}, \bibinfo {author} {\bibfnamefont {M.}~\bibnamefont {Kohmoto}},
  \bibinfo {author} {\bibfnamefont {M.~P.}\ \bibnamefont {Nightingale}},\ and\
  \bibinfo {author} {\bibfnamefont {M.}~\bibnamefont {den Nijs}},\ }\bibfield
  {title} {\bibinfo {title} {Quantized hall conductance in a two-dimensional
  periodic potential},\ }\href {https://doi.org/10.1103/PhysRevLett.49.405}
  {\bibfield  {journal} {\bibinfo  {journal} {Physical Review Letters}\
  }\textbf {\bibinfo {volume} {49}},\ \bibinfo {pages} {405} (\bibinfo {year}
  {1982})}\BibitemShut {NoStop}%
\bibitem [{\citenamefont {Wen}(2017)}]{wen:17}%
  \BibitemOpen
  \bibfield  {author} {\bibinfo {author} {\bibfnamefont {X.~G.}\ \bibnamefont
  {Wen}},\ }\bibfield  {title} {\bibinfo {title} {Colloquium: Zoo of
  quantum-topological phases of matter},\ }\href
  {https://doi.org/10.1103/RevModPhys.89.041004} {\bibfield  {journal}
  {\bibinfo  {journal} {Reviews of Modern Physics}\ }\textbf {\bibinfo {volume}
  {89}} (\bibinfo {year} {2017})}\BibitemShut {NoStop}%
\bibitem [{\citenamefont {Tokura}\ \emph {et~al.}(2019)\citenamefont {Tokura},
  \citenamefont {Yasuda},\ and\ \citenamefont {Tsukazaki}}]{tokura:19}%
  \BibitemOpen
  \bibfield  {author} {\bibinfo {author} {\bibfnamefont {Y.}~\bibnamefont
  {Tokura}}, \bibinfo {author} {\bibfnamefont {K.}~\bibnamefont {Yasuda}},\
  and\ \bibinfo {author} {\bibfnamefont {A.}~\bibnamefont {Tsukazaki}},\
  }\bibfield  {title} {\bibinfo {title} {Magnetic topological insulators},\
  }\href {https://doi.org/10.1038/s42254-018-0011-5} {\bibfield  {journal}
  {\bibinfo  {journal} {Nature Reviews Physics}\ }\textbf {\bibinfo {volume}
  {1}} (\bibinfo {year} {2019})}\BibitemShut {NoStop}%
\bibitem [{\citenamefont {Wieder}\ \emph {et~al.}(2022)\citenamefont {Wieder},
  \citenamefont {Bradlyn}, \citenamefont {Cano}, \citenamefont {Vergniory},
  \citenamefont {Elcoro}, \citenamefont {Soluyanov}, \citenamefont {Felser},
  \citenamefont {Neupert}, \citenamefont {Regnault},\ and\ \citenamefont
  {Bernevig}}]{wieder:22}%
  \BibitemOpen
  \bibfield  {author} {\bibinfo {author} {\bibfnamefont {B.~J.}\ \bibnamefont
  {Wieder}}, \bibinfo {author} {\bibfnamefont {B.}~\bibnamefont {Bradlyn}},
  \bibinfo {author} {\bibfnamefont {J.}~\bibnamefont {Cano}}, \bibinfo {author}
  {\bibfnamefont {Z.~J. W. M.~G.}\ \bibnamefont {Vergniory}}, \bibinfo {author}
  {\bibfnamefont {L.}~\bibnamefont {Elcoro}}, \bibinfo {author} {\bibfnamefont
  {A.~A.}\ \bibnamefont {Soluyanov}}, \bibinfo {author} {\bibfnamefont
  {C.}~\bibnamefont {Felser}}, \bibinfo {author} {\bibfnamefont
  {T.}~\bibnamefont {Neupert}}, \bibinfo {author} {\bibfnamefont
  {N.}~\bibnamefont {Regnault}},\ and\ \bibinfo {author} {\bibfnamefont
  {B.~A.}\ \bibnamefont {Bernevig}},\ }\bibfield  {title} {\bibinfo {title}
  {Topological materials discovery from crystal symmetry},\ }\href
  {https://doi.org/10.1038/s41578-021-00380-2} {\bibfield  {journal} {\bibinfo
  {journal} {Nature Reviews Materials}\ }\textbf {\bibinfo {volume} {7}}
  (\bibinfo {year} {2022})}\BibitemShut {NoStop}%
\bibitem [{\citenamefont {von Klitzing}\ \emph {et~al.}(1980)\citenamefont {von
  Klitzing}, \citenamefont {Dorda},\ and\ \citenamefont
  {Pepper}}]{klitzing:80}%
  \BibitemOpen
  \bibfield  {author} {\bibinfo {author} {\bibfnamefont {K.}~\bibnamefont {von
  Klitzing}}, \bibinfo {author} {\bibfnamefont {G.}~\bibnamefont {Dorda}},\
  and\ \bibinfo {author} {\bibfnamefont {M.}~\bibnamefont {Pepper}},\
  }\bibfield  {title} {\bibinfo {title} {New method for high-accuracy
  determination of the fine- structure constant based on quantized hall
  resistance},\ }\href {https://doi.org/10.1103/physrevlett.45.494} {\bibfield
  {journal} {\bibinfo  {journal} {Physical Review Letters}\ }\textbf {\bibinfo
  {volume} {45}},\ \bibinfo {pages} {494} (\bibinfo {year} {1980})}\BibitemShut
  {NoStop}%
\bibitem [{\citenamefont {Haldane}(1988)}]{haldane:88}%
  \BibitemOpen
  \bibfield  {author} {\bibinfo {author} {\bibfnamefont {F.~D.~M.}\
  \bibnamefont {Haldane}},\ }\bibfield  {title} {\bibinfo {title} {Model for a
  quantum hall effect without landau levels: Condensed-matter realization of
  the 'parity anomaly'},\ }\href {https://doi.org/10.1103/PhysRevLett.61.2015}
  {\bibfield  {journal} {\bibinfo  {journal} {Physical Review Letters}\
  }\textbf {\bibinfo {volume} {61}} (\bibinfo {year} {1988})}\BibitemShut
  {NoStop}%
\bibitem [{\citenamefont {Kane}\ and\ \citenamefont {Mele}(2005)}]{kane:05}%
  \BibitemOpen
  \bibfield  {author} {\bibinfo {author} {\bibfnamefont {C.~L.}\ \bibnamefont
  {Kane}}\ and\ \bibinfo {author} {\bibfnamefont {E.~J.}\ \bibnamefont
  {Mele}},\ }\bibfield  {title} {\bibinfo {title} {Z$_2$ topological order and
  the quantum spin hall effect},\ }\href
  {https://doi.org/10.1103/PhysRevLett.95.146802} {\bibfield  {journal}
  {\bibinfo  {journal} {Physical Review Letters}\ }\textbf {\bibinfo {volume}
  {95}} (\bibinfo {year} {2005})}\BibitemShut {NoStop}%
\bibitem [{\citenamefont {Bernevig}\ \emph {et~al.}(2006)\citenamefont
  {Bernevig}, \citenamefont {Hughes},\ and\ \citenamefont
  {Zhang}}]{bernevig:06}%
  \BibitemOpen
  \bibfield  {author} {\bibinfo {author} {\bibfnamefont {B.~A.}\ \bibnamefont
  {Bernevig}}, \bibinfo {author} {\bibfnamefont {T.~L.}\ \bibnamefont
  {Hughes}},\ and\ \bibinfo {author} {\bibfnamefont {S.~C.}\ \bibnamefont
  {Zhang}},\ }\bibfield  {title} {\bibinfo {title} {Quantum spin hall effect
  and topological phase transition in hgte quantum wells},\ }\href
  {https://doi.org/10.1126/science.1133734} {\bibfield  {journal} {\bibinfo
  {journal} {Science}\ }\textbf {\bibinfo {volume} {314}},\ \bibinfo {pages}
  {1757} (\bibinfo {year} {2006})}\BibitemShut {NoStop}%
\bibitem [{\citenamefont {K\"{o}nig}\ \emph {et~al.}(2007)\citenamefont
  {K\"{o}nig}, \citenamefont {Wiedmann}, \citenamefont {Br\"{u}ne},
  \citenamefont {Roth}, \citenamefont {Buhmann}, \citenamefont {Molenkamp},
  \citenamefont {Qi},\ and\ \citenamefont {Zhang}}]{konig:07}%
  \BibitemOpen
  \bibfield  {author} {\bibinfo {author} {\bibfnamefont {M.}~\bibnamefont
  {K\"{o}nig}}, \bibinfo {author} {\bibfnamefont {S.}~\bibnamefont {Wiedmann}},
  \bibinfo {author} {\bibfnamefont {C.}~\bibnamefont {Br\"{u}ne}}, \bibinfo
  {author} {\bibfnamefont {A.}~\bibnamefont {Roth}}, \bibinfo {author}
  {\bibfnamefont {H.}~\bibnamefont {Buhmann}}, \bibinfo {author} {\bibfnamefont
  {L.~W.}\ \bibnamefont {Molenkamp}}, \bibinfo {author} {\bibfnamefont {X.-L.}\
  \bibnamefont {Qi}},\ and\ \bibinfo {author} {\bibfnamefont {S.-C.}\
  \bibnamefont {Zhang}},\ }\bibfield  {title} {\bibinfo {title} {Quantum spin
  hall insulator state in {H}g{T}e quantum wells},\ }\href
  {https://doi.org/10.1126/science.1148047} {\bibfield  {journal} {\bibinfo
  {journal} {Science}\ }\textbf {\bibinfo {volume} {318}},\ \bibinfo {pages}
  {766} (\bibinfo {year} {2007})}\BibitemShut {NoStop}%
\bibitem [{\citenamefont {Fu}\ \emph {et~al.}(2007)\citenamefont {Fu},
  \citenamefont {Kane},\ and\ \citenamefont {Mele}}]{fu:07}%
  \BibitemOpen
  \bibfield  {author} {\bibinfo {author} {\bibfnamefont {L.}~\bibnamefont
  {Fu}}, \bibinfo {author} {\bibfnamefont {C.~L.}\ \bibnamefont {Kane}},\ and\
  \bibinfo {author} {\bibfnamefont {E.~J.}\ \bibnamefont {Mele}},\ }\bibfield
  {title} {\bibinfo {title} {Topological insulators in three dimensions},\
  }\href {https://doi.org/10.1103/PhysRevLett.98.106803} {\bibfield  {journal}
  {\bibinfo  {journal} {Physical Review Letters}\ }\textbf {\bibinfo {volume}
  {98}} (\bibinfo {year} {2007})}\BibitemShut {NoStop}%
\bibitem [{\citenamefont {Moore}\ and\ \citenamefont
  {Balents}(2007)}]{moore:07}%
  \BibitemOpen
  \bibfield  {author} {\bibinfo {author} {\bibfnamefont {J.~E.}\ \bibnamefont
  {Moore}}\ and\ \bibinfo {author} {\bibfnamefont {L.}~\bibnamefont
  {Balents}},\ }\bibfield  {title} {\bibinfo {title} {Topological invariants of
  time-reversal-invariant band structures},\ }\href
  {https://doi.org/10.1103/PhysRevB.75.121306} {\bibfield  {journal} {\bibinfo
  {journal} {Physical Review B}\ }\textbf {\bibinfo {volume} {75}} (\bibinfo
  {year} {2007})}\BibitemShut {NoStop}%
\bibitem [{\citenamefont {Hsieh}\ \emph {et~al.}(2008)\citenamefont {Hsieh},
  \citenamefont {Qian}, \citenamefont {Wray}, \citenamefont {Xia},
  \citenamefont {Hor}, \citenamefont {Cava},\ and\ \citenamefont
  {Hasan}}]{hsieh:08}%
  \BibitemOpen
  \bibfield  {author} {\bibinfo {author} {\bibfnamefont {D.}~\bibnamefont
  {Hsieh}}, \bibinfo {author} {\bibfnamefont {D.}~\bibnamefont {Qian}},
  \bibinfo {author} {\bibfnamefont {L.}~\bibnamefont {Wray}}, \bibinfo {author}
  {\bibfnamefont {Y.}~\bibnamefont {Xia}}, \bibinfo {author} {\bibfnamefont
  {Y.~S.}\ \bibnamefont {Hor}}, \bibinfo {author} {\bibfnamefont {R.~J.}\
  \bibnamefont {Cava}},\ and\ \bibinfo {author} {\bibfnamefont {M.~Z.}\
  \bibnamefont {Hasan}},\ }\bibfield  {title} {\bibinfo {title} {A topological
  {D}irac insulator in a quantum spin hall phase},\ }\href
  {https://doi.org/10.1038/nature06843} {\bibfield  {journal} {\bibinfo
  {journal} {Nature}\ }\textbf {\bibinfo {volume} {452}} (\bibinfo {year}
  {2008})}\BibitemShut {NoStop}%
\bibitem [{\citenamefont {Roy}(2009)}]{roy:09}%
  \BibitemOpen
  \bibfield  {author} {\bibinfo {author} {\bibfnamefont {R.}~\bibnamefont
  {Roy}},\ }\bibfield  {title} {\bibinfo {title} {Topological phases and the
  quantum spin hall effect in three dimensions},\ }\href
  {https://doi.org/10.1103/PhysRevB.79.195322} {\bibfield  {journal} {\bibinfo
  {journal} {Physical Review B}\ }\textbf {\bibinfo {volume} {79}} (\bibinfo
  {year} {2009})}\BibitemShut {NoStop}%
\bibitem [{\citenamefont {Moore}(2010)}]{moore:10}%
  \BibitemOpen
  \bibfield  {author} {\bibinfo {author} {\bibfnamefont {J.~E.}\ \bibnamefont
  {Moore}},\ }\bibfield  {title} {\bibinfo {title} {The birth of topological
  insulators},\ }\href {https://doi.org/10.1038/nature08916} {\bibfield
  {journal} {\bibinfo  {journal} {Nature}\ }\textbf {\bibinfo {volume} {464}},\
  \bibinfo {pages} {194} (\bibinfo {year} {2010})}\BibitemShut {NoStop}%
\bibitem [{\citenamefont {Haldane}\ and\ \citenamefont
  {Raghu}(2008)}]{haldane:08}%
  \BibitemOpen
  \bibfield  {author} {\bibinfo {author} {\bibfnamefont {F.~D.~M.}\
  \bibnamefont {Haldane}}\ and\ \bibinfo {author} {\bibfnamefont
  {S.}~\bibnamefont {Raghu}},\ }\bibfield  {title} {\bibinfo {title} {Possible
  realization of directional optical waveguides in photonic crystals with
  broken time-reversal symmetry},\ }\href
  {https://doi.org/10.1103/PhysRevLett.100.013904} {\bibfield  {journal}
  {\bibinfo  {journal} {Physical Review Letters}\ }\textbf {\bibinfo {volume}
  {100}} (\bibinfo {year} {2008})}\BibitemShut {NoStop}%
\bibitem [{\citenamefont {Ozawa}\ \emph {et~al.}(2019)\citenamefont {Ozawa},
  \citenamefont {Price}, \citenamefont {Amo}, \citenamefont {Goldman},
  \citenamefont {Hafezi}, \citenamefont {Lu}, \citenamefont {Rechtsman},
  \citenamefont {Simon}, \citenamefont {Zilberberg},\ and\ \citenamefont
  {Carusotto}}]{ozawa:19}%
  \BibitemOpen
  \bibfield  {author} {\bibinfo {author} {\bibfnamefont {T.}~\bibnamefont
  {Ozawa}}, \bibinfo {author} {\bibfnamefont {H.~M.}\ \bibnamefont {Price}},
  \bibinfo {author} {\bibfnamefont {A.}~\bibnamefont {Amo}}, \bibinfo {author}
  {\bibfnamefont {N.}~\bibnamefont {Goldman}}, \bibinfo {author} {\bibfnamefont
  {M.}~\bibnamefont {Hafezi}}, \bibinfo {author} {\bibfnamefont
  {L.}~\bibnamefont {Lu}}, \bibinfo {author} {\bibfnamefont {M.~C.}\
  \bibnamefont {Rechtsman}}, \bibinfo {author} {\bibfnamefont {D.~S.~J.}\
  \bibnamefont {Simon}}, \bibinfo {author} {\bibfnamefont {O.}~\bibnamefont
  {Zilberberg}},\ and\ \bibinfo {author} {\bibfnamefont {I.}~\bibnamefont
  {Carusotto}},\ }\bibfield  {title} {\bibinfo {title} {Topological
  photonics},\ }\href {https://doi.org/10.1103/RevModPhys.91.015006} {\bibfield
   {journal} {\bibinfo  {journal} {Reviews of Modern Physics}\ }\textbf
  {\bibinfo {volume} {91}} (\bibinfo {year} {2019})}\BibitemShut {NoStop}%
\bibitem [{\citenamefont {Zhang}\ \emph {et~al.}(2023)\citenamefont {Zhang},
  \citenamefont {Zangeneh-Nejad}, \citenamefont {Chen}, \citenamefont {Lu},\
  and\ \citenamefont {Christensen}}]{zhang:23}%
  \BibitemOpen
  \bibfield  {author} {\bibinfo {author} {\bibfnamefont {X.}~\bibnamefont
  {Zhang}}, \bibinfo {author} {\bibfnamefont {F.}~\bibnamefont
  {Zangeneh-Nejad}}, \bibinfo {author} {\bibfnamefont {Z.-G.}\ \bibnamefont
  {Chen}}, \bibinfo {author} {\bibfnamefont {M.-H.}\ \bibnamefont {Lu}},\ and\
  \bibinfo {author} {\bibfnamefont {J.}~\bibnamefont {Christensen}},\
  }\bibfield  {title} {\bibinfo {title} {A second wave of topological phenomena
  in photonics and acoustics},\ }\href
  {https://doi.org/10.1038/s41586-023-06163-9} {\bibfield  {journal} {\bibinfo
  {journal} {Nature}\ }\textbf {\bibinfo {volume} {618}},\ \bibinfo {pages}
  {687} (\bibinfo {year} {2023})}\BibitemShut {NoStop}%
\bibitem [{\citenamefont {Zhu}\ \emph {et~al.}(2023)\citenamefont {Zhu},
  \citenamefont {Deng}, \citenamefont {Liu}, \citenamefont {Lu}, \citenamefont
  {Wang}, \citenamefont {Lin}, \citenamefont {Huang}, \citenamefont {Jiang},\
  and\ \citenamefont {Liu}}]{zhu:23}%
  \BibitemOpen
  \bibfield  {author} {\bibinfo {author} {\bibfnamefont {W.}~\bibnamefont
  {Zhu}}, \bibinfo {author} {\bibfnamefont {W.}~\bibnamefont {Deng}}, \bibinfo
  {author} {\bibfnamefont {Y.}~\bibnamefont {Liu}}, \bibinfo {author}
  {\bibfnamefont {J.}~\bibnamefont {Lu}}, \bibinfo {author} {\bibfnamefont
  {H.-X.}\ \bibnamefont {Wang}}, \bibinfo {author} {\bibfnamefont {Z.-K.}\
  \bibnamefont {Lin}}, \bibinfo {author} {\bibfnamefont {X.}~\bibnamefont
  {Huang}}, \bibinfo {author} {\bibfnamefont {J.-H.}\ \bibnamefont {Jiang}},\
  and\ \bibinfo {author} {\bibfnamefont {Z.}~\bibnamefont {Liu}},\ }\bibfield
  {title} {\bibinfo {title} {Topological phononic metamaterials},\ }\href
  {https://doi.org/10.1088/1361-6633/aceeee} {\bibfield  {journal} {\bibinfo
  {journal} {Reports on Progress in Physics}\ }\textbf {\bibinfo {volume} {86}}
  (\bibinfo {year} {2023})}\BibitemShut {NoStop}%
\bibitem [{\citenamefont {Xue}\ \emph {et~al.}(2022)\citenamefont {Xue},
  \citenamefont {Yang},\ and\ \citenamefont {Zhang}}]{xue:22}%
  \BibitemOpen
  \bibfield  {author} {\bibinfo {author} {\bibfnamefont {H.}~\bibnamefont
  {Xue}}, \bibinfo {author} {\bibfnamefont {Y.}~\bibnamefont {Yang}},\ and\
  \bibinfo {author} {\bibfnamefont {B.}~\bibnamefont {Zhang}},\ }\bibfield
  {title} {\bibinfo {title} {Topological acoustics},\ }\href
  {https://doi.org/10.1038/s41578-022-00465-6} {\bibfield  {journal} {\bibinfo
  {journal} {Nature Reviews Materials}\ }\textbf {\bibinfo {volume} {7}},\
  \bibinfo {pages} {974} (\bibinfo {year} {2022})}\BibitemShut {NoStop}%
\bibitem [{\citenamefont {Ma}\ \emph {et~al.}(2019)\citenamefont {Ma},
  \citenamefont {Xiao},\ and\ \citenamefont {Chan}}]{ma:19}%
  \BibitemOpen
  \bibfield  {author} {\bibinfo {author} {\bibfnamefont {G.}~\bibnamefont
  {Ma}}, \bibinfo {author} {\bibfnamefont {M.}~\bibnamefont {Xiao}},\ and\
  \bibinfo {author} {\bibfnamefont {C.~T.}\ \bibnamefont {Chan}},\ }\bibfield
  {title} {\bibinfo {title} {Topological phases in acoustic and mechanical
  systems},\ }\href {https://doi.org/10.1038/s42254-019-0030-x} {\bibfield
  {journal} {\bibinfo  {journal} {Nature Reviews Physics}\ }\textbf {\bibinfo
  {volume} {1}},\ \bibinfo {pages} {281} (\bibinfo {year} {2019})}\BibitemShut
  {NoStop}%
\bibitem [{\citenamefont {Huber}(2016)}]{huber:16}%
  \BibitemOpen
  \bibfield  {author} {\bibinfo {author} {\bibfnamefont {S.~D.}\ \bibnamefont
  {Huber}},\ }\bibfield  {title} {\bibinfo {title} {Topological mechanics},\
  }\href {https://doi.org/10.1038/nphys3801} {\bibfield  {journal} {\bibinfo
  {journal} {Nature Physics}\ }\textbf {\bibinfo {volume} {12}},\ \bibinfo
  {pages} {621} (\bibinfo {year} {2016})}\BibitemShut {NoStop}%
\bibitem [{\citenamefont {Shah}\ \emph {et~al.}(2024)\citenamefont {Shah},
  \citenamefont {Brendel}, \citenamefont {Peano},\ and\ \citenamefont
  {Marquardt}}]{shah:24}%
  \BibitemOpen
  \bibfield  {author} {\bibinfo {author} {\bibfnamefont {T.}~\bibnamefont
  {Shah}}, \bibinfo {author} {\bibfnamefont {C.}~\bibnamefont {Brendel}},
  \bibinfo {author} {\bibfnamefont {V.}~\bibnamefont {Peano}},\ and\ \bibinfo
  {author} {\bibfnamefont {F.}~\bibnamefont {Marquardt}},\ }\bibfield  {title}
  {\bibinfo {title} {Colloquium: Topologically protected transport in
  engineered mechanical systems},\ }\href
  {https://doi.org/10.1103/RevModPhys.96.021002} {\bibfield  {journal}
  {\bibinfo  {journal} {Reviews of Modern Physics}\ }\textbf {\bibinfo {volume}
  {96}} (\bibinfo {year} {2024})}\BibitemShut {NoStop}%
\bibitem [{\citenamefont {Khanikaev}\ and\ \citenamefont
  {Alu}(2024)}]{khanikaev:24}%
  \BibitemOpen
  \bibfield  {author} {\bibinfo {author} {\bibfnamefont {A.~B.}\ \bibnamefont
  {Khanikaev}}\ and\ \bibinfo {author} {\bibfnamefont {A.}~\bibnamefont
  {Alu}},\ }\bibfield  {title} {\bibinfo {title} {Topological photonics:
  robustness and beyond},\ }\href {https://doi.org/10.1038/s41467-024-45194-2}
  {\bibfield  {journal} {\bibinfo  {journal} {Nature Communications}\ }\textbf
  {\bibinfo {volume} {15}} (\bibinfo {year} {2024})}\BibitemShut {NoStop}%
\bibitem [{\citenamefont {Zhao}(2025)}]{zhao:18}%
  \BibitemOpen
  \bibfield  {author} {\bibinfo {author} {\bibfnamefont {E.~H.}\ \bibnamefont
  {Zhao}},\ }\bibfield  {title} {\bibinfo {title} {Topological circuits of
  inductors and capacitors},\ }\href
  {https://doi.org/10.1016/j.aop.2018.10.006} {\bibfield  {journal} {\bibinfo
  {journal} {Annals of Physics}\ }\textbf {\bibinfo {volume} {399}},\ \bibinfo
  {pages} {289} (\bibinfo {year} {2025})}\BibitemShut {NoStop}%
\bibitem [{\citenamefont {Dong}\ \emph {et~al.}(2021)\citenamefont {Dong},
  \citenamefont {Juricic},\ and\ \citenamefont {Roy}}]{dong:21}%
  \BibitemOpen
  \bibfield  {author} {\bibinfo {author} {\bibfnamefont {J.}~\bibnamefont
  {Dong}}, \bibinfo {author} {\bibfnamefont {V.}~\bibnamefont {Juricic}},\ and\
  \bibinfo {author} {\bibfnamefont {B.}~\bibnamefont {Roy}},\ }\bibfield
  {title} {\bibinfo {title} {Topolectric circuits: Theory and construction},\
  }\href {https://doi.org/10.1103/PhysRevResearch.3.023056} {\bibfield
  {journal} {\bibinfo  {journal} {Physical Review Research}\ }\textbf {\bibinfo
  {volume} {3}} (\bibinfo {year} {2021})}\BibitemShut {NoStop}%
\bibitem [{\citenamefont {Kotwal}\ \emph {et~al.}(2021)\citenamefont {Kotwal},
  \citenamefont {Moseley}, \citenamefont {Stegmaier}, , \citenamefont {Imhoff},
  \citenamefont {Brand}, \citenamefont {Thomale}, \citenamefont
  {Ronellenfitsch},\ and\ \citenamefont {Dunkel}}]{kotwal:21}%
  \BibitemOpen
  \bibfield  {author} {\bibinfo {author} {\bibfnamefont {T.}~\bibnamefont
  {Kotwal}}, \bibinfo {author} {\bibfnamefont {F.}~\bibnamefont {Moseley}},
  \bibinfo {author} {\bibfnamefont {A.}~\bibnamefont {Stegmaier}}, , \bibinfo
  {author} {\bibfnamefont {S.}~\bibnamefont {Imhoff}}, \bibinfo {author}
  {\bibfnamefont {H.}~\bibnamefont {Brand}}, \bibinfo {author} {\bibfnamefont
  {R.}~\bibnamefont {Thomale}}, \bibinfo {author} {\bibfnamefont
  {H.}~\bibnamefont {Ronellenfitsch}},\ and\ \bibinfo {author} {\bibfnamefont
  {J.}~\bibnamefont {Dunkel}},\ }\bibfield  {title} {\bibinfo {title} {Active
  topolectrical circuits},\ }\href {https://doi.org/10.1073/pnas.2106411118}
  {\bibfield  {journal} {\bibinfo  {journal} {Proceedings of the National
  Academy of Sciences of the United States of America}\ }\textbf {\bibinfo
  {volume} {118}} (\bibinfo {year} {2021})}\BibitemShut {NoStop}%
\bibitem [{\citenamefont {Yang}\ \emph {et~al.}(2024)\citenamefont {Yang},
  \citenamefont {Song}, \citenamefont {Cao},\ and\ \citenamefont
  {Yan}}]{yang:24}%
  \BibitemOpen
  \bibfield  {author} {\bibinfo {author} {\bibfnamefont {H.}~\bibnamefont
  {Yang}}, \bibinfo {author} {\bibfnamefont {L.}~\bibnamefont {Song}}, \bibinfo
  {author} {\bibfnamefont {Y.}~\bibnamefont {Cao}},\ and\ \bibinfo {author}
  {\bibfnamefont {P.}~\bibnamefont {Yan}},\ }\bibfield  {title} {\bibinfo
  {title} {Circuit realization of topological physics},\ }\href
  {https://doi.org/10.1016/j.physrep.2024.09.007} {\bibfield  {journal}
  {\bibinfo  {journal} {Physics Reports-Review Section of Physics Letters}\
  }\textbf {\bibinfo {volume} {1093}},\ \bibinfo {pages} {1} (\bibinfo {year}
  {2024})}\BibitemShut {NoStop}%
\bibitem [{\citenamefont {Sahin}\ \emph {et~al.}(2025)\citenamefont {Sahin},
  \citenamefont {Jalil},\ and\ \citenamefont {Lee}}]{sahin:25}%
  \BibitemOpen
  \bibfield  {author} {\bibinfo {author} {\bibfnamefont {H.}~\bibnamefont
  {Sahin}}, \bibinfo {author} {\bibfnamefont {M.~B.~A.}\ \bibnamefont
  {Jalil}},\ and\ \bibinfo {author} {\bibfnamefont {C.~H.}\ \bibnamefont
  {Lee}},\ }\bibfield  {title} {\bibinfo {title} {Topolectrical
  circuits—recent experimental advances and developments},\ }\href
  {https://doi.org/10.1063/5.0265293} {\bibfield  {journal} {\bibinfo
  {journal} {{APL} Electronic Devices}\ }\textbf {\bibinfo {volume} {1}}
  (\bibinfo {year} {2025})}\BibitemShut {NoStop}%
\bibitem [{\citenamefont {Chen}\ \emph {et~al.}(2025)\citenamefont {Chen},
  \citenamefont {Zhang}, \citenamefont {Zou}, \citenamefont {Sun},\ and\
  \citenamefont {Zhang}}]{chen:25}%
  \BibitemOpen
  \bibfield  {author} {\bibinfo {author} {\bibfnamefont {T.}~\bibnamefont
  {Chen}}, \bibinfo {author} {\bibfnamefont {W.}~\bibnamefont {Zhang}},
  \bibinfo {author} {\bibfnamefont {D.}~\bibnamefont {Zou}}, \bibinfo {author}
  {\bibfnamefont {Y.}~\bibnamefont {Sun}},\ and\ \bibinfo {author}
  {\bibfnamefont {X.}~\bibnamefont {Zhang}},\ }\bibfield  {title} {\bibinfo
  {title} {Engineering topological states and quantum-inspired information
  processing using classical circuits},\ }\href
  {https://doi.org/10.1002/qute.202400448} {\bibfield  {journal} {\bibinfo
  {journal} {Advanced Quantum Technologies}\ }\textbf {\bibinfo {volume} {8}}
  (\bibinfo {year} {2025})}\BibitemShut {NoStop}%
\bibitem [{\citenamefont {Ningyuan}\ \emph {et~al.}(2015)\citenamefont
  {Ningyuan}, \citenamefont {Owens}, \citenamefont {Sommer}, \citenamefont
  {Schuster},\ and\ \citenamefont {Simon}}]{ningyuan:15}%
  \BibitemOpen
  \bibfield  {author} {\bibinfo {author} {\bibfnamefont {J.}~\bibnamefont
  {Ningyuan}}, \bibinfo {author} {\bibfnamefont {C.}~\bibnamefont {Owens}},
  \bibinfo {author} {\bibfnamefont {A.}~\bibnamefont {Sommer}}, \bibinfo
  {author} {\bibfnamefont {D.}~\bibnamefont {Schuster}},\ and\ \bibinfo
  {author} {\bibfnamefont {J.}~\bibnamefont {Simon}},\ }\bibfield  {title}
  {\bibinfo {title} {Time- and site-resolved dynamics in a topological
  circuit},\ }\href {https://doi.org/10.1103/PhysRevX.5.021031} {\bibfield
  {journal} {\bibinfo  {journal} {Physical Review X}\ }\textbf {\bibinfo
  {volume} {5}} (\bibinfo {year} {2015})}\BibitemShut {NoStop}%
\bibitem [{\citenamefont {Albert}\ \emph {et~al.}(2015)\citenamefont {Albert},
  \citenamefont {Glazman},\ and\ \citenamefont {Jiang}}]{albert:15}%
  \BibitemOpen
  \bibfield  {author} {\bibinfo {author} {\bibfnamefont {V.~V.}\ \bibnamefont
  {Albert}}, \bibinfo {author} {\bibfnamefont {L.~I.}\ \bibnamefont
  {Glazman}},\ and\ \bibinfo {author} {\bibfnamefont {L.}~\bibnamefont
  {Jiang}},\ }\bibfield  {title} {\bibinfo {title} {Topological properties of
  linear circuit lattices},\ }\href
  {https://doi.org/10.1103/PhysRevLett.114.173902} {\bibfield  {journal}
  {\bibinfo  {journal} {Physical Review Letters}\ }\textbf {\bibinfo {volume}
  {114}} (\bibinfo {year} {2015})}\BibitemShut {NoStop}%
\bibitem [{\citenamefont {Haenel}\ \emph {et~al.}(2019)\citenamefont {Haenel},
  \citenamefont {Branch},\ and\ \citenamefont {Franz}}]{haenel:19}%
  \BibitemOpen
  \bibfield  {author} {\bibinfo {author} {\bibfnamefont {R.}~\bibnamefont
  {Haenel}}, \bibinfo {author} {\bibfnamefont {T.}~\bibnamefont {Branch}},\
  and\ \bibinfo {author} {\bibfnamefont {M.}~\bibnamefont {Franz}},\ }\bibfield
   {title} {\bibinfo {title} {Chern insulators for electromagnetic waves in
  electrical circuit networks},\ }\href
  {https://doi.org/10.1103/PhysRevB.99.235110} {\bibfield  {journal} {\bibinfo
  {journal} {Physical Review B}\ }\textbf {\bibinfo {volume} {99}} (\bibinfo
  {year} {2019})}\BibitemShut {NoStop}%
\bibitem [{\citenamefont {Hofmann}\ \emph {et~al.}(2019)\citenamefont
  {Hofmann}, \citenamefont {Helbig}, \citenamefont {Lee}, \citenamefont
  {Greiter},\ and\ \citenamefont {Thomale}}]{hofmann:19}%
  \BibitemOpen
  \bibfield  {author} {\bibinfo {author} {\bibfnamefont {T.}~\bibnamefont
  {Hofmann}}, \bibinfo {author} {\bibfnamefont {T.}~\bibnamefont {Helbig}},
  \bibinfo {author} {\bibfnamefont {C.~H.}\ \bibnamefont {Lee}}, \bibinfo
  {author} {\bibfnamefont {M.}~\bibnamefont {Greiter}},\ and\ \bibinfo {author}
  {\bibfnamefont {R.}~\bibnamefont {Thomale}},\ }\bibfield  {title} {\bibinfo
  {title} {Chiral voltage propagation and calibration in a topolectrical chern
  circuit},\ }\href {https://doi.org/10.1103/PhysRevLett.122.247702} {\bibfield
   {journal} {\bibinfo  {journal} {Physical Review Letters}\ }\textbf {\bibinfo
  {volume} {122}} (\bibinfo {year} {2019})}\BibitemShut {NoStop}%
\bibitem [{\citenamefont {Lee}\ \emph {et~al.}(2018)\citenamefont {Lee},
  \citenamefont {Imhof}, \citenamefont {Berger}, \citenamefont {Bayer},
  \citenamefont {Brehm}, \citenamefont {Molenkamp}, \citenamefont {Kiessling},\
  and\ \citenamefont {Thomale}}]{lee:18}%
  \BibitemOpen
  \bibfield  {author} {\bibinfo {author} {\bibfnamefont {C.~H.}\ \bibnamefont
  {Lee}}, \bibinfo {author} {\bibfnamefont {S.}~\bibnamefont {Imhof}}, \bibinfo
  {author} {\bibfnamefont {C.}~\bibnamefont {Berger}}, \bibinfo {author}
  {\bibfnamefont {F.}~\bibnamefont {Bayer}}, \bibinfo {author} {\bibfnamefont
  {J.}~\bibnamefont {Brehm}}, \bibinfo {author} {\bibfnamefont {L.~W.}\
  \bibnamefont {Molenkamp}}, \bibinfo {author} {\bibfnamefont {T.}~\bibnamefont
  {Kiessling}},\ and\ \bibinfo {author} {\bibfnamefont {R.}~\bibnamefont
  {Thomale}},\ }\bibfield  {title} {\bibinfo {title} {Topolectrical circuits},\
  }\href {https://doi.org/10.1038/s42005-018-0035-2} {\bibfield  {journal}
  {\bibinfo  {journal} {Communications Physics}\ }\textbf {\bibinfo {volume}
  {1}} (\bibinfo {year} {2018})}\BibitemShut {NoStop}%
\bibitem [{\citenamefont {Lu}\ \emph {et~al.}(2019)\citenamefont {Lu},
  \citenamefont {Jia}, \citenamefont {Su}, \citenamefont {Owens}, \citenamefont
  {Juzeliunas}, \citenamefont {Schuster},\ and\ \citenamefont {Simon}}]{lu:19}%
  \BibitemOpen
  \bibfield  {author} {\bibinfo {author} {\bibfnamefont {Y.}~\bibnamefont
  {Lu}}, \bibinfo {author} {\bibfnamefont {N.}~\bibnamefont {Jia}}, \bibinfo
  {author} {\bibfnamefont {L.}~\bibnamefont {Su}}, \bibinfo {author}
  {\bibfnamefont {C.}~\bibnamefont {Owens}}, \bibinfo {author} {\bibfnamefont
  {G.}~\bibnamefont {Juzeliunas}}, \bibinfo {author} {\bibfnamefont {D.~I.}\
  \bibnamefont {Schuster}},\ and\ \bibinfo {author} {\bibfnamefont
  {J.}~\bibnamefont {Simon}},\ }\bibfield  {title} {\bibinfo {title} {Probing
  the berry curvature and fermi arcs of a weyl circuit},\ }\href
  {https://doi.org/10.1103/PhysRevB.99.020302} {\bibfield  {journal} {\bibinfo
  {journal} {Physical Review B}\ }\textbf {\bibinfo {volume} {99}} (\bibinfo
  {year} {2019})}\BibitemShut {NoStop}%
\bibitem [{\citenamefont {Rafi-Ul-Islam}\ \emph {et~al.}(2020)\citenamefont
  {Rafi-Ul-Islam}, \citenamefont {Siu}, \citenamefont {Sun},\ and\
  \citenamefont {Jalil}}]{Rafi-Ul-Islam:20}%
  \BibitemOpen
  \bibfield  {author} {\bibinfo {author} {\bibfnamefont {S.~M.}\ \bibnamefont
  {Rafi-Ul-Islam}}, \bibinfo {author} {\bibfnamefont {Z.~B.}\ \bibnamefont
  {Siu}}, \bibinfo {author} {\bibfnamefont {C.}~\bibnamefont {Sun}},\ and\
  \bibinfo {author} {\bibfnamefont {M.~B.~A.}\ \bibnamefont {Jalil}},\
  }\bibfield  {title} {\bibinfo {title} {Realization of weyl semimetal phases
  in topoelectrical circuits},\ }\href
  {https://doi.org/10.1088/1367-2630/ab6eaf} {\bibfield  {journal} {\bibinfo
  {journal} {New Journal of Physics}\ }\textbf {\bibinfo {volume} {22}}
  (\bibinfo {year} {2020})}\BibitemShut {NoStop}%
\bibitem [{\citenamefont {Benalcazar}\ \emph {et~al.}(2017)\citenamefont
  {Benalcazar}, \citenamefont {Bernevig},\ and\ \citenamefont
  {Hughes}}]{benalcazar:17}%
  \BibitemOpen
  \bibfield  {author} {\bibinfo {author} {\bibfnamefont {W.~A.}\ \bibnamefont
  {Benalcazar}}, \bibinfo {author} {\bibfnamefont {B.~A.}\ \bibnamefont
  {Bernevig}},\ and\ \bibinfo {author} {\bibfnamefont {T.~L.}\ \bibnamefont
  {Hughes}},\ }\bibfield  {title} {\bibinfo {title} {Quantized electric
  multipole insulators},\ }\href {https://doi.org/10.1126/science.aah6442}
  {\bibfield  {journal} {\bibinfo  {journal} {Science}\ }\textbf {\bibinfo
  {volume} {357}} (\bibinfo {year} {2017})}\BibitemShut {NoStop}%
\bibitem [{\citenamefont {Schindler}\ \emph {et~al.}(2018)\citenamefont
  {Schindler}, \citenamefont {Cook}, \citenamefont {Vergniory}, \citenamefont
  {Wang}, \citenamefont {Parkin}, \citenamefont {Bernevig},\ and\ \citenamefont
  {Neupert}}]{schindler:18}%
  \BibitemOpen
  \bibfield  {author} {\bibinfo {author} {\bibfnamefont {F.}~\bibnamefont
  {Schindler}}, \bibinfo {author} {\bibfnamefont {A.~M.}\ \bibnamefont {Cook}},
  \bibinfo {author} {\bibfnamefont {M.~G.}\ \bibnamefont {Vergniory}}, \bibinfo
  {author} {\bibfnamefont {Z.}~\bibnamefont {Wang}}, \bibinfo {author}
  {\bibfnamefont {S.~S.~P.}\ \bibnamefont {Parkin}}, \bibinfo {author}
  {\bibfnamefont {B.~A.}\ \bibnamefont {Bernevig}},\ and\ \bibinfo {author}
  {\bibfnamefont {T.}~\bibnamefont {Neupert}},\ }\bibfield  {title} {\bibinfo
  {title} {Higher-order topological insulators},\ }\href
  {https://doi.org/10.1126/sciadv.aat0346} {\bibfield  {journal} {\bibinfo
  {journal} {Science Advances}\ }\textbf {\bibinfo {volume} {4}} (\bibinfo
  {year} {2018})}\BibitemShut {NoStop}%
\bibitem [{\citenamefont {Xie}\ \emph {et~al.}(2021)\citenamefont {Xie},
  \citenamefont {Wang}, \citenamefont {Zhang}, \citenamefont {Zhan},
  \citenamefont {Jiang}, \citenamefont {Lu},\ and\ \citenamefont
  {Chen}}]{xie:21}%
  \BibitemOpen
  \bibfield  {author} {\bibinfo {author} {\bibfnamefont {B.}~\bibnamefont
  {Xie}}, \bibinfo {author} {\bibfnamefont {H.-X.}\ \bibnamefont {Wang}},
  \bibinfo {author} {\bibfnamefont {X.}~\bibnamefont {Zhang}}, \bibinfo
  {author} {\bibfnamefont {P.}~\bibnamefont {Zhan}}, \bibinfo {author}
  {\bibfnamefont {J.-H.}\ \bibnamefont {Jiang}}, \bibinfo {author}
  {\bibfnamefont {M.}~\bibnamefont {Lu}},\ and\ \bibinfo {author}
  {\bibfnamefont {Y.}~\bibnamefont {Chen}},\ }\bibfield  {title} {\bibinfo
  {title} {Higher-order band topology},\ }\href
  {https://doi.org/10.1038/s42254-021-00323-4} {\bibfield  {journal} {\bibinfo
  {journal} {Nature Reviews Physics}\ }\textbf {\bibinfo {volume} {3}},\
  \bibinfo {pages} {520} (\bibinfo {year} {2021})}\BibitemShut {NoStop}%
\bibitem [{\citenamefont {Imhof}\ \emph {et~al.}(2018)\citenamefont {Imhof},
  \citenamefont {Berger}, \citenamefont {Bayer}, \citenamefont {Brehm},
  \citenamefont {Molenkamp}, \citenamefont {Kiessling}, \citenamefont
  {Schindler}, \citenamefont {Lee}, \citenamefont {Greiter}, \citenamefont
  {Neupert},\ and\ \citenamefont {Thomale}}]{imhof:18}%
  \BibitemOpen
  \bibfield  {author} {\bibinfo {author} {\bibfnamefont {S.}~\bibnamefont
  {Imhof}}, \bibinfo {author} {\bibfnamefont {C.}~\bibnamefont {Berger}},
  \bibinfo {author} {\bibfnamefont {F.}~\bibnamefont {Bayer}}, \bibinfo
  {author} {\bibfnamefont {J.}~\bibnamefont {Brehm}}, \bibinfo {author}
  {\bibfnamefont {L.~W.}\ \bibnamefont {Molenkamp}}, \bibinfo {author}
  {\bibfnamefont {T.}~\bibnamefont {Kiessling}}, \bibinfo {author}
  {\bibfnamefont {F.}~\bibnamefont {Schindler}}, \bibinfo {author}
  {\bibfnamefont {C.~H.}\ \bibnamefont {Lee}}, \bibinfo {author} {\bibfnamefont
  {M.}~\bibnamefont {Greiter}}, \bibinfo {author} {\bibfnamefont
  {T.}~\bibnamefont {Neupert}},\ and\ \bibinfo {author} {\bibfnamefont
  {R.}~\bibnamefont {Thomale}},\ }\bibfield  {title} {\bibinfo {title}
  {Topolectrical-circuit realization of topological corner modes},\ }\href
  {https://doi.org/10.1038/s41567-018-0246-1} {\bibfield  {journal} {\bibinfo
  {journal} {Nature Physics}\ }\textbf {\bibinfo {volume} {14}} (\bibinfo
  {year} {2018})}\BibitemShut {NoStop}%
\bibitem [{\citenamefont {Serra-Garcia}\ \emph {et~al.}(2019)\citenamefont
  {Serra-Garcia}, \citenamefont {S\"{u}sstrunk},\ and\ \citenamefont
  {Huber}}]{serra-garcia:19}%
  \BibitemOpen
  \bibfield  {author} {\bibinfo {author} {\bibfnamefont {M.}~\bibnamefont
  {Serra-Garcia}}, \bibinfo {author} {\bibfnamefont {R.}~\bibnamefont
  {S\"{u}sstrunk}},\ and\ \bibinfo {author} {\bibfnamefont {S.~D.}\
  \bibnamefont {Huber}},\ }\bibfield  {title} {\bibinfo {title} {Observation of
  quadrupole transitions and edge mode topology in an lc circuit network},\
  }\href {https://doi.org/10.1103/PhysRevB.99.020304} {\bibfield  {journal}
  {\bibinfo  {journal} {Physical Review B}\ }\textbf {\bibinfo {volume} {99}}
  (\bibinfo {year} {2019})}\BibitemShut {NoStop}%
\bibitem [{\citenamefont {Lv}\ \emph {et~al.}(2021)\citenamefont {Lv},
  \citenamefont {Chen}, \citenamefont {Li}, \citenamefont {amd B.~Zhou},
  \citenamefont {Dong}, \citenamefont {Zhao}, \citenamefont {Li}, \citenamefont
  {Wang}, \citenamefont {Tao}, \citenamefont {Shi},\ and\ \citenamefont
  {Xu}}]{lv:21}%
  \BibitemOpen
  \bibfield  {author} {\bibinfo {author} {\bibfnamefont {B.}~\bibnamefont
  {Lv}}, \bibinfo {author} {\bibfnamefont {R.}~\bibnamefont {Chen}}, \bibinfo
  {author} {\bibfnamefont {R.}~\bibnamefont {Li}}, \bibinfo {author}
  {\bibfnamefont {C.~G.}\ \bibnamefont {amd B.~Zhou}}, \bibinfo {author}
  {\bibfnamefont {G.}~\bibnamefont {Dong}}, \bibinfo {author} {\bibfnamefont
  {C.}~\bibnamefont {Zhao}}, \bibinfo {author} {\bibfnamefont {Y.}~\bibnamefont
  {Li}}, \bibinfo {author} {\bibfnamefont {Y.}~\bibnamefont {Wang}}, \bibinfo
  {author} {\bibfnamefont {H.}~\bibnamefont {Tao}}, \bibinfo {author}
  {\bibfnamefont {J.}~\bibnamefont {Shi}},\ and\ \bibinfo {author}
  {\bibfnamefont {D.-H.}\ \bibnamefont {Xu}},\ }\bibfield  {title} {\bibinfo
  {title} {Realization of quasicrystalline quadrupole topological insulators in
  electrical circuits},\ }\href {https://doi.org/10.1038/s42005-021-00610-7}
  {\bibfield  {journal} {\bibinfo  {journal} {Communications Physics}\ }\textbf
  {\bibinfo {volume} {4}} (\bibinfo {year} {2021})}\BibitemShut {NoStop}%
\bibitem [{\citenamefont {Bao}\ \emph {et~al.}(2019)\citenamefont {Bao},
  \citenamefont {Zou}, \citenamefont {Zhang}, \citenamefont {He}, \citenamefont
  {Sun},\ and\ \citenamefont {Zhang}}]{bao:19}%
  \BibitemOpen
  \bibfield  {author} {\bibinfo {author} {\bibfnamefont {J.}~\bibnamefont
  {Bao}}, \bibinfo {author} {\bibfnamefont {D.}~\bibnamefont {Zou}}, \bibinfo
  {author} {\bibfnamefont {W.}~\bibnamefont {Zhang}}, \bibinfo {author}
  {\bibfnamefont {W.}~\bibnamefont {He}}, \bibinfo {author} {\bibfnamefont
  {H.}~\bibnamefont {Sun}},\ and\ \bibinfo {author} {\bibfnamefont
  {X.}~\bibnamefont {Zhang}},\ }\bibfield  {title} {\bibinfo {title}
  {Topoelectrical circuit octupole insulator with topologically protected
  corner states},\ }\href {https://doi.org/10.1103/PhysRevB.100.201406}
  {\bibfield  {journal} {\bibinfo  {journal} {Physical Review B}\ }\textbf
  {\bibinfo {volume} {100}} (\bibinfo {year} {2019})}\BibitemShut {NoStop}%
\bibitem [{\citenamefont {Liu}\ \emph {et~al.}(2020)\citenamefont {Liu},
  \citenamefont {Ma}, \citenamefont {Zhang}, \citenamefont {Zhang},
  \citenamefont {Yang}, \citenamefont {You}, \citenamefont {Gao}, \citenamefont
  {Xiang}, \citenamefont {Cui},\ and\ \citenamefont {Zhang}}]{liu:20}%
  \BibitemOpen
  \bibfield  {author} {\bibinfo {author} {\bibfnamefont {S.}~\bibnamefont
  {Liu}}, \bibinfo {author} {\bibfnamefont {S.}~\bibnamefont {Ma}}, \bibinfo
  {author} {\bibfnamefont {Q.}~\bibnamefont {Zhang}}, \bibinfo {author}
  {\bibfnamefont {L.}~\bibnamefont {Zhang}}, \bibinfo {author} {\bibfnamefont
  {C.}~\bibnamefont {Yang}}, \bibinfo {author} {\bibfnamefont {O.}~\bibnamefont
  {You}}, \bibinfo {author} {\bibfnamefont {W.}~\bibnamefont {Gao}}, \bibinfo
  {author} {\bibfnamefont {Y.}~\bibnamefont {Xiang}}, \bibinfo {author}
  {\bibfnamefont {T.~J.}\ \bibnamefont {Cui}},\ and\ \bibinfo {author}
  {\bibfnamefont {S.}~\bibnamefont {Zhang}},\ }\bibfield  {title} {\bibinfo
  {title} {Octupole corner state in a three-dimensional topological circuit},\
  }\href {https://doi.org/10.1038/s41377-020-00381-w} {\bibfield  {journal}
  {\bibinfo  {journal} {Light-Science \& Applications}\ }\textbf {\bibinfo
  {volume} {1}} (\bibinfo {year} {2020})}\BibitemShut {NoStop}%
\bibitem [{\citenamefont {Zhang}\ \emph {et~al.}(2020)\citenamefont {Zhang},
  \citenamefont {Zou}, \citenamefont {Bao}, \citenamefont {He}, \citenamefont
  {Pei}, \citenamefont {Sun},\ and\ \citenamefont {Zhang}}]{zhang:20}%
  \BibitemOpen
  \bibfield  {author} {\bibinfo {author} {\bibfnamefont {W.}~\bibnamefont
  {Zhang}}, \bibinfo {author} {\bibfnamefont {D.}~\bibnamefont {Zou}}, \bibinfo
  {author} {\bibfnamefont {J.}~\bibnamefont {Bao}}, \bibinfo {author}
  {\bibfnamefont {W.}~\bibnamefont {He}}, \bibinfo {author} {\bibfnamefont
  {Q.}~\bibnamefont {Pei}}, \bibinfo {author} {\bibfnamefont {H.}~\bibnamefont
  {Sun}},\ and\ \bibinfo {author} {\bibfnamefont {X.}~\bibnamefont {Zhang}},\
  }\bibfield  {title} {\bibinfo {title} {Topolectrical-circuit realization of a
  four-dimensional hexadecapole insulator},\ }\href
  {https://doi.org/10.1103/PhysRevB.102.100102} {\bibfield  {journal} {\bibinfo
   {journal} {Physical Review B}\ }\textbf {\bibinfo {volume} {102}} (\bibinfo
  {year} {2020})}\BibitemShut {NoStop}%
\bibitem [{\citenamefont {Ezawa}(2019{\natexlab{a}})}]{ezawa:19:a}%
  \BibitemOpen
  \bibfield  {author} {\bibinfo {author} {\bibfnamefont {M.}~\bibnamefont
  {Ezawa}},\ }\bibfield  {title} {\bibinfo {title} {Non-hermitian higher-order
  topological states in nonreciprocal and reciprocal systems with their
  electric-circuit realization},\ }\href
  {https://doi.org/10.1103/PhysRevB.99.201411} {\bibfield  {journal} {\bibinfo
  {journal} {Physical Review B}\ }\textbf {\bibinfo {volume} {99}} (\bibinfo
  {year} {2019}{\natexlab{a}})}\BibitemShut {NoStop}%
\bibitem [{\citenamefont {Zhang}\ \emph {et~al.}(2021)\citenamefont {Zhang},
  \citenamefont {Zou}, \citenamefont {Pei}, \citenamefont {He}, \citenamefont
  {Bao}, \citenamefont {Sun},\ and\ \citenamefont {Zhang}}]{zhang:21}%
  \BibitemOpen
  \bibfield  {author} {\bibinfo {author} {\bibfnamefont {W.}~\bibnamefont
  {Zhang}}, \bibinfo {author} {\bibfnamefont {D.}~\bibnamefont {Zou}}, \bibinfo
  {author} {\bibfnamefont {Q.}~\bibnamefont {Pei}}, \bibinfo {author}
  {\bibfnamefont {W.}~\bibnamefont {He}}, \bibinfo {author} {\bibfnamefont
  {J.}~\bibnamefont {Bao}}, \bibinfo {author} {\bibfnamefont {H.}~\bibnamefont
  {Sun}},\ and\ \bibinfo {author} {\bibfnamefont {X.}~\bibnamefont {Zhang}},\
  }\bibfield  {title} {\bibinfo {title} {Experimental observation of
  higher-order topological anderson insulators},\ }\href
  {https://doi.org/10.1103/PhysRevLett.126.146802} {\bibfield  {journal}
  {\bibinfo  {journal} {Physical Review Letters}\ }\textbf {\bibinfo {volume}
  {126}} (\bibinfo {year} {2021})}\BibitemShut {NoStop}%
\bibitem [{\citenamefont {Ezawa}(2018)}]{ezawa:18}%
  \BibitemOpen
  \bibfield  {author} {\bibinfo {author} {\bibfnamefont {M.}~\bibnamefont
  {Ezawa}},\ }\bibfield  {title} {\bibinfo {title} {Higher-order topological
  electric circuits and topological corner resonance on the breathing kagome
  and pyrochlore lattices},\ }\href
  {https://doi.org/10.1103/PhysRevB.98.201402} {\bibfield  {journal} {\bibinfo
  {journal} {Physical Review B}\ }\textbf {\bibinfo {volume} {98}} (\bibinfo
  {year} {2018})}\BibitemShut {NoStop}%
\bibitem [{\citenamefont {Ezawa}(2019{\natexlab{b}})}]{ezawa:19:b}%
  \BibitemOpen
  \bibfield  {author} {\bibinfo {author} {\bibfnamefont {M.}~\bibnamefont
  {Ezawa}},\ }\bibfield  {title} {\bibinfo {title} {Braiding of majorana-like
  corner states in electric circuits and its non-hermitian generalization},\
  }\href {https://doi.org/10.1103/PhysRevB.100.045407} {\bibfield  {journal}
  {\bibinfo  {journal} {Physical Review B}\ }\textbf {\bibinfo {volume} {100}}
  (\bibinfo {year} {2019}{\natexlab{b}})}\BibitemShut {NoStop}%
\bibitem [{\citenamefont {Song}\ \emph {et~al.}(2020)\citenamefont {Song},
  \citenamefont {Yang}, \citenamefont {Cao},\ and\ \citenamefont
  {Yan}}]{song:20}%
  \BibitemOpen
  \bibfield  {author} {\bibinfo {author} {\bibfnamefont {L.}~\bibnamefont
  {Song}}, \bibinfo {author} {\bibfnamefont {H.}~\bibnamefont {Yang}}, \bibinfo
  {author} {\bibfnamefont {Y.}~\bibnamefont {Cao}},\ and\ \bibinfo {author}
  {\bibfnamefont {P.}~\bibnamefont {Yan}},\ }\bibfield  {title} {\bibinfo
  {title} {Realization of the square-root higher-order topological insulator in
  electric circuits},\ }\href {https://doi.org/10.1021/acs.nanolett.0c03049}
  {\bibfield  {journal} {\bibinfo  {journal} {Nano Letters}\ }\textbf {\bibinfo
  {volume} {20}},\ \bibinfo {pages} {7566} (\bibinfo {year}
  {2020})}\BibitemShut {NoStop}%
\bibitem [{\citenamefont {Rafi-Ul-Islam}\ \emph {et~al.}(109)\citenamefont
  {Rafi-Ul-Islam}, \citenamefont {Siu}, \citenamefont {Sahin},\ and\
  \citenamefont {Jalil}}]{rafi-ul-islam:24}%
  \BibitemOpen
  \bibfield  {author} {\bibinfo {author} {\bibfnamefont {S.~M.}\ \bibnamefont
  {Rafi-Ul-Islam}}, \bibinfo {author} {\bibfnamefont {Z.~B.}\ \bibnamefont
  {Siu}}, \bibinfo {author} {\bibfnamefont {H.}~\bibnamefont {Sahin}},\ and\
  \bibinfo {author} {\bibfnamefont {M.~B.~A.}\ \bibnamefont {Jalil}},\
  }\bibfield  {title} {\bibinfo {title} {Chiral surface and hinge states in
  higher-order weyl semimetallic circuits},\ }\href
  {https://doi.org/10.1103/PhysRevB.109.085430} {\bibfield  {journal} {\bibinfo
   {journal} {Physical Review B}\ }\textbf {\bibinfo {volume} {8}} (\bibinfo
  {year} {109})}\BibitemShut {NoStop}%
\bibitem [{\citenamefont {Luo}\ \emph {et~al.}(2025)\citenamefont {Luo},
  \citenamefont {Song}, \citenamefont {Yang}, \citenamefont {Yan},\ and\
  \citenamefont {Cao}}]{luo:25}%
  \BibitemOpen
  \bibfield  {author} {\bibinfo {author} {\bibfnamefont {S.}~\bibnamefont
  {Luo}}, \bibinfo {author} {\bibfnamefont {L.}~\bibnamefont {Song}}, \bibinfo
  {author} {\bibfnamefont {H.}~\bibnamefont {Yang}}, \bibinfo {author}
  {\bibfnamefont {P.}~\bibnamefont {Yan}},\ and\ \bibinfo {author}
  {\bibfnamefont {Y.}~\bibnamefont {Cao}},\ }\bibfield  {title} {\bibinfo
  {title} {Topolectrical circuit simulation of two-dimensional graphyne
  structures},\ }\href {https://doi.org/10.1103/PhysRevB.111.024113} {\bibfield
   {journal} {\bibinfo  {journal} {Physical Review B}\ }\textbf {\bibinfo
  {volume} {111}} (\bibinfo {year} {2025})}\BibitemShut {NoStop}%
\bibitem [{\citenamefont {Song}\ \emph {et~al.}(2025)\citenamefont {Song},
  \citenamefont {Yang}, \citenamefont {Cao},\ and\ \citenamefont
  {Yan}}]{song:25}%
  \BibitemOpen
  \bibfield  {author} {\bibinfo {author} {\bibfnamefont {L.}~\bibnamefont
  {Song}}, \bibinfo {author} {\bibfnamefont {H.}~\bibnamefont {Yang}}, \bibinfo
  {author} {\bibfnamefont {Y.}~\bibnamefont {Cao}},\ and\ \bibinfo {author}
  {\bibfnamefont {P.}~\bibnamefont {Yan}},\ }\bibfield  {title} {\bibinfo
  {title} {Realization of the square-root dirac semimetal in electrical
  circuits},\ }\href {https://doi.org/10.1063/5.0251551} {\bibfield  {journal}
  {\bibinfo  {journal} {Journal of Applied Physics}\ }\textbf {\bibinfo
  {volume} {137}} (\bibinfo {year} {2025})}\BibitemShut {NoStop}%
\bibitem [{\citenamefont {Peterson}\ \emph {et~al.}(2018)\citenamefont
  {Peterson}, \citenamefont {Benalcazar}, \citenamefont {Hughes},\ and\
  \citenamefont {Bahl}}]{peterson:18}%
  \BibitemOpen
  \bibfield  {author} {\bibinfo {author} {\bibfnamefont {C.~W.}\ \bibnamefont
  {Peterson}}, \bibinfo {author} {\bibfnamefont {W.~A.}\ \bibnamefont
  {Benalcazar}}, \bibinfo {author} {\bibfnamefont {T.~L.}\ \bibnamefont
  {Hughes}},\ and\ \bibinfo {author} {\bibfnamefont {G.}~\bibnamefont {Bahl}},\
  }\bibfield  {title} {\bibinfo {title} {A quantized microwave quadrupole
  insulator with topologically protected corner states},\ }\href
  {https://doi.org/10.1038/nature25777} {\bibfield  {journal} {\bibinfo
  {journal} {Nature}\ }\textbf {\bibinfo {volume} {555}} (\bibinfo {year}
  {2018})}\BibitemShut {NoStop}%
\bibitem [{\citenamefont {Serra-Garcia}\ \emph {et~al.}(2018)\citenamefont
  {Serra-Garcia}, \citenamefont {Peri}, \citenamefont {S\"{u}sstrunk},
  \citenamefont {Bilal}, \citenamefont {Larsen}, \citenamefont {Villanueva},\
  and\ \citenamefont {Huber}}]{serra-garcia:18}%
  \BibitemOpen
  \bibfield  {author} {\bibinfo {author} {\bibfnamefont {M.}~\bibnamefont
  {Serra-Garcia}}, \bibinfo {author} {\bibfnamefont {V.}~\bibnamefont {Peri}},
  \bibinfo {author} {\bibfnamefont {R.}~\bibnamefont {S\"{u}sstrunk}}, \bibinfo
  {author} {\bibfnamefont {O.~R.}\ \bibnamefont {Bilal}}, \bibinfo {author}
  {\bibfnamefont {T.}~\bibnamefont {Larsen}}, \bibinfo {author} {\bibfnamefont
  {L.~G.}\ \bibnamefont {Villanueva}},\ and\ \bibinfo {author} {\bibfnamefont
  {S.~D.}\ \bibnamefont {Huber}},\ }\bibfield  {title} {\bibinfo {title}
  {Observation of a phononic quadrupole topological insulator},\ }\href
  {https://doi.org/10.1038/nature25156} {\bibfield  {journal} {\bibinfo
  {journal} {Nature}\ }\textbf {\bibinfo {volume} {555}} (\bibinfo {year}
  {2018})}\BibitemShut {NoStop}%
\bibitem [{\citenamefont {Mittal}\ \emph {et~al.}(2019)\citenamefont {Mittal},
  \citenamefont {Orre}, \citenamefont {Zhu}, \citenamefont {Gorlach},
  \citenamefont {Poddubny},\ and\ \citenamefont {Hafezi}}]{mittal:19}%
  \BibitemOpen
  \bibfield  {author} {\bibinfo {author} {\bibfnamefont {S.}~\bibnamefont
  {Mittal}}, \bibinfo {author} {\bibfnamefont {V.~V.}\ \bibnamefont {Orre}},
  \bibinfo {author} {\bibfnamefont {G.}~\bibnamefont {Zhu}}, \bibinfo {author}
  {\bibfnamefont {M.~A.}\ \bibnamefont {Gorlach}}, \bibinfo {author}
  {\bibfnamefont {A.}~\bibnamefont {Poddubny}},\ and\ \bibinfo {author}
  {\bibfnamefont {M.}~\bibnamefont {Hafezi}},\ }\bibfield  {title} {\bibinfo
  {title} {Photonic quadrupole topological phases},\ }\href
  {https://doi.org/10.1038/s41566-019-0452-0} {\bibfield  {journal} {\bibinfo
  {journal} {Nature Photonics}\ }\textbf {\bibinfo {volume} {13}} (\bibinfo
  {year} {2019})}\BibitemShut {NoStop}%
\bibitem [{\citenamefont {Xue}\ \emph {et~al.}(2019)\citenamefont {Xue},
  \citenamefont {Yang}, \citenamefont {Gao}, \citenamefont {Chong},\ and\
  \citenamefont {Zhang}}]{xue:19}%
  \BibitemOpen
  \bibfield  {author} {\bibinfo {author} {\bibfnamefont {H.}~\bibnamefont
  {Xue}}, \bibinfo {author} {\bibfnamefont {Y.}~\bibnamefont {Yang}}, \bibinfo
  {author} {\bibfnamefont {F.}~\bibnamefont {Gao}}, \bibinfo {author}
  {\bibfnamefont {Y.}~\bibnamefont {Chong}},\ and\ \bibinfo {author}
  {\bibfnamefont {B.}~\bibnamefont {Zhang}},\ }\bibfield  {title} {\bibinfo
  {title} {Acoustic higher-order topological insulator on a kagome lattice},\
  }\href {https://doi.org/10.1038/s41563-018-0251-x} {\bibfield  {journal}
  {\bibinfo  {journal} {Nature Materials}\ }\textbf {\bibinfo {volume} {18}}
  (\bibinfo {year} {2019})}\BibitemShut {NoStop}%
\bibitem [{\citenamefont {Cerjan}\ \emph {et~al.}(2020)\citenamefont {Cerjan},
  \citenamefont {Ju\"{u}rgensen}, \citenamefont {Benalcazar}, \citenamefont
  {Mukherjee},\ and\ \citenamefont {Rechtsman}}]{cerjan:20}%
  \BibitemOpen
  \bibfield  {author} {\bibinfo {author} {\bibfnamefont {A.}~\bibnamefont
  {Cerjan}}, \bibinfo {author} {\bibfnamefont {M.}~\bibnamefont
  {Ju\"{u}rgensen}}, \bibinfo {author} {\bibfnamefont {W.~A.}\ \bibnamefont
  {Benalcazar}}, \bibinfo {author} {\bibfnamefont {S.}~\bibnamefont
  {Mukherjee}},\ and\ \bibinfo {author} {\bibfnamefont {M.~C.}\ \bibnamefont
  {Rechtsman}},\ }\bibfield  {title} {\bibinfo {title} {Observation of a
  higher-order topological bound state in the continuum},\ }\href
  {https://doi.org/10.1103/PhysRevLett.125.213901} {\bibfield  {journal}
  {\bibinfo  {journal} {Physical Review Letters}\ }\textbf {\bibinfo {volume}
  {125}} (\bibinfo {year} {2020})}\BibitemShut {NoStop}%
\bibitem [{\citenamefont {Zhen}\ \emph {et~al.}()\citenamefont {Zhen},
  \citenamefont {Hsu}, \citenamefont {Lu}, \citenamefont {Stone},\ and\
  \citenamefont {Soljacic}}]{zhen:14}%
  \BibitemOpen
  \bibfield  {author} {\bibinfo {author} {\bibfnamefont {B.}~\bibnamefont
  {Zhen}}, \bibinfo {author} {\bibfnamefont {C.~W.}\ \bibnamefont {Hsu}},
  \bibinfo {author} {\bibfnamefont {L.}~\bibnamefont {Lu}}, \bibinfo {author}
  {\bibfnamefont {A.~D.}\ \bibnamefont {Stone}},\ and\ \bibinfo {author}
  {\bibfnamefont {M.}~\bibnamefont {Soljacic}},\ }\bibfield  {title} {\bibinfo
  {title} {Topological nature of optical bound states in the continuum},\
  }\href {https://doi.org/10.1103/PhysRevLett.113.257401} {\bibfield  {journal}
  {\bibinfo  {journal} {Physical Review Letters}\ }\textbf {\bibinfo {volume}
  {113}}}\BibitemShut {NoStop}%
\bibitem [{\citenamefont {Hsu}\ \emph {et~al.}(2016)\citenamefont {Hsu},
  \citenamefont {Zhen}, \citenamefont {Stone}, \citenamefont {Joannopoulos},\
  and\ \citenamefont {Soljacic}}]{hsu:16}%
  \BibitemOpen
  \bibfield  {author} {\bibinfo {author} {\bibfnamefont {C.~W.}\ \bibnamefont
  {Hsu}}, \bibinfo {author} {\bibfnamefont {B.}~\bibnamefont {Zhen}}, \bibinfo
  {author} {\bibfnamefont {A.~D.}\ \bibnamefont {Stone}}, \bibinfo {author}
  {\bibfnamefont {J.~D.}\ \bibnamefont {Joannopoulos}},\ and\ \bibinfo {author}
  {\bibfnamefont {M.}~\bibnamefont {Soljacic}},\ }\bibfield  {title} {\bibinfo
  {title} {Bound states in the continuum},\ }\href
  {https://doi.org/10.1038/natrevmats.2016.48} {\bibfield  {journal} {\bibinfo
  {journal} {Nature Reviews Materials}\ }\textbf {\bibinfo {volume} {1}}
  (\bibinfo {year} {2016})}\BibitemShut {NoStop}%
\bibitem [{\citenamefont {Kupriianov}\ \emph {et~al.}(2019)\citenamefont
  {Kupriianov}, \citenamefont {Xu}, \citenamefont {Sayanskiy}, \citenamefont
  {Dmitriev}, \citenamefont {Kivshar},\ and\ \citenamefont
  {Tuz}}]{kupriianov:19}%
  \BibitemOpen
  \bibfield  {author} {\bibinfo {author} {\bibfnamefont {A.~S.}\ \bibnamefont
  {Kupriianov}}, \bibinfo {author} {\bibfnamefont {Y.}~\bibnamefont {Xu}},
  \bibinfo {author} {\bibfnamefont {A.}~\bibnamefont {Sayanskiy}}, \bibinfo
  {author} {\bibfnamefont {V.}~\bibnamefont {Dmitriev}}, \bibinfo {author}
  {\bibfnamefont {Y.~S.}\ \bibnamefont {Kivshar}},\ and\ \bibinfo {author}
  {\bibfnamefont {V.~R.}\ \bibnamefont {Tuz}},\ }\bibfield  {title} {\bibinfo
  {title} {Metasurface engineering through bound states in the continuum},\
  }\href {https://doi.org/10.1103/PhysRevApplied.12.014024} {\bibfield
  {journal} {\bibinfo  {journal} {Physical Review Applied}\ }\textbf {\bibinfo
  {volume} {12}} (\bibinfo {year} {2019})}\BibitemShut {NoStop}%
\bibitem [{\citenamefont {Bogdanov}\ \emph {et~al.}(2019)\citenamefont
  {Bogdanov}, \citenamefont {Koshelev}, \citenamefont {Kapitanova},
  \citenamefont {Rybin}, \citenamefont {Gladyshev}, \citenamefont {Sadrieva},
  \citenamefont {Samusev}, \citenamefont {Kivshar},\ and\ \citenamefont
  {Limonov}}]{bogdanov:19}%
  \BibitemOpen
  \bibfield  {author} {\bibinfo {author} {\bibfnamefont {A.~A.}\ \bibnamefont
  {Bogdanov}}, \bibinfo {author} {\bibfnamefont {K.~L.}\ \bibnamefont
  {Koshelev}}, \bibinfo {author} {\bibfnamefont {P.~V.}\ \bibnamefont
  {Kapitanova}}, \bibinfo {author} {\bibfnamefont {M.~V.}\ \bibnamefont
  {Rybin}}, \bibinfo {author} {\bibfnamefont {S.~A.}\ \bibnamefont
  {Gladyshev}}, \bibinfo {author} {\bibfnamefont {Z.~F.}\ \bibnamefont
  {Sadrieva}}, \bibinfo {author} {\bibfnamefont {K.~B.}\ \bibnamefont
  {Samusev}}, \bibinfo {author} {\bibfnamefont {Y.~S.}\ \bibnamefont
  {Kivshar}},\ and\ \bibinfo {author} {\bibfnamefont {M.~F.}\ \bibnamefont
  {Limonov}},\ }\bibfield  {title} {\bibinfo {title} {Bound states in the
  continuum and fano resonances in the strong mode coupling regime},\ }\href
  {https://doi.org/10.1117/1.Ap.1.1.016001} {\bibfield  {journal} {\bibinfo
  {journal} {Advanced Photonics}\ }\textbf {\bibinfo {volume} {1}} (\bibinfo
  {year} {2019})}\BibitemShut {NoStop}%
\bibitem [{\citenamefont {Benalcazar}\ and\ \citenamefont
  {Cerjan}(2020)}]{benalcazar:20}%
  \BibitemOpen
  \bibfield  {author} {\bibinfo {author} {\bibfnamefont {W.~A.}\ \bibnamefont
  {Benalcazar}}\ and\ \bibinfo {author} {\bibfnamefont {A.}~\bibnamefont
  {Cerjan}},\ }\bibfield  {title} {\bibinfo {title} {Bound states in the
  continuum of higher-order topological insulators},\ }\href
  {https://doi.org/10.1103/PhysRevB.101.161116} {\bibfield  {journal} {\bibinfo
   {journal} {Physical Review B}\ }\textbf {\bibinfo {volume} {101}} (\bibinfo
  {year} {2020})}\BibitemShut {NoStop}%
\bibitem [{\citenamefont {Azzam}\ and\ \citenamefont
  {Kildishev}(2021)}]{azzam:21}%
  \BibitemOpen
  \bibfield  {author} {\bibinfo {author} {\bibfnamefont {S.~I.}\ \bibnamefont
  {Azzam}}\ and\ \bibinfo {author} {\bibfnamefont {A.}~\bibnamefont
  {Kildishev}},\ }\bibfield  {title} {\bibinfo {title} {Photonic bound states
  in the continuum: From basics to applications},\ }\href
  {https://doi.org/10.1002/adom.202001469} {\bibfield  {journal} {\bibinfo
  {journal} {Advanced Optical Materials}\ }\textbf {\bibinfo {volume} {9}}
  (\bibinfo {year} {2021})}\BibitemShut {NoStop}%
\bibitem [{\citenamefont {Koshelev}\ \emph {et~al.}(2023)\citenamefont
  {Koshelev}, \citenamefont {Sadrieva}, \citenamefont {Shcherbakov},
  \citenamefont {Kivshar},\ and\ \citenamefont {Bogdanov}}]{koshelev:23}%
  \BibitemOpen
  \bibfield  {author} {\bibinfo {author} {\bibfnamefont {K.~L.}\ \bibnamefont
  {Koshelev}}, \bibinfo {author} {\bibfnamefont {Z.~F.}\ \bibnamefont
  {Sadrieva}}, \bibinfo {author} {\bibfnamefont {A.~A.}\ \bibnamefont
  {Shcherbakov}}, \bibinfo {author} {\bibfnamefont {Y.~S.}\ \bibnamefont
  {Kivshar}},\ and\ \bibinfo {author} {\bibfnamefont {A.~A.}\ \bibnamefont
  {Bogdanov}},\ }\bibfield  {title} {\bibinfo {title} {Bound states in the
  continuum in photonic structures},\ }\href
  {https://doi.org/10.3367/UFNe.2021.12.039120} {\bibfield  {journal} {\bibinfo
   {journal} {Physics-Uspekhi}\ }\textbf {\bibinfo {volume} {66}},\ \bibinfo
  {pages} {494} (\bibinfo {year} {2023})}\BibitemShut {NoStop}%
\bibitem [{\citenamefont {Rafi-Ul-Islam}\ \emph {et~al.}(2022)\citenamefont
  {Rafi-Ul-Islam}, \citenamefont {Siu}, \citenamefont {Sahin},\ and\
  \citenamefont {Jalil}}]{rafi-ul-islam:22}%
  \BibitemOpen
  \bibfield  {author} {\bibinfo {author} {\bibfnamefont {S.~M.}\ \bibnamefont
  {Rafi-Ul-Islam}}, \bibinfo {author} {\bibfnamefont {Z.~B.}\ \bibnamefont
  {Siu}}, \bibinfo {author} {\bibfnamefont {H.}~\bibnamefont {Sahin}},\ and\
  \bibinfo {author} {\bibfnamefont {M.~B.~A.}\ \bibnamefont {Jalil}},\
  }\bibfield  {title} {\bibinfo {title} {Type-ii corner modes in topolectrical
  circuits},\ }\href {https://doi.org/10.1103/PhysRevB.106.245128} {\bibfield
  {journal} {\bibinfo  {journal} {Physical Review B}\ } (\bibinfo {year}
  {2022})}\BibitemShut {NoStop}%
\bibitem [{\citenamefont {Olekhno}\ \emph {et~al.}(2022)\citenamefont
  {Olekhno}, \citenamefont {Rozenblit}, \citenamefont {Kachin}, \citenamefont
  {Dmitriev}, \citenamefont {Burmistrov}, \citenamefont {Seregin},
  \citenamefont {Zhirihin},\ and\ \citenamefont {Gorlach}}]{olekhno:22}%
  \BibitemOpen
  \bibfield  {author} {\bibinfo {author} {\bibfnamefont {N.~A.}\ \bibnamefont
  {Olekhno}}, \bibinfo {author} {\bibfnamefont {A.~D.}\ \bibnamefont
  {Rozenblit}}, \bibinfo {author} {\bibfnamefont {V.~I.}\ \bibnamefont
  {Kachin}}, \bibinfo {author} {\bibfnamefont {A.~A.}\ \bibnamefont
  {Dmitriev}}, \bibinfo {author} {\bibfnamefont {O.~I.}\ \bibnamefont
  {Burmistrov}}, \bibinfo {author} {\bibfnamefont {P.~S.}\ \bibnamefont
  {Seregin}}, \bibinfo {author} {\bibfnamefont {D.~V.}\ \bibnamefont
  {Zhirihin}},\ and\ \bibinfo {author} {\bibfnamefont {M.~A.}\ \bibnamefont
  {Gorlach}},\ }\bibfield  {title} {\bibinfo {title} {Experimental realization
  of topological corner states in long-range-coupled electrical circuits},\
  }\href {https://doi.org/10.1103/PhysRevB.105.L081107} {\bibfield  {journal}
  {\bibinfo  {journal} {Physical Review B}\ }\textbf {\bibinfo {volume} {105}}
  (\bibinfo {year} {2022})}\BibitemShut {NoStop}%
\bibitem [{\citenamefont {Liu}\ and\ \citenamefont
  {Wakabayashi}(2017)}]{liu:17}%
  \BibitemOpen
  \bibfield  {author} {\bibinfo {author} {\bibfnamefont {F.}~\bibnamefont
  {Liu}}\ and\ \bibinfo {author} {\bibfnamefont {K.}~\bibnamefont
  {Wakabayashi}},\ }\bibfield  {title} {\bibinfo {title} {Novel topological
  phase with a zero berry curvature},\ }\href
  {https://doi.org/10.1103/PhysRevLett.118.076803} {\bibfield  {journal}
  {\bibinfo  {journal} {Physical Review Letters}\ }\textbf {\bibinfo {volume}
  {118}} (\bibinfo {year} {2017})}\BibitemShut {NoStop}%
\bibitem [{\citenamefont {Liu}\ \emph {et~al.}(2024)\citenamefont {Liu},
  \citenamefont {Cao}, \citenamefont {Xu}, \citenamefont {Xu}, \citenamefont
  {Li},\ and\ \citenamefont {Huang}}]{liu:24:prl}%
  \BibitemOpen
  \bibfield  {author} {\bibinfo {author} {\bibfnamefont {Z.}~\bibnamefont
  {Liu}}, \bibinfo {author} {\bibfnamefont {P.-C.}\ \bibnamefont {Cao}},
  \bibinfo {author} {\bibfnamefont {L.}~\bibnamefont {Xu}}, \bibinfo {author}
  {\bibfnamefont {G.}~\bibnamefont {Xu}}, \bibinfo {author} {\bibfnamefont
  {Y.}~\bibnamefont {Li}},\ and\ \bibinfo {author} {\bibfnamefont
  {J.}~\bibnamefont {Huang}},\ }\bibfield  {title} {\bibinfo {title}
  {Higher-order topological in-bulk corner state in pure diffusion systems},\
  }\href {https://doi.org/10.1103/PhysRevLett.132.176302} {\bibfield  {journal}
  {\bibinfo  {journal} {Physical Review Letters}\ }\textbf {\bibinfo {volume}
  {132}} (\bibinfo {year} {2024})}\BibitemShut {NoStop}%
\bibitem [{\citenamefont {Xie}\ \emph {et~al.}(2018)\citenamefont {Xie},
  \citenamefont {Wang}, \citenamefont {Wang}, \citenamefont {Zhu},
  \citenamefont {Jiang}, \citenamefont {Lu},\ and\ \citenamefont
  {Chen}}]{xie:18}%
  \BibitemOpen
  \bibfield  {author} {\bibinfo {author} {\bibfnamefont {B.-Y.}\ \bibnamefont
  {Xie}}, \bibinfo {author} {\bibfnamefont {H.-F.}\ \bibnamefont {Wang}},
  \bibinfo {author} {\bibfnamefont {H.-X.}\ \bibnamefont {Wang}}, \bibinfo
  {author} {\bibfnamefont {X.-Y.}\ \bibnamefont {Zhu}}, \bibinfo {author}
  {\bibfnamefont {J.-H.}\ \bibnamefont {Jiang}}, \bibinfo {author}
  {\bibfnamefont {M.-H.}\ \bibnamefont {Lu}},\ and\ \bibinfo {author}
  {\bibfnamefont {Y.-F.}\ \bibnamefont {Chen}},\ }\bibfield  {title} {\bibinfo
  {title} {Second-order photonic topological insulator with corner states},\
  }\href {https://doi.org/10.1103/PhysRevB.98.205147} {\bibfield  {journal}
  {\bibinfo  {journal} {Physical Review B}\ }\textbf {\bibinfo {volume} {98}},\
  \bibinfo {pages} {2469} (\bibinfo {year} {2018})}\BibitemShut {NoStop}%
\bibitem [{\citenamefont {Qian}\ \emph {et~al.}(2024)\citenamefont {Qian},
  \citenamefont {Zhang}, \citenamefont {Sun},\ and\ \citenamefont
  {Zhang}}]{qian:24}%
  \BibitemOpen
  \bibfield  {author} {\bibinfo {author} {\bibfnamefont {L.}~\bibnamefont
  {Qian}}, \bibinfo {author} {\bibfnamefont {W.}~\bibnamefont {Zhang}},
  \bibinfo {author} {\bibfnamefont {H.}~\bibnamefont {Sun}},\ and\ \bibinfo
  {author} {\bibfnamefont {X.}~\bibnamefont {Zhang}},\ }\bibfield  {title}
  {\bibinfo {title} {Non-abelian topological bound states in the continuum},\
  }\href {https://doi.org/10.1103/PhysRevLett.132.046601} {\bibfield  {journal}
  {\bibinfo  {journal} {Physical Review Letters}\ }\textbf {\bibinfo {volume}
  {132}} (\bibinfo {year} {2024})}\BibitemShut {NoStop}%
\bibitem [{\citenamefont {Liu}\ \emph {et~al.}(2019)\citenamefont {Liu},
  \citenamefont {Gao}, \citenamefont {Zhang}, \citenamefont {Ma}, \citenamefont
  {Zhang}, \citenamefont {Liu}, \citenamefont {Xiang}, \citenamefont {Cui},\
  and\ \citenamefont {Zhang}}]{liu:19}%
  \BibitemOpen
  \bibfield  {author} {\bibinfo {author} {\bibfnamefont {S.}~\bibnamefont
  {Liu}}, \bibinfo {author} {\bibfnamefont {W.}~\bibnamefont {Gao}}, \bibinfo
  {author} {\bibfnamefont {Q.}~\bibnamefont {Zhang}}, \bibinfo {author}
  {\bibfnamefont {S.}~\bibnamefont {Ma}}, \bibinfo {author} {\bibfnamefont
  {L.}~\bibnamefont {Zhang}}, \bibinfo {author} {\bibfnamefont
  {C.}~\bibnamefont {Liu}}, \bibinfo {author} {\bibfnamefont {Y.~J.}\
  \bibnamefont {Xiang}}, \bibinfo {author} {\bibfnamefont {T.~J.}\ \bibnamefont
  {Cui}},\ and\ \bibinfo {author} {\bibfnamefont {S.}~\bibnamefont {Zhang}},\
  }\bibfield  {title} {\bibinfo {title} {Topologically protected edge state in
  two-dimensional su-schrieffer-heeger circuit},\ }\href
  {https://doi.org/10.34133/2019/8609875} {\bibfield  {journal} {\bibinfo
  {journal} {Research}\ } (\bibinfo {year} {2019})}\BibitemShut {NoStop}%
\bibitem [{\citenamefont {Robnik}(1986)}]{robnik:86}%
  \BibitemOpen
  \bibfield  {author} {\bibinfo {author} {\bibfnamefont {M.}~\bibnamefont
  {Robnik}},\ }\bibfield  {title} {\bibinfo {title} {A simple separable
  hamiltonian having bound states in the continuum},\ }\href
  {https://doi.org/10.1088/0305-4470/19/18/029} {\bibfield  {journal} {\bibinfo
   {journal} {Journal of Physics {A}-Mathematical and General}\ }\textbf
  {\bibinfo {volume} {19}} (\bibinfo {year} {1986})}\BibitemShut {NoStop}%
\bibitem [{\citenamefont {Franca}\ \emph {et~al.}(2024)\citenamefont {Franca},
  \citenamefont {Seidemann}, \citenamefont {Hassler}, \citenamefont {van~den
  Brink},\ and\ \citenamefont {Fulga}}]{franca:24}%
  \BibitemOpen
  \bibfield  {author} {\bibinfo {author} {\bibfnamefont {S.}~\bibnamefont
  {Franca}}, \bibinfo {author} {\bibfnamefont {T.}~\bibnamefont {Seidemann}},
  \bibinfo {author} {\bibfnamefont {F.}~\bibnamefont {Hassler}}, \bibinfo
  {author} {\bibfnamefont {J.}~\bibnamefont {van~den Brink}},\ and\ \bibinfo
  {author} {\bibfnamefont {I.~C.}\ \bibnamefont {Fulga}},\ }\bibfield  {title}
  {\bibinfo {title} {Impedance spectroscopy of chiral symmetric topoelectrical
  circuits},\ }\href {https://doi.org/10.1103/PhysRevB.109.L241103} {\bibfield
  {journal} {\bibinfo  {journal} {Physical Review B}\ }\textbf {\bibinfo
  {volume} {109}} (\bibinfo {year} {2024})}\BibitemShut {NoStop}%
\bibitem [{\citenamefont {Li}\ \emph {et~al.}(2020)\citenamefont {Li},
  \citenamefont {Zhirihin}, \citenamefont {Gorlach}, \citenamefont {Ni},
  \citenamefont {Filonov}, \citenamefont {Slobozhanyuk}, \citenamefont {Alu},\
  and\ \citenamefont {Khanikaev}}]{li:2020}%
  \BibitemOpen
  \bibfield  {author} {\bibinfo {author} {\bibfnamefont {M.}~\bibnamefont
  {Li}}, \bibinfo {author} {\bibfnamefont {D.}~\bibnamefont {Zhirihin}},
  \bibinfo {author} {\bibfnamefont {M.}~\bibnamefont {Gorlach}}, \bibinfo
  {author} {\bibfnamefont {X.}~\bibnamefont {Ni}}, \bibinfo {author}
  {\bibfnamefont {D.}~\bibnamefont {Filonov}}, \bibinfo {author} {\bibfnamefont
  {A.}~\bibnamefont {Slobozhanyuk}}, \bibinfo {author} {\bibfnamefont
  {A.}~\bibnamefont {Alu}},\ and\ \bibinfo {author} {\bibfnamefont {A.~B.}\
  \bibnamefont {Khanikaev}},\ }\bibfield  {title} {\bibinfo {title}
  {Higher-order topological states in photonic kagome crystals with long-range
  interactions},\ }\href {https://doi.org/10.1038/s41566-019-0561-9} {\bibfield
   {journal} {\bibinfo  {journal} {Nature Photonics}\ }\textbf {\bibinfo
  {volume} {14}} (\bibinfo {year} {2020})}\BibitemShut {NoStop}%
\bibitem [{\citenamefont {Xiong}\ \emph {et~al.}(2024)\citenamefont {Xiong},
  \citenamefont {Wang}, \citenamefont {Zhang}, \citenamefont {Zhang},
  \citenamefont {Cheng},\ and\ \citenamefont {Liu}}]{xiong:24}%
  \BibitemOpen
  \bibfield  {author} {\bibinfo {author} {\bibfnamefont {W.}~\bibnamefont
  {Xiong}}, \bibinfo {author} {\bibfnamefont {S.}~\bibnamefont {Wang}},
  \bibinfo {author} {\bibfnamefont {Z.}~\bibnamefont {Zhang}}, \bibinfo
  {author} {\bibfnamefont {H.}~\bibnamefont {Zhang}}, \bibinfo {author}
  {\bibfnamefont {Y.}~\bibnamefont {Cheng}},\ and\ \bibinfo {author}
  {\bibfnamefont {X.}~\bibnamefont {Liu}},\ }\bibfield  {title} {\bibinfo
  {title} {Observation of multiple off-site corner states induced by
  next-nearest-neighbor coupling in a sonic crystal},\ }\href
  {https://doi.org/10.1103/PhysRevB.109.024305} {\bibfield  {journal} {\bibinfo
   {journal} {Physical Review B}\ }\textbf {\bibinfo {volume} {109}} (\bibinfo
  {year} {2024})}\BibitemShut {NoStop}%
\bibitem [{\citenamefont {Helbig}\ \emph {et~al.}(2020)\citenamefont {Helbig},
  \citenamefont {Hofmann}, \citenamefont {Imhof}, \citenamefont {Abdelghany},
  \citenamefont {Kiessling}, \citenamefont {Molenkamp}, \citenamefont {Lee},
  \citenamefont {Szameit}, \citenamefont {Greiter},\ and\ \citenamefont
  {Thomale}}]{helbig:20}%
  \BibitemOpen
  \bibfield  {author} {\bibinfo {author} {\bibfnamefont {T.}~\bibnamefont
  {Helbig}}, \bibinfo {author} {\bibfnamefont {T.}~\bibnamefont {Hofmann}},
  \bibinfo {author} {\bibfnamefont {S.}~\bibnamefont {Imhof}}, \bibinfo
  {author} {\bibfnamefont {M.}~\bibnamefont {Abdelghany}}, \bibinfo {author}
  {\bibfnamefont {T.}~\bibnamefont {Kiessling}}, \bibinfo {author}
  {\bibfnamefont {L.~W.}\ \bibnamefont {Molenkamp}}, \bibinfo {author}
  {\bibfnamefont {C.~H.}\ \bibnamefont {Lee}}, \bibinfo {author} {\bibfnamefont
  {A.}~\bibnamefont {Szameit}}, \bibinfo {author} {\bibfnamefont
  {M.}~\bibnamefont {Greiter}},\ and\ \bibinfo {author} {\bibfnamefont
  {R.}~\bibnamefont {Thomale}},\ }\bibfield  {title} {\bibinfo {title}
  {Generalized bulk-boundary correspondence in non-hermitian topolectrical
  circuits},\ }\href {https://doi.org/10.1038/s41567-020-0922-9} {\bibfield
  {journal} {\bibinfo  {journal} {Nature Physics}\ }\textbf {\bibinfo {volume}
  {16}} (\bibinfo {year} {2020})}\BibitemShut {NoStop}%
\end{thebibliography}

%

\end{document}




\title{{\small Supplementary Material}\\\texorpdfstring{Higher-order topological bound states in the continuum in a topoelectrical lattice with long-range coupling}{Higher-order topological bound states in the continuum in a topoelectrical lattice with long-range coupling}
}



\author{Araceli \surname{Guti\'{e}rrez--Llorente}}
\email[]{araceli.gutierrez@urjc.es}
\affiliation{Universidad Rey Juan Carlos, Escuela Superior de Ciencias Experimentales y Tecnolog\'{i}a, Madrid 28933, Spain}



\maketitle 
\tableofcontents

\newpage

\section{2D SSH circuit}

\subsection{Localization and amplitude decay of corner-localized states in a bipartite lattice}

\begin{figure}[htb]
 \includegraphics[keepaspectratio=true, width=0.95\linewidth]{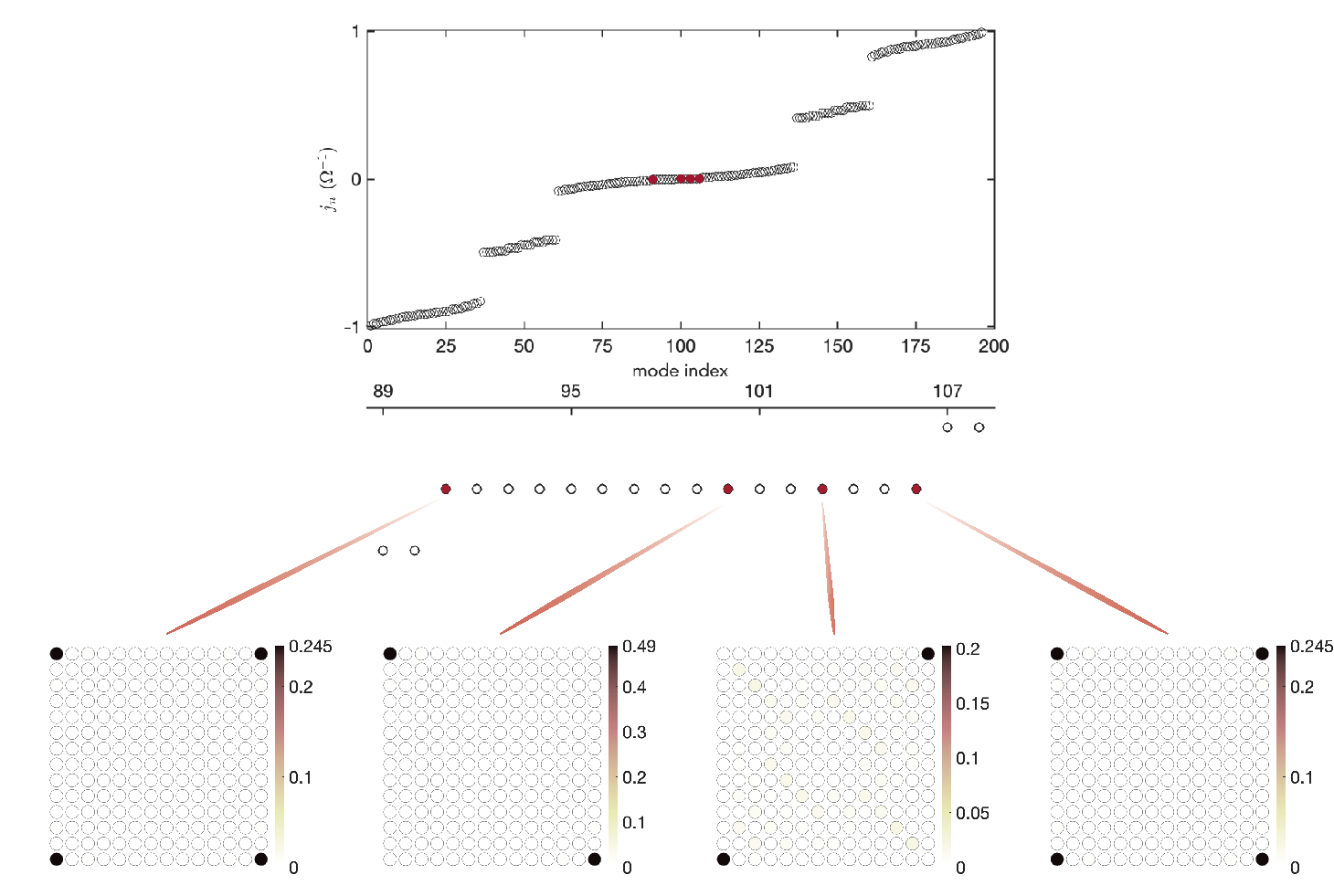}\caption{Top panel: Enlarged view of the four corner-localized modes embedded in the continuum. Bottom panel: Normalized spatial distribution of the amplitude for each corner-localized eigenmode, $|\psi(\omega)|^2$. The colourmap quantifies the degree of localization across the circuit layout ($N=196$ nodes). Two modes exhibit sharp localization at the four corners. These modes are symmetric combinations of corner states confined to different sublattices within the bipartite lattice.  Two other modes are primarily confined to the corners along the diagonals, each supported by a single sublattice and favouring one diagonal, with one exhibiting comparatively weaker confinement.}
 \label{Fig_SI_002}
\end{figure}

\begin{figure}[htb]
 \includegraphics[keepaspectratio=true, width=0.95\linewidth]{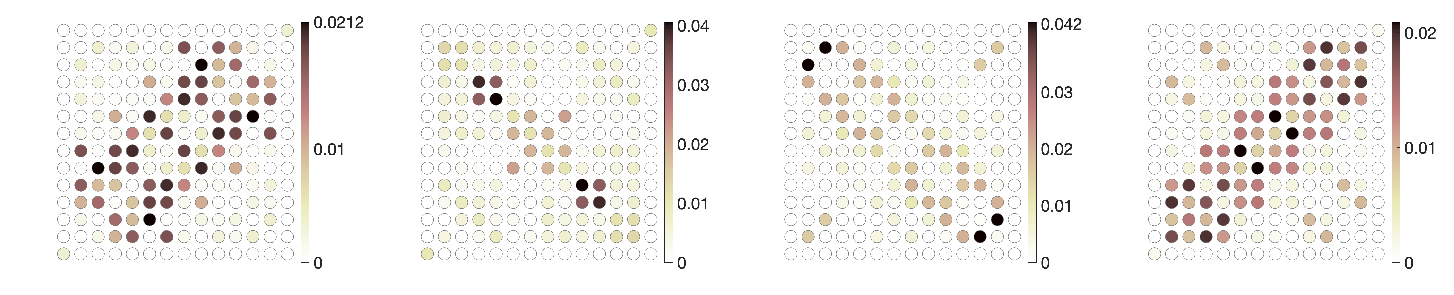}\caption{Normalized spatial distribution of the amplitude for selected bulk modes with zero admittance, $|\psi(\omega)|^2$. The colourmap quantifies the degree of localization across the circuit layout ($N=196$ nodes).  In contrast to corner-localized modes, which exhibit sharp confinement at specific nodes, these bulk modes are more spatially extended, reflecting the delocalized nature of the continuum.}
 \label{Fig_SI_003}
\end{figure}


\begin{table}[hbt]
\caption{\label{tab:table_S001}Inner product values between corner-localized modes and bulk modes. The computed values, ranging from $10^{-16}$ to $10^{-18}$) confirm the spatial and spectral isolation of the corner modes within numerical precision, consistent with their topological protection and bound-in-the-continuum character.}
\renewcommand{\arraystretch}{1.3}
\begin{tabular}{|c||c|c|c|c|c|c|c|c|c|c|c|c|c|c|c|c||}
\toprule
Corner \textbackslash Bulk & \#92 & \#93 & \#94 & \#95 & \#96 & \#97 & \#98 & \#99 & \#101 & \#102 & \#104 & \#105 \\
\hline\hline
\#91 & 1.79e-16 & 1.86e-16 & 3.36e-17 & 4.27e-17 & 1.24e-16 & 6.12e-16 & 5.09e-17 & 1.20e-16 & 1.26e-16 & 1.39e-16 & 6.71e-17 & 2.66e-17 \\
\hline
\#100 & 3.02e-16 & 6.31e-17 & 4.34e-17 & 1.02e-16 & 2.69e-16 & 7.27e-17 & 2.18e-16 & 3.28e-17 & 2.86e-16 & 3.82e-16 & 2.39e-16 & 1.71e-16 \\
\hline
\#103 & 2.46e-18 & 4.42e-16 & 5.64e-16 & 7.73e-16 & 1.71e-16 & 6.39e-17 & 1.63e-16 & 2.64e-16 & 1.57e-16 & 3.24e-16 & 2.11e-16 & 6.15e-17 \\
\hline
\#106 & 5.68e-17 & 3.21e-17 & 1.74e-16 & 4.86e-17 & 2.26e-17 & 5.25e-16 & 1.35e-16 & 9.90e-17 & 4.63e-17 & 3.98e-17 & 2.54e-18 & 4.68e-18 \\
\bottomrule
\end{tabular}
\end{table}

\clearpage

\begin{figure}[htb]
\includegraphics[keepaspectratio=true, width=0.65\linewidth]{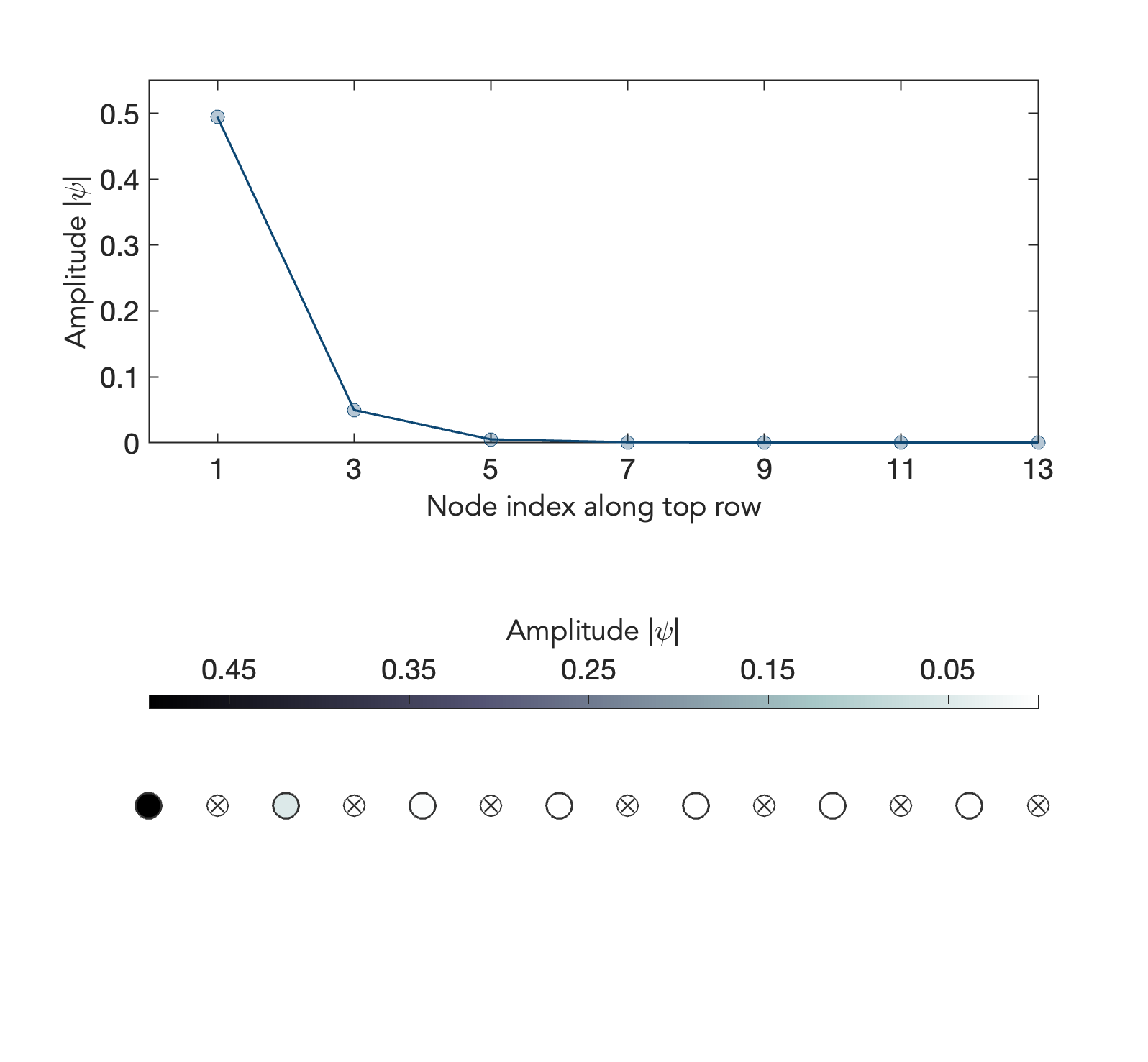}\caption{Upper panel: Linear-scale plot of the amplitude $|\psi|$ of a corner-localized state in the 2D SSH lattice ($N=196$ nodes, coupling ratio $\lambda = 0.1$), localized at the top-left corner and supported by one of the sublattices within the bipartite lattice. The decay is consistent with the expected exponential localization $\psi(n) \propto \lambda^n $, where $n$ denotes the number of unit-cell steps from the corner. Node indices are indicated on the horizontal axis. Lower panel: Schematic of the first row, where circles represent all nodes, filled markers indicate nodes of the supporting sublattice, and crosses mark nodes of the opposite sublattice. The color map encodes the wavefunction amplitude $|\psi|$. The eigenmode of the full lattice localized at all four corners is a symmetric combination of such corner-localized states, each confined to a different sublattice. See also Fig.~\ref{Fig_map_and_decay}.}
 \label{Fig_combined}
\end{figure}

\begin{figure}[htb]
 \includegraphics[keepaspectratio=true, width=0.95\linewidth]{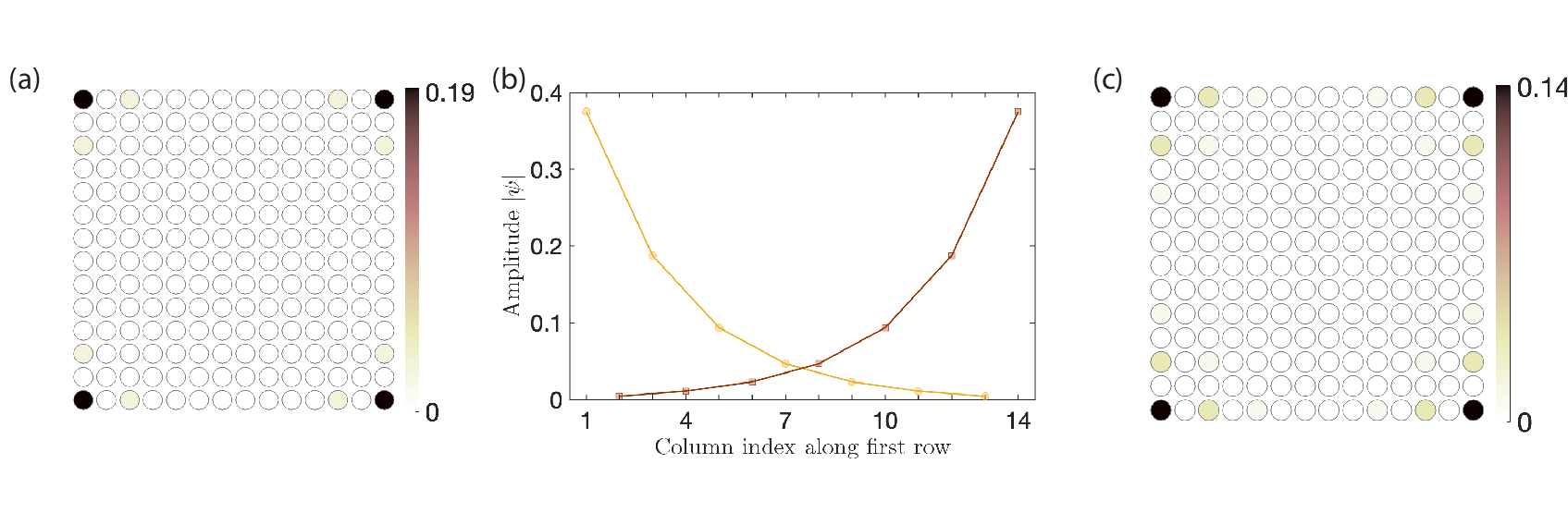}\caption{
(a) Normalized spatial distribution of the amplitude for a corner mode with zero admittance, $|\psi(\omega)|^2$, in a 2D SSH lattice with coupling ratio $\lambda = 0.35$. The colourmap quantifies the degree of localization across the circuit layout ($N=196$ nodes). (b) Amplitude profiles $|\psi|$ along the first row for two corner-localized states corresponding to the mode shown in panel (a) with $\lambda = 0.35$. These states are localized at the top-left and top-right corners of the lattice. Circles and squares denotate nodes belonging to opposite sublattices.  (c) Normalized spatial distribution of the amplitude $|\psi(\omega)|^2$ for a corner mode with zero admittance for coupling ratio $\lambda = 0.5$.}
\label{Fig_map_and_decay}
\end{figure}

\clearpage
\subsection{Circuit Laplacian under periodic boundaries}

\begin{figure}[htb]
\includegraphics[keepaspectratio=true, width=0.31\linewidth]{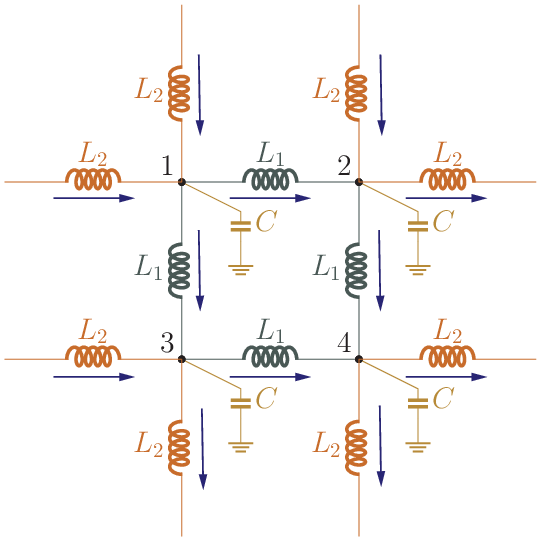}\caption{Unit cell of the 2D SSH circuit illustrating current directions (blue arrows) for nodal analysis in terms of node voltages, defined with respect to a common ground. $L_1$ and $L_2$ represent the intracell and intercell couplings, respectively.}
\label{Fig_uc_current_directions}
\end{figure}

The unit cell of the 2D SSH circuit shown in Fig.~\ref{Fig_uc_current_directions} consists of four nodes labeled 1-4. The node voltages denoted by $V_1, \:\:i=1,\dots,4$ are measured with respect to the common ground. The branch voltages across the inductors can be expressed in terms of these node voltages. Applying Kirchhoff’s Current Law (KCL) at each node $V_i$, with currents leaving the node taken as positive (as indicated in Fig.~\ref{Fig_uc_current_directions}) yields a set of coupled equations for the node voltages.  Using Bloch’s theorem, the node voltages are represented as Bloch waves, enabling the periodic lattice to be described by a finite matrix equation for the unit cell, which depends on the Bloch wavevector ${\bf{k}}=(k_x, k_y)$.

\setlength{\jot}{12pt}
\begin{align}
I_{1} &= \frac{V_1-V_2}{i\omega L_1} + \frac{V_1-V_3}{i\omega L_1} - \frac{V_2 e^{-ik_x}-V_1}{i\omega L_2} -\frac{V_3 e^{-ik_y}-V_1}{i\omega L_2} + i\omega C \:V_{1} \label{eq:I1}\\
I_{2} &= -\frac{V_1-V_2}{i\omega L_1} + \frac{V_2-V_4}{i\omega L_1} + \frac{V_2 -V_1 e^{ik_x}}{i\omega L_2} -\frac{V_4 e^{-ik_y}-V_2}{i\omega L_2}    + i\omega C \:V_{2} \label{eq:I2}\\
I_{3}&= -\frac{V_1-V_3}{i\omega L_1} + \frac{V_3-V_4}{i\omega L_1}+ \frac{V_3 -V_1 e^{ik_y}}{i\omega L_2} - \frac{V_4 e^{-ik_x}-V_3}{i\omega L_2}   + i\omega C \: V_{3} \label{eq:I3}\\
I_{4} &= - \frac{V_2-V_4}{i\omega L_1} -\frac{V_3-V_4}{i\omega L_1} + \frac{V_4 -V_3 e^{ik_x}}{i\omega L_2} + \frac{V_4-V_2 e^{ik_y}}{i\omega L_2}  + i\omega C \: V_{4} \label{eq:I4}
\end{align} 

The matrix equation, derived from Eq.~\ref{eq:I1} to Eq.~\ref{eq:I4}, takes the form $\mathbf{I} = J(\omega, {\bf{k}}) \mathbf{V}$, where $\mathbf{I}$ is the column vector representing the algebraic sum of all branch currents flowing at each node node, $\mathbf{V}$ is the column vector of node voltages, and $J(\omega)$ denotes the circuit Laplacian or admittance matrix: 

\setlength{\arraycolsep}{12pt} 
\begin{equation}\label{eq:matrix_I_V}
\begin{bmatrix}
I_1\\
I_2\\
I_3\\
I_4\\
\end{bmatrix}
= J(\omega, {\bf{k}})
\begin{bmatrix}
V_1\\
V_2\\
V_3\\
V_4\\
\end{bmatrix}
\end{equation}

The admittance matrix is given by

\setlength{\arraycolsep}{12pt} 
\begin{equation}\label{eq:J}
J(\omega, {\bf{k}})= i \omega 
\begin{bmatrix}
C - \frac{2}{\omega^2 L_1} - \frac{2}{\omega^2 L_2} & -\frac{1}{i\omega L_1} - \frac{e^{-ik_x}}{i\omega L_2}  & -\frac{1}{i\omega L_1} - \frac{e^{-ik_y}}{i\omega L_2} & 0\\[6pt]
-\frac{1}{i\omega L_1} - \frac{e^{ik_x}}{i\omega L_2} & C - \frac{2}{\omega^2 L_1} - \frac{2}{\omega^2 L_2} & 0 & -\frac{1}{i\omega L_1} - \frac{e^{-ik_y}}{i\omega L_2}\\[6pt]
-\frac{1}{i\omega L_1} - \frac{e^{ik_y}}{i\omega L_2} & 0 & C - \frac{2}{\omega^2 L_1} - \frac{2}{\omega^2 L_2} & -\frac{1}{i\omega L_1} - \frac{e^{-ik_x}}{i\omega L_2}\\[6pt]
0 & -\frac{1}{i\omega L_1} - \frac{e^{ik_y}}{i\omega L_2} & -\frac{1}{i\omega L_1} - \frac{e^{ik_x}}{i\omega L_2} & C - \frac{2}{\omega^2 L_1} - \frac{2}{\omega^2 L_2}\\
\end{bmatrix}
\end{equation}\\

This matrix can be expressed as

\begin{equation}\label{eq_laplacian_halmiltonian}
J(\omega, {\bf{k}}) = i \omega \left[ \left(C - \frac{2}{\omega^2 L_1} - \frac{2}{\omega^2 L_2}\right) \mathbb{I} +  H(\omega, {\bf{k}}) \right]  
\end{equation}

where $\mathbb{I}$ is the $4\times 4$ identity matrix and $H(\omega, {\bf{k}})$ is given by the Hermitian matrix

\setlength{\arraycolsep}{12pt} 
\begin{equation}\label{eq:hamiltonian_01}
H(\omega, {\bf{k}})=  
\begin{bmatrix}
0 & \frac{1}{\omega^2 L_1} + \frac{e^{-ik_x}}{\omega^2 L_2}  & \frac{1}{\omega^2 L_1} + \frac{e^{-ik_y}}{\omega^2 L_2} & 0\\[6pt]
\frac{1}{\omega^2 L_1} + \frac{e^{ik_x}}{\omega^2 L_2} & 0 & 0 & \frac{1}{i\omega^2 L_1} + \frac{e^{-ik_y}}{\omega^2 L_2}\\[6pt]
\frac{1}{\omega^2 L_1} + \frac{e^{ik_y}}{\omega^2 L_2} & 0 & 0 & \frac{1}{\omega^2 L_1} + \frac{e^{-ik_x}}{\omega^2 L_2}\\[6pt]
0 & \frac{1}{\omega^2 L_1} + \frac{e^{ik_y}}{\omega^2 L_2} & \frac{1}{\omega^2 L_1} + \frac{e^{ik_x}}{\omega^2 L_2} & 0\\
\end{bmatrix}
\end{equation}

At frequency $\omega_0$ given by

\begin{equation}\label{eq_res_frq}
\omega_0=\sqrt{\frac{2(L_1 + L_2)}{C L_1 L_2}}
\end{equation}

the diagonal terms in Eq.~\ref{eq_laplacian_halmiltonian} vanish, and the Laplacian becomes becomes purely off-diagonal, resembling a tight-binding Hamiltonian

\setlength{\arraycolsep}{12pt} 
\begin{equation}\label{eq:laplacian_02}
J(\omega_0, {\bf{k}})=  i \sqrt{\frac{C}{2L_1(\lambda +1)}}
\begin{bmatrix}
0 & 1 + \lambda e^{-ik_x}   & 1 + \lambda e^{-ik_y} & 0\\[6pt]
 1 + \lambda e^{ik_x} & 0 & 0 & 1 + \lambda e^{-ik_y}\\[6pt]
1 + \lambda e^{ik_y} & 0 & 0 & 1 + \lambda e^{-ik_x}\\[6pt]
0 & 1 + \lambda e^{ik_y} & 1 + \lambda e^{ik_x} & 0\\
\end{bmatrix}
\end{equation}

where $\lambda = L_1/L_2$. Therefore, the circuit Laplacian, whose eigenvalues have dimensions of admittance (see Eq.~\ref{eq:laplacian_02}), mimics the Hamiltonian of the SSH model.

\subsection{Circuit Laplacian with open boundaries}

\begin{figure}[htb]
\centering
\includegraphics[width=0.6\linewidth]{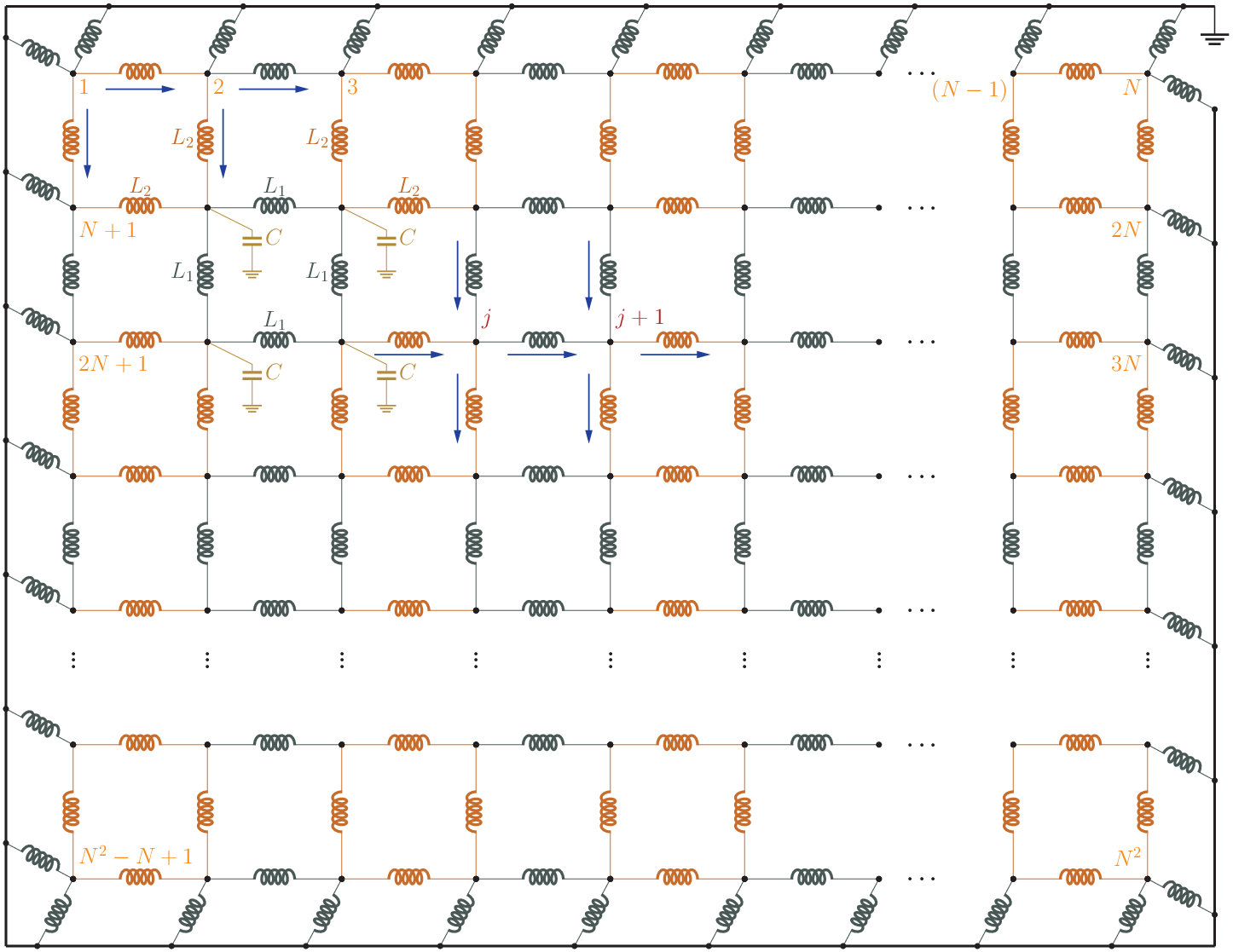}
\caption{2D SSH circuit with open boundaries and an $N\times N$ node configuration. Inductors $L_1$ (green) and $L_2$ (brown) represent intracell and intercell couplings, respectively. All nodes, including edge and corner nodes, are grounded via capacitors $C$ (yellow), which are omitted for clarity. Node labels $1,\:2,\:3,\dots,\:N-1,\:N,\: \dots,\:N^2$ are shown in orange, and nodes $j$ and $j+1$ are highlighted to illustrate the nodal analysis equations (Eq.~\ref{eq:nodal_j} and Eq.\ref{eq:nodal_j+1}).  Blue arrows indicate current directions used in nodal analysis.}
\label{Fig_circuit_NN_finite_w_currents}
\end{figure}

\clearpage

Using nodal analysis, the input currents at the first two nodes of the circuit are given by:

\begin{equation}
\begin{aligned}
I_1 &= \frac{V_1-V_2}{i\omega L_2} + \frac{V_1-V_{N+1}}{i \omega L_2} + V_1 \left( \frac{2}{i \omega L_1} + i \omega C\right) = \left(\frac{2}{i \omega L_1} + \frac{2}{i \omega L_2} + i \omega C \right)V_1 -\frac{1}{i \omega L_2} V_2 -\frac{1}{i \omega L_2} V_{N+1}\\[1.2ex]
&=i\omega\:\left[ \left(C- \frac{2}{ \omega^2 L_1} - \frac{2}{ \omega^2 L_2}  \right)V_1 + \frac{1}{ \omega^2 L_2} V_2 +\frac{1}{ \omega^2 L_2} V_{N+1}\right]
\end{aligned}\label{eq:nodal_I1}
\end{equation}

\begin{equation}
\begin{aligned}
I_2 &= -\frac{V_1-V_2}{i\omega L_2} +\frac{V_2-V_3}{i\omega L_1} + \frac{V_2-V_{N+2}}{i\omega L_2} + V_2 \left( \frac{1}{i \omega L_1} + i \omega C\right) \\[1.2ex]
&= -\frac{1}{i\omega L_2}V_1 + \left(\frac{2}{i \omega L_1} + \frac{2}{i \omega L_2} + i \omega C \right)V_2 -\frac{1}{i\omega L_1}V_3 -\frac{1}{i\omega L_2}V_{N+2}\\[1.2ex]
&=i\omega\:\left[ \frac{1}{\omega^2 L_2}V_1 + \left(\frac{2}{ \omega^2 L_1} + \frac{2}{ \omega^2 L_2} + C \right)V_2 + \frac{1}{\omega^2 L_1}V_3 + \frac{1}{\omega^2 L_2}V_{N+2}\right]
\end{aligned}\label{eq:nodal_I2}
\end{equation}\\

Similarly, for nodes $j$ and $j+1$ highlighted in Fig.~\ref{Fig_circuit_NN_finite_w_currents}

\begin{equation}
\begin{aligned}
I_{j} &= -\frac{V_{j-N}-V_{j}}{i \omega L_1} -\frac{V_{j-1}-V_{j}}{i \omega L_2} + \frac{V_{j}-V_{j+1}}{i \omega L_1} + \frac{V_{j}-V_{j+N}}{i \omega L_2} + i \omega C\: V_{j}\\[1.2ex]
&= -\frac{V_{j-N}}{i \omega L_1} -\frac{V_{j-1}}{i \omega L_2} + \left[\frac{2}{i \omega L_1} + \frac{2}{i \omega L_2} + i \omega C \right]V_{j} -\frac{V_{j+1}}{i \omega L_1} -\frac{V_{j+N}}{i \omega L_2} \\[1.2ex]
&=i\omega\:\left[ \frac{1}{\omega^2 L_1} V_{j-N} + \frac{1}{ \omega^2 L_2} V_{j-1} +  \left(C-\frac{2}{ \omega^2 L_1} + \frac{2}{ \omega^2 L_2} \right)V_{j} + \frac{1}{ \omega^2 L_1} V_{j+1} + \frac{1}{ \omega^2 L_2} V_{j+N}\right]
\end{aligned}\label{eq:nodal_j}
\end{equation}

\begin{equation}
\begin{aligned}
I_{j+1} &= -\frac{V_{j+1-N}-V_{j+1}}{i \omega L_1} -\frac{V_{j}-V_{j+1}}{i \omega L_1} + \frac{V_{j+1}-V_{j+2}}{i \omega L_2} + \frac{V_{j+1}-V_{j+1+N}}{i \omega L_2} + i \omega C\: V_{j+1}\\[1.2ex]
&= -\frac{V_{j+1-N}}{i \omega L_1} -\frac{V_{j}}{i \omega L_1} + \left[\frac{2}{i \omega L_1} + \frac{2}{i \omega L_2} + i \omega C \right]V_{j+1} -\frac{V_{j+2}}{i \omega L_2} -\frac{V_{j+1+N}}{i \omega L_2}\\[1.2ex]
&=i\omega\:\left[ \frac{1}{\omega^2 L_1} V_{j+1-N} + \frac{1}{ \omega^2 L_1} V_{j}  +  \left(C - \frac{2}{\omega^2 L_1} - \frac{2}{\omega^2 L_2} \right)V_{j+1} + \frac{1}{\omega^2 L_2} V_{j+2} + \frac{1}{\omega^2 L_2} V_{j+1+N} \right]
\end{aligned}\label{eq:nodal_j+1}
\end{equation}\\

The circuit Laplacian, $J(\omega)$, is an $N^2\times N^2$ matrix that can be expressed as $J(\omega)=D-A$, where $D$ is a diagonal matrix whose terms represent the total admittance from each node to ground and to the rest of the circuit, and the matrix $A$ is the admittance adjacency matrix, which is symmetric and has zeros along its diagonal:

\begin{equation}
\mathbf{J}(\omega) =
\begin{bmatrix}
d_1 & a_{12} & a_{13} & \cdots & a_{1N^2} \\
a_{12} & d_2 & a_{23} & \cdots &  a_{2N^2} \\
a_{13} & a_{23} & d_3 & \cdots & a_{3N^2} \\
\vdots & \vdots & \vdots & \ddots & \vdots \\
a_{1N^2} & a_{2N^2} & a_{3N^2} & \cdots & d_{N^{2}}
\end{bmatrix}
\end{equation}

where

\begin{equation}
d_i=C-\frac{2}{\omega^2 L_1} -\frac{2}{\omega^2 L_2}, \;\;\; i=1,\:2,\dots,\:N^2
\end{equation}

\begin{equation}
a_{ij} =
\begin{cases}
\frac{1}{\omega^2 L_k} & \text{if nodes } i,j \text{ are connected by inductor } L_k, k=1,\:2 \\
0 & \text{if nodes } i,j \text{ are not connected}
\end{cases}\end{equation}

At the characteristic frequency $\omega_0=\sqrt{\frac{2(L_1 + L_2)}{C L_1 L_2}}$, the diagonal terms of $J(\omega)$ cancel, leaving only the off-diagonal admittance contributions.

\clearpage
\subsection{Analytical derivation of corner modes}

\begin{figure}[htb]
\centering
\includegraphics[width=0.25\linewidth]{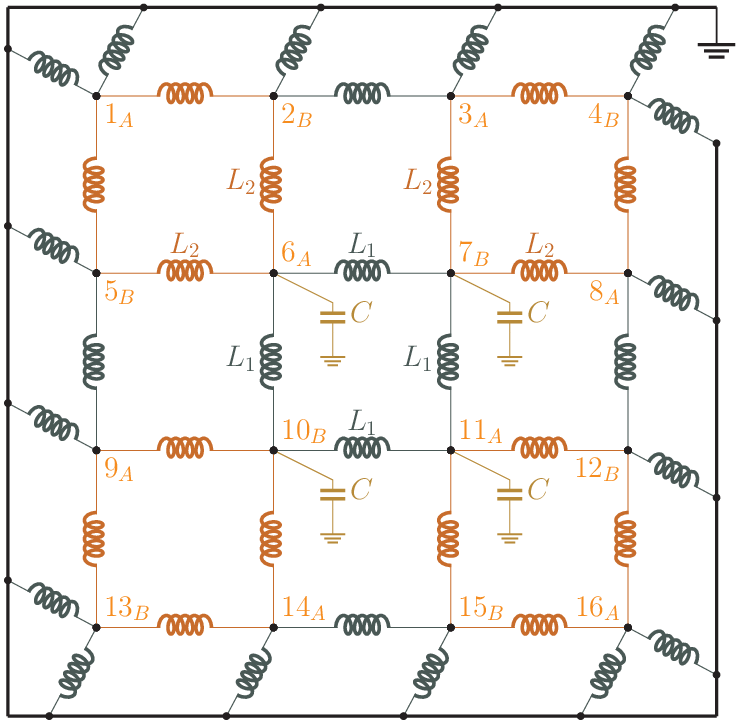}
\caption{2D SSH circuit with $4\times4$ node configuration. Inductors $L_1$ (green) and $L_2$ (brown) represent intracell and intercell couplings, respectively. All nodes, including edge and corner nodes, are grounded via capacitors $C$ (yellow), which are omitted for clarity. Node labels $1_i, \dots, 16_i$ are shown in orange, where the subscript $i \in \{A, \:B\}$ denotes the sublattice.  The circuit forms a bipartite lattice, where couplings occur only between nodes belonging to different sublattices.}
\label{Fig_circuit_4x4}
\end{figure}

This section exploits the chiral symmetry of the Laplacian, introduced previously, to identify zero-admittance states localized at the corners.  This analytical approach complements numerical simulations and provides deeper insight into the system's topological properties.  The circuit configuration with $4\times4$ nodes is shown in Fig.~\ref{Fig_circuit_4x4}, and its Laplacian $J_{4\times4}$ is given by:

\begin{equation}\label{eq:J_4x4}
\setlength{\arraycolsep}{6pt}        
\renewcommand{\arraystretch}{0.9}    
J_{4x4}(\lambda)= \frac{1}{{\omega_{0}}^2\:L_1}\:
\begin{pmatrix}
0 & \lambda & 0 & 0 & \lambda & 0 & 0 & 0 & 0 & 0 & 0 & 0 & 0 & 0 & 0 & 0 \\
\lambda & 0 & 1 & 0 & 0 & \lambda & 0 & 0 & 0 & 0 & 0 & 0 & 0 & 0 & 0 & 0 \\
0 & 1 & 0 & \lambda & 0 & 0 & \lambda & 0 & 0 & 0 & 0 & 0 & 0 & 0 & 0 & 0 \\
0 & 0 & \lambda & 0 & 0 & 0 & 0 & \lambda & 0 & 0 & 0 & 0 & 0 & 0 & 0 & 0 \\
\lambda & 0 & 0 & 0 & 0 & \lambda & 0 & 0 & 1 & 0 & 0 & 0 & 0 & 0 & 0 & 0 \\
0 & \lambda & 0 & 0 & \lambda & 0 & 1 & 0 & 0 & 1 & 0 & 0 & 0 & 0 & 0 & 0 \\
0 & 0 & \lambda & 0 & 0 & 1 & 0 & \lambda & 0 & 0 & 1 & 0 & 0 & 0 & 0 & 0 \\
0 & 0 & 0 & \lambda & 0 & 0 & \lambda & 0 & 0 & 0 & 0 & 1 & 0 & 0 & 0 & 0 \\
0 & 0 & 0 & 0 & 1 & 0 & 0 & 0 & 0 & \lambda & 0 & 0 & \lambda & 0 & 0 & 0 \\
0 & 0 & 0 & 0 & 0 & 1 & 0 & 0 & \lambda & 0 & 1 & 0 & 0 & \lambda & 0 & 0 \\
0 & 0 & 0 & 0 & 0 & 0 & 1 & 0 & 0 & 1 & 0 & \lambda & 0 & 0 & \lambda & 0 \\
0 & 0 & 0 & 0 & 0 & 0 & 0 & 1 & 0 & 0 & \lambda & 0 & 0 & 0 & 0 & \lambda \\
0 & 0 & 0 & 0 & 0 & 0 & 0 & 0 & \lambda & 0 & 0 & 0 & 0 & \lambda & 0 & 0 \\
0 & 0 & 0 & 0 & 0 & 0 & 0 & 0 & 0 & \lambda & 0 & 0 & \lambda & 0 & 1 & 0 \\
0 & 0 & 0 & 0 & 0 & 0 & 0 & 0 & 0 & 0 & \lambda & 0 & 0 & 1 & 0 & \lambda \\
0 & 0 & 0 & 0 & 0 & 0 & 0 & 0 & 0 & 0 & 0 & \lambda & 0 & 0 & \lambda & 0
\end{pmatrix}
\end{equation}

Due to chiral symmetry, the circuit Laplacian (analogous to the tight-binding Hamiltonian) can be expressed in a block-off-diagonal form when the nodes are reordered by sublattice. Specifically, if all nodes belonging to sublattice $A$ are grouped first, and those of sublattice $B$ second, the matrix $J_{4\times4}$ takes the form

\begin{equation}\label{eq:J_4x4_chiral}
J_{4\times4}=
\begin{pmatrix}
0 & Q\\
Q^{T} & 0
\end{pmatrix}
\end{equation}

where $Q$ is an $8\times8$ matrix containing the intracell and intercell couplings between sublattices $A$ and $B$. In this representation, all connectivity is encoded in the off-diagonal blocks. This structure leads to eigenvalues that occur in $\pm$ pairs and  zero-admittance modes that reside entirely on one sublattice.

Zero-admittance modes (eigenvalue zero) lie in the kernel of $Q$ or $Q^{T}$. Therefore, the problem reduces to calculating the kernel subspaces of these two blocks. For the $4\times 4$ lattice considered, both $\ker(Q)$ and $\ker(Q^{T}$ have dimension 2. The basis vectors (which are not unique) of these subspaces span the set of zero-admittance states, and any corner mode can be constructed as a linear combination of these basis vectors, chosen to maximize localization at a specific corner.

The eigenvalue problem $J_{4\times4}\psi=0$ becomes

\begin{align}
Q\psi_{B} &=0 \label{eq:psi_B}\\
Q^{T}\psi_{A} &=0 \label{eq:psi_A}
\end{align}

The kernel, $\ker(Q)$, is two-dimensional, spanned by the vectors: $\{ {\psi_{1}^{B}}, \: {\psi_{2}^{B}}  \}$. Similarly, the kernel $\ker(Q^{T})$ is spanned by the vectors: $\{ {\psi_{1}^{A}}, \: {\psi_{2}^{A}}  \}$. Therefore, the kernel of $Q$ corresponds to vectors $\psi_{B}$ (modes that live on sublattice B), and the kernel of $Q^{T}$ corresponds to vectors $\psi_{A}$ (modes that live on sublattice A).

These vectors, expressed in the original node ordering, where sublattices A and B alternates, $1_{A}, \: 2_{B}, \: 3_{A},\dots,\:16_{A}$ are 16-component states corresponding to the full lattice (Fig.~\ref{Fig_circuit_4x4} as reference):

\begin{equation}
{\psi_{1}}^{B} =
(0 \quad 0 \quad 0 \quad -1 \quad 0 \quad 0 \quad 1 \quad 0 \quad 0 \quad -1 \quad 0 \quad 0 \quad 1 \quad 0 \quad 0 \quad 0)
\label{eq:psi_1_B}
\end{equation}

\begin{equation}
{\psi_{2}}^{B} =
(0 \quad -1 \quad 0 \quad 0 \quad 1 \quad 0 \quad \frac{1}{\lambda} \quad 0 \quad 0 \quad -\frac{1}{\lambda} \quad 0 \quad -1 \quad 0 \quad 0 \quad 1 \quad 0)
\label{eq:psi_2_B}
\end{equation}

\begin{equation}
{\psi_{1}}^{A} =
(0 \quad 0 \quad -1 \quad 0 \quad 0 \quad \frac{1}{\lambda} \quad 0 \quad 1 \quad -1 \quad 0 \quad -\frac{1}{\lambda} \quad 0 \quad 0 \quad 1 \quad 0 \quad 0)
\label{eq:psi_1_A}
\end{equation}

\begin{equation}
{\psi_{2}}^{A} =
(-1 \quad 0 \quad 0 \quad 0 \quad 0 \quad 1 \quad 0 \quad 0 \quad 0 \quad 0 \quad -1 \quad 0 \quad 0 \quad 0 \quad 0 \quad 1)
\label{eq:psi_2_A}
\end{equation}

To construct a mode localized at a specific corner, the unit vector corresponding to that corner node is expressed as a linear combination of the kernel basis vectors. Solving for the coefficients yields the combination that maximizes the amplitude at that corner, and the resulting vector is then normalized.

This method yields analytical expressions for the corner-state wavefunctions corresponding to the circuit corners $\psi_{1_{A}}$, $\psi_{4_{B}}$, $\psi_{13_{B}}$, and $\psi_{16_{A}}$ (Fig.~\ref{Fig_circuit_4x4}) as functions of the coupling ratio $\lambda$, 

\begin{align}
\psi_{1_{A}} &= c_{1} \: {\psi_{1}}^{A} + c_{2} \: {\psi_{2}}^{A} \label{eq:psi_corner_1_A}\\
\psi_{16_{A}} &= -c_{1} \: {\psi_{1}}^{A} - c_{2} \: {\psi_{2}}^{A} \label{eq:psi_corner_16_A}\\
\psi_{4_{B}} &= c_{2} \: {\psi_{1}}^{B} + c_{1} \: {\psi_{2}}^{B} \label{eq:psi_corner_4_B}\\
\psi_{13_{B}} &= -c_{2} \: {\psi_{1}}^{B} - c_{1} \: {\psi_{2}}^{B} \label{eq:psi_corner_13_B}
\end{align}

where the coefficients $c_1$ and $c_2$ are given by

\begin{align}
c_1 = \frac{\lambda}{2(4\lambda^2+1)}\label{eq:c_1}\\
c_1 = \frac{\lambda}{2(4\lambda^2+1)}\label{eq:c_2}
\end{align}

This corner states live entirely on one sublattice ($A$ or $B$) and decay away from the corner as expected for localized modes. 

Analyzing the expressions for the corner-state wavefunctions in Eq.~\ref{eq:psi_corner_1_A} to Eq.~\ref{eq:psi_corner_13_B} reveals that the localization strength depends on the coupling ratio $\lambda$.  Figure 4 in the main manuscript shows the approximated analytical expressions for the limit $\lambda\ll1$.  Under these conditions, the amplitudes of the wavefunction decay rapidly away from the corner, consistent with an exponential decay profile, as expected for topologically protected corner modes in the nontrivial phase. Conversely, when $\lambda>1$ the corner localization is lost.

\begin{figure}[htb]
\centering
\includegraphics[width=0.55\linewidth]{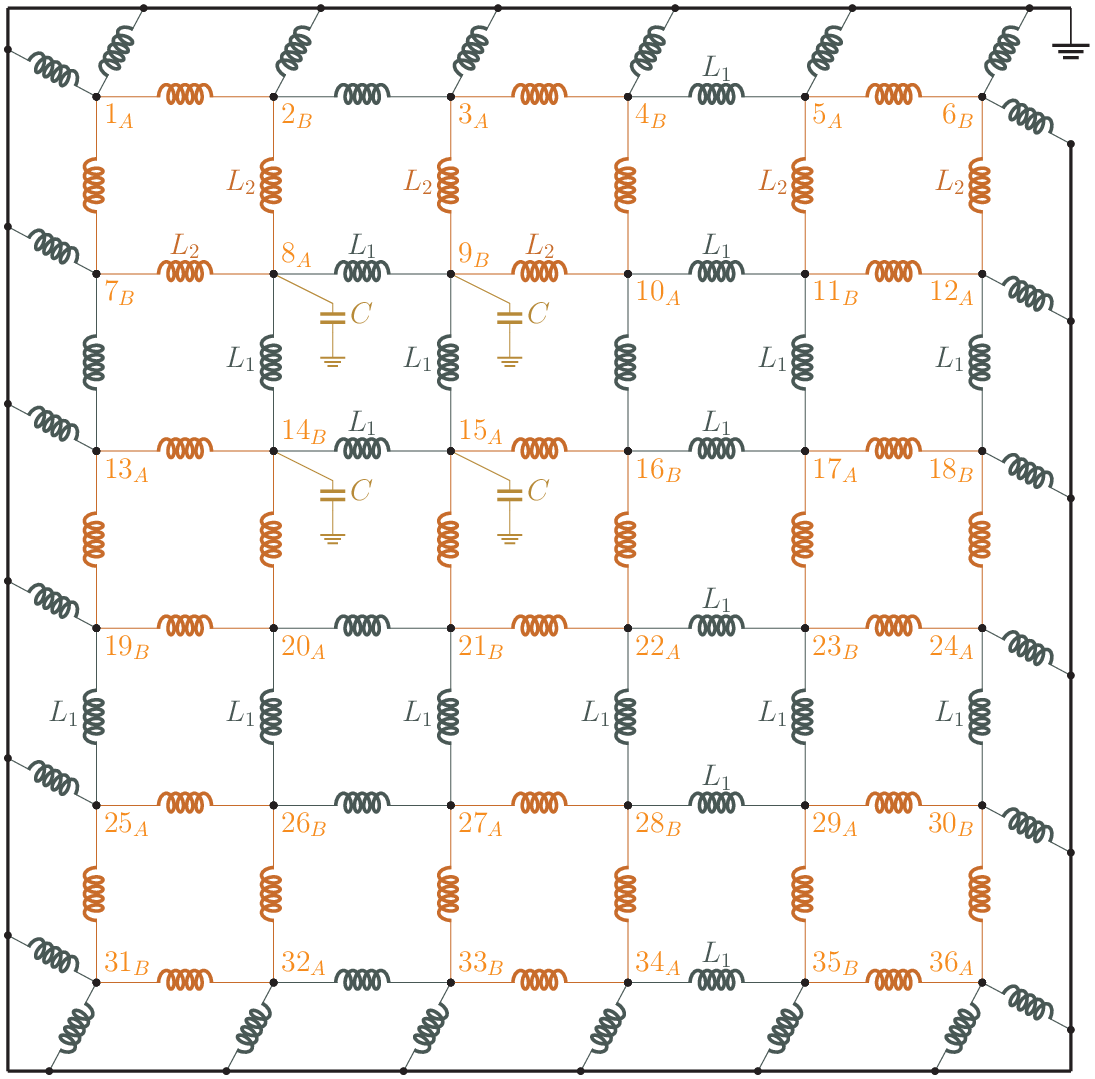}
\caption{2D SSH circuit with $6\times 6$ node configuration. Inductors $L_1$ (green) and $L_2$ (brown) represent intracell and intercell couplings, respectively. All nodes, including edge and corner nodes, are grounded via capacitors $C$ (yellow), which are omitted for clarity. Node labels $1_i, \dots, 36_i$ are shown in orange, where the subscript $i \in \{A, \:B\}$ denotes the sublattice.  The circuit forms a bipartite lattice, where couplings occur only between nodes belonging to different sublattices.}
\label{Fig_circuit_6x6}
\end{figure}

The same analytical approach can be extended to a larger lattice, such as a $6\times6$ layout (Fig.~\ref{Fig_circuit_6x6}). The circuit Laplacian retains its block-off-diagonal structure under chiral symmetry, and the zero-admittance modes are again determined by the kernels of the corresponding $Q$ and $Q^{T}$ blocks (Eq.~\ref{eq:J_4x4_chiral}). Although the dimension of these kernels increases with system size, the procedure for constructing corner-localized states (expressing the corner unit vector as a linear combination of the kernel basis vectors) remains essentially the same, allowing explicit analytical expressions for corner modes.

In this case, the kernels of $Q$ and $Q^T$ have dimension 3. 
Vectors of their respective basis for each sublattice (nodes $1_A, 3_A, 5_A, 8_A, \dots$ for sublattice $A$; nodes $2_B, 4_B, 6_B, 7_B \dots$ for sublattice $B$) are given by:

\begin{equation}\label{eq:psi_1_A_6x6}
{\psi_{1}}^{A} =
(
0 \quad 0 \: -1 \quad 0 \quad \frac{1}{\lambda} \quad 1 \quad 0 \quad -1 \quad -\frac{1}{\lambda} \quad \frac{1}{\lambda} \quad 1 \quad 0 \quad -1 \quad -\frac{1}{\lambda} \quad 0 \quad 1 \quad 0 \quad 0)
\end{equation}

\begin{equation}
{\psi_{2}}^{A} =
(0 \quad -1 \quad 0 \quad \frac{1}{\lambda} \quad 1 \quad 0 \quad -1 \quad -\frac{1}{\lambda} \quad -1 \quad 1 \quad \frac{1}{\lambda} \quad 1 \quad 0 \quad -1 \quad -\frac{1}{\lambda} \quad 0 \quad 1 \quad 0)
\label{eq:psi_2_A_6x6}
\end{equation}

\begin{equation}
{\psi_{3}}^{A} =
(-1 \quad 0 \quad 0 \quad 1 \quad 0 \quad 0 \quad 0 \quad -1 \quad 0 \quad 0 \quad 1 \quad 0 \quad 0 \quad 0 \quad -1 \quad 0 \quad 0 \quad 1)
\label{eq:psi_3_A_6x6}
\end{equation}

\begin{equation}
{\psi_{1}}^{B} =
(0 \quad 0 \quad -1 \quad 0 \quad 0 \quad  1 \quad 0 \quad -1 \quad 0 \quad 0 \quad 1 \quad 0 \quad -1 \quad 0 \quad 0 \quad 1 \quad 0 \quad 0 )
\end{equation}\label{eq:psi_1_B_6x6}

\begin{equation}
{\psi_{2}}^{B} =
(0 \quad -1 \quad 0 \quad 0 \quad 1 \quad \frac{1}{\lambda} \quad -1 \quad -\frac{1}{\lambda} \quad -1 \quad 1 \quad \frac{1}{\lambda} \quad 1 \quad -\frac{1}{\lambda} \quad -1 \quad 0 \quad 0 \quad 1 \quad 0)
\label{eq:psi_2_B_6x6}\\
\end{equation}

\begin{equation}
{\psi_{3}}^{B} =
(-1 \quad 0 \quad 0 \quad 1 \quad \frac{1}{\lambda} \quad 0 \quad -\frac{1}{\lambda} \quad -1 \quad 0 \quad 0 \quad 1 \quad \frac{1}{\lambda} \quad 0 \quad -\frac{1}{\lambda} \quad -1 \quad 0 \quad 0 \quad 1)
\label{eq:psi_3_B_6x6}
\end{equation}

The corner-state wavefunctions corresponding to the circuit corners $\psi_{1_{A}}$, $\psi_{6_{B}}$, $\psi_{31_{B}}$, and $\psi_{36_{A}}$ (Fig.~\ref{Fig_circuit_6x6}) can be expressesd as

\begin{align}
\psi_{1_{A}} &= c_{1} \: {\psi_{1}}^{A} + c_{2} \: {\psi_{2}}^{A} + c_{3} \: {\psi_{3}}^{A} \label{eq:psi_corner_1_A_6x6}\\
\psi_{36_{A}} &= -c_{1} \: {\psi_{1}}^{A} - c_{2} \: {\psi_{2}}^{A} - c_{3} \: {\psi_{3}}^{A} \label{eq:psi_corner_36_A_6x6}\\
\psi_{6_{B}} &= c_{3} \: {\psi_{1}}^{B} + c_{2} \: {\psi_{2}}^{B}  + c_{1} \: {\psi_{3}}^{B} \label{eq:psi_corner_6_B_6x6}\\
\psi_{31_{B}} &= -c_{3} \: {\psi_{1}}^{B} - c_{2} \: {\psi_{2}}^{B} -  c_{1} \: {\psi_{3}}^{B} \label{eq:psi_corner_31_B_6x6}
\end{align}

where the coefficients $c_1$, $c_2$, and $c_3$ are are defined as

\begin{align}
c_1 &= \frac{2\lambda^2 (\lambda^2 -1)}{32 \lambda^4 +13 \lambda^2 +4}\label{eq:c_1_6x6}\\[1.2ex]
c_2 &= \frac{\lambda (3\lambda^2 + 4)}{2(32 \lambda^4 +13 \lambda^2 +4)}\label{eq:c_2_6x6}\\[1.2ex]
c_3 &= -\frac{12\lambda^4 +5 \lambda^2 +4}{2(32 \lambda^4 +13 \lambda^2 +4)}\label{eq:c_3_6x6}
\end{align}

As an example, the amplitude of the corner mode $\psi_{1A}$ can be calculated at different nodes of sublattice $A$:

\begin{equation}
|\psi_{1A}(\mbox{node }1_A)| = \left| \frac{12\lambda^4 +5 \lambda^2 +4}{2(32 \lambda^4 +13 \lambda^2 +4)} \right| \xrightarrow{\:\lambda \ll 1 \:} \approx \frac{1}{2}
\end{equation}

\begin{equation}
|\psi_{1A}(\mbox{node }3_A)| = \left|-\frac{12\lambda^4 +5 \lambda^2 +4}{2(32 \lambda^4 +13 \lambda^2 +4)}\right| \xrightarrow{\:\lambda \ll 1 \:} \approx \frac{\lambda}{2}
\end{equation}

\begin{equation}
|\psi_{1A}(\mbox{node }5_A)| = \left|  -\frac{\lambda (3\lambda^2 + 4)}{2(32 \lambda^4 +13 \lambda^2 +4)}   \right| \xrightarrow{\:\lambda \ll 1 \:} \approx \frac{\lambda^2}{2}
\end{equation}

\begin{equation}
|\psi_{1A}(\mbox{node }8_A)| = \left| \frac{1}{\lambda} \: \frac{\lambda (3\lambda^2 + 4)}{2(32 \lambda^4 +13 \lambda^2 +4)} -\frac{12\lambda^4 +5 \lambda^2 +4}{2(32 \lambda^4 +13 \lambda^2 +4)}  \right| \xrightarrow{\:\lambda \ll 1 \:} \approx \frac{\lambda^2}{4}
\end{equation}

\begin{equation}
|\psi_{1A}(\mbox{node }10_A)| = \left| \frac{1}{\lambda} \: \frac{2\lambda^2 (\lambda^2 -1)}{32 \lambda^4 +13 \lambda^2 +4} + \frac{\lambda (3\lambda^2 + 4)}{2(32 \lambda^4 +13 \lambda^2 +4)} \right| \xrightarrow{\:\lambda \ll 1 \:} \approx \frac{7\lambda^3}{8}
\end{equation}

These approximations agree with the theoretical prediction of an exponentially decaying corner mode amplitude away from the corner.

\clearpage
\subsection{Implementation of disorder}

Disorder is introduced in the 2D SSH circuit by randomizing the inductance nominal values $L_1$ and $L_2$ of the links between nodes.  This disorder is applied only to the off-diagonal elements of the admittance matrix, which represent the couplings between nodes. Each inductance corresponding to links of the circuit's nodes is modified by a random factor:

\begin{equation}
L_{1,\text{disorder}}(i,j)=L_1 (1+\delta(2\cdot \mbox{rand}-1))
\end{equation}

\begin{equation}
L_{2,\text{disorder}}(i,j)=L_2 (1+\delta(2\cdot \mbox{rand}-1))
\end{equation}

where $\delta$ is the disorder strength and $\mbox{rand}$ generates a number uniformly in $[0,1]$. This introduces a uniform random variation for each link such that  

\begin{equation}
L_{1,\text{disorder}}(i,j) \in [L_1\cdot (1-\delta),\; L_1\cdot (1+\delta)]
\end{equation}

\begin{equation}
L_{2,\text{disorder}}(i,j) \in [L_2\cdot (1-\delta),\; L_2\cdot (1+\delta)]
\end{equation}

Therefore, each inductance is randomly scaled within $\pm \delta$ of its nominal value.  The matrix is symmetrized to ensure reciprocal connections.  This creates a uniform distribution of inductance values around the nominal value, simulating imperfections or variability in physical components.

In addition to off-diagonal disorder, diagonal disorder is introduced by randomizing the capacitance values associated with onsite terms (connections to ground) in the topoelectrical circuit. Each capacitor is scaled by a random factor within the range $[1-\delta, 1+\delta]$, where $\delta$ denotes the disorder strength:

\begin{equation}
C_{\text{disorder}}(i,i)=C (1+\delta(2\cdot \mbox{rand}-1))
\end{equation}

which results in a uniform distribution of capacitance values around the nominal value $C$:

\begin{equation}
C_{\text{disorder}}(i,i) \in [C\cdot (1-\delta),\; C\cdot (1+\delta)]
\end{equation}

This variation affects only the diagonal elements of the admittance matrix, simulating imperfections in the onsite grounding components.

\begin{figure}[htb]
 \includegraphics[keepaspectratio=true, width=0.45\linewidth]{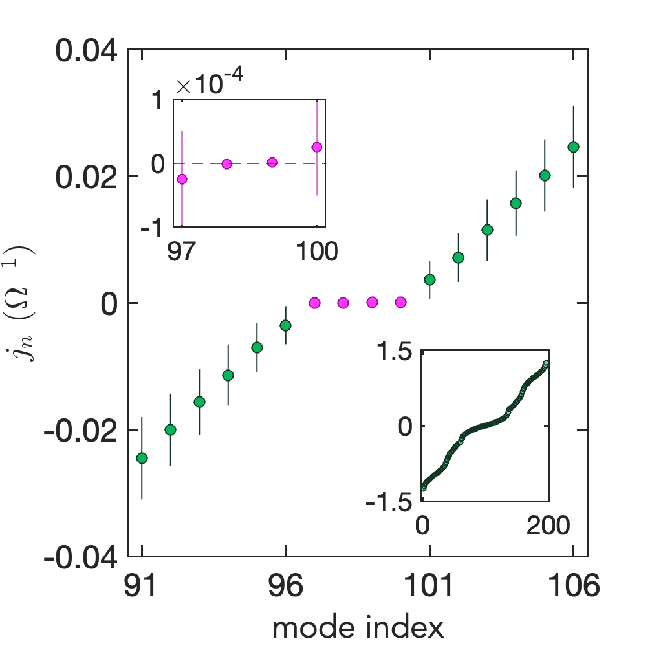}\label{Fig_SI_005}\caption{Numerically computed eigenvalues for the 2D SSH circuit comprising 196 nodes, under 40\% off-diagonal disorder introduced by randomly varying the inductance values of the $L_1$ and $L_2$ couplings. For each mode, the eigenvalue is averaged over 100 disorder simulations. Error bars represent the standard deviation across these simulations. Zero-admittance corner modes are highlighted (purple solid circles). Insets: enlarged view of the corner-localized states at zero-admittance (top left); full admittance spectrum (bottom right).}
\end{figure}

\clearpage

\section{2D SSH circuit with NNN interaction}

\subsection{\texorpdfstring{Corner modes for parameters $\alpha=0.1$, $\beta=1$}{Corner modes for parameters alpha=0.1, beta=1}}

\begin{figure}[htb]
 \includegraphics[keepaspectratio=true, width=\linewidth]{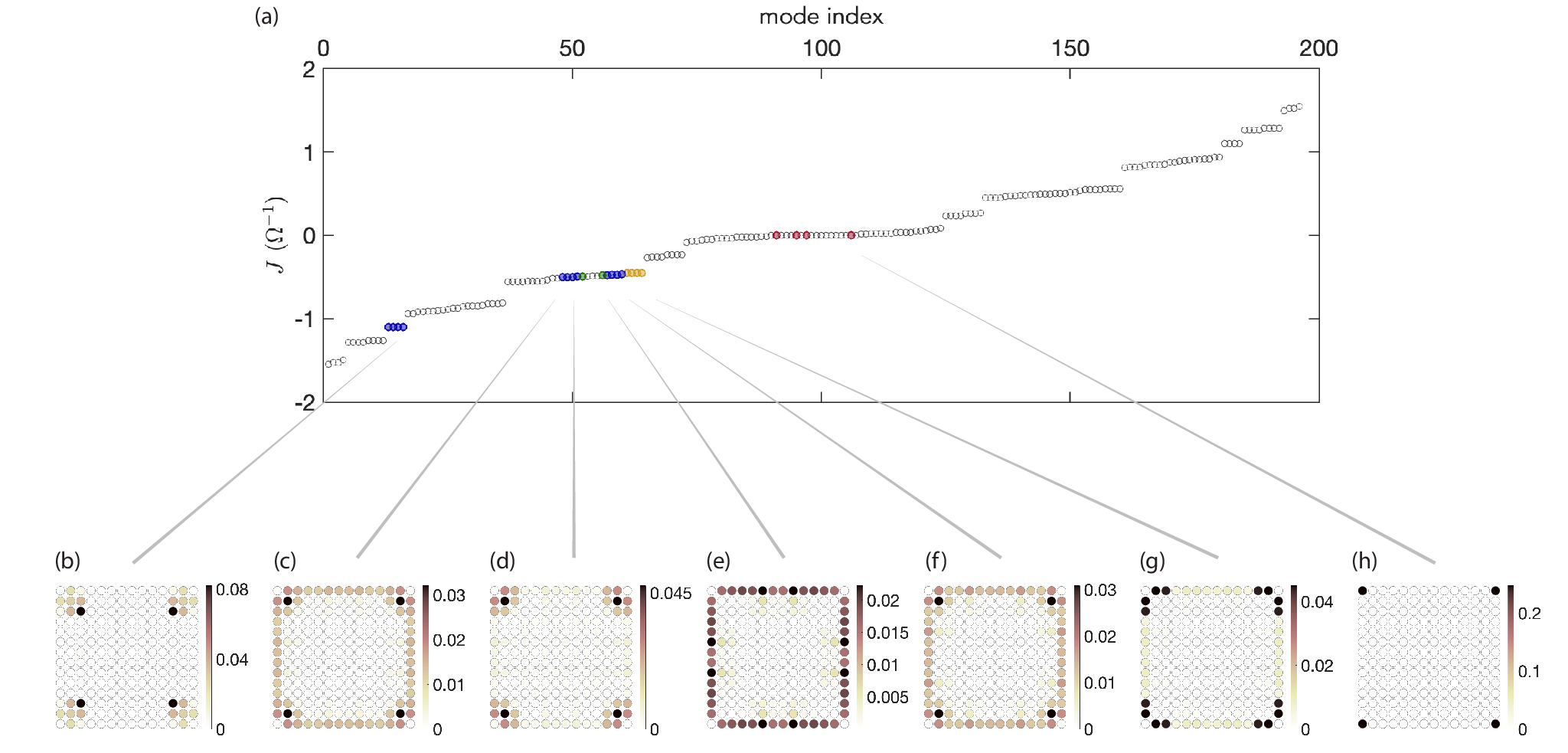}\caption{(a) Calculated eigenvalue spectra for a circuit with $14 \times 14$ sites including NNN coupling and parameters $\alpha=0.1$, $\beta=1$. Compared to the case with $\alpha=0.1$, $\beta=0.005$ shown in the main manuscript, modes localized at the corners of the first inner nested square of the lattice remain present (c, d, f) although some have acquired a pronounced edge-like character (c, f). Additionally, modes localized at the corners of the second inner nested square emerge (b). In contrast to the case with $\beta=0.005$, where edge modes are absent, edge modes emerge in this eigenvalue spectrum (e).  Corner states forming a V-shaped configuration that straddles the corner, referred to as type II in the main manuscript, are still observed (g). Corner states embedded in the continuum and pinned to zero admittance are preserved (h).}\label{Fig_SI_006}
\end{figure}